\documentclass[preprint,10pt]{elsarticle}
\usepackage{amsmath,amssymb}
\usepackage{mathtools}
\usepackage{bm}
\usepackage{bbold}
\usepackage{fullpage}
\usepackage{todonotes}
\usepackage{subfig}
\usepackage{graphicx}
\usepackage{multirow}

\newcommand{\protein}[1]{x_{#1}}
\newcommand{\prot}[1]{\delta x_{#1}}
\newcommand{\protnondim}[1]{\delta \tilde{x}_{#1}}
\newcommand{\y}[1]{y_{#1}}
\newcommand{\km}[3]{k^-_{{#1},#2#3}}
\newcommand{\kp}[3]{k^+_{#1#2,{#3}}}
\newcommand{\lam}[2]{\lambda_{#1#2}}

\newcommand{\T}{{\rm T}}

\newcommand{\lblock}[2]{L^{\rm{{#1}},\rm{{#2}}}}
\newcommand{\lblockb}[2]{\bm{L}^{\rm{{#1}},\rm{{#2}}}}
\newcommand{\eps}{\epsilon}
\newcommand{\noise}[1]{\eta_{#1}}
\newcommand{\order}{\mathcal{O}}
\newcommand{\timescale}[1]{\tau_{#1}}
\newcommand{\timeerror}{\Delta}

\newcommand{\bdx}{\bm{\delta x}}
\renewcommand{\sp}{s'} 
\newcommand{\spp}{s''} 
\newcommand{\Ns}{N^\text{s}}
\newcommand{\bdxs}{\bm{\delta x}^\text{s}}
\newcommand{\bdxb}{\bm{\delta x}^\text{b}}
\newcommand{\dxany}{\delta x}
\newcommand{\dx}{\delta x}
\newcommand{\dt}{\Delta t}
\newcommand{\del}{\delta}

\newcommand{\superss}{{\rm{s},\rm{s}}}
\newcommand{\supersb}{{\rm{s},\rm{b}}}
\newcommand{\superbs}{{\rm{b},\rm{s}}}
\newcommand{\superbb}{{\rm{b},\rm{b}}}
\newcommand{\supersss}{{\rm{ss},\rm{s}}}

\journal{Journal of Theoretical Biology}

\begin{document}

\begin{frontmatter}



\title{
Memory effects in biochemical networks as the natural counterpart of extrinsic noise}


\author[label1]{Katy J. Rubin}
\author[label2]{Katherine Lawler}
\author[label1]{Peter Sollich\corref{cor1}}
\author[label3]{Tony Ng}
\address[label1]{Department of Mathematics, King's College London, Strand, London, WC2R 2LS, UK}
\address[label2]{Institute for Mathematical and Molecular Biomedicine, King’s College London, Hodgkin Building, London, SE1 1UL, UK}
\address[label3]{Richard Dimbleby Department of Cancer Research, Division of Cancer Studies, King’s College London, London, SE1 1UL, UK and UCL Cancer Institute, Paul O'Gorman Building, University College London, London, WC1E 6DD, UK}
\cortext[cor1]{Department of Mathematics, King's College London, Strand, London, WC2R 2LS, UK. Tel:+44 20 78482875.\\ peter.sollich@kcl.ac.uk}
\begin{abstract}

We show that in the generic situation where a biological network,
e.g.\ a protein interaction network, is in fact a subnetwork embedded
in a larger ``bulk'' network, the presence of the bulk causes not just
extrinsic noise but also {\em memory effects}. This means that the
dynamics of the subnetwork will depend not only on its present state,
but also its past. We use projection techniques to get explicit
expressions for the {\em memory functions} that encode such memory
effects, for generic protein interaction networks involving binary and
unary reactions such as complex formation and phosphorylation,
respectively. Remarkably, in the limit of low intrinsic copy-number
noise such expressions can be obtained even for nonlinear dependences
on the past. We illustrate the method with examples from a protein
interaction network around epidermal growth factor receptor (EGFR),
which is relevant to cancer signalling. These examples demonstrate that inclusion
of memory terms is not only important conceptually but also leads to
substantially higher quantitative accuracy in the predicted subnetwork
dynamics.

\end{abstract}


\end{frontmatter}


\section{Introduction}
Biological networks are often complex and models are required to try
and understand their behaviour \cite{Bhalla2003}. This has stimulated an ongoing research effort into the construction of reduced models that allow
one to focus on subnetworks of a larger system. Such subnetworks may carry
out biologically important functions, or be of interest because they capture parts of the system where there is 
less uncertainty in the network structure or dynamical 
parameters such as reaction rates. 
The example network considered here is epidermal growth factor receptor (EGFR) signalling, which is a relatively small and well-studied network \citep{Kholodenko99} and contains a number of subnetworks, such as Src homology and collagen domain protein (Shc) and Shc-interacting proteins. 
An
understanding of the properties of such subnetworks can then in turn be used to help rationalise the behaviour of a larger
network \cite{Ackermann2012, Conradi2007, Shojaie2010}.

\begin{figure}[!ht]
  \centering
  \subfloat[]{\includegraphics[scale=1]{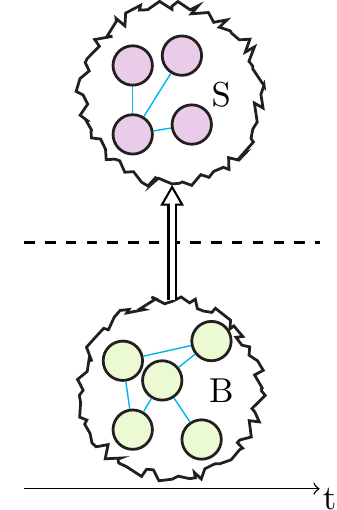}}\qquad
  \subfloat[]{\includegraphics[scale=1]{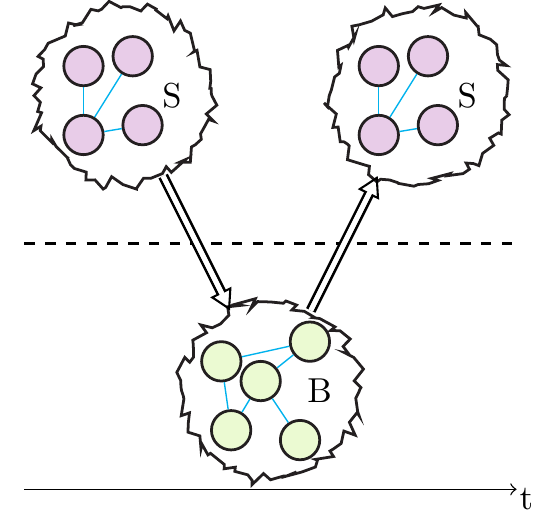}}
  \caption{Extrinsic noise versus memory. (a) Extrinsic noise on the subnetwork S arises from fluctuations of the bulk B that are uncontrolled and generally uncontrollable via experimental conditions. (b) Memory effects arise because the behaviour of S in the past will generically influence B, and this effect will feed back to S at a later time: the time evolution of S depends on its own past. 
  }
  \label{fig:memorypic}
\end{figure}

The above considerations motivate the analysis of subnetwork dynamics by
model reduction, where one starts from a description of a large
network and reduces this to an effective description of the
subnetwork. Further motivation comes from the fact that almost any
biological network that we choose to model is incomplete, and in reality is a
subnetwork embedded in a larger ``bulk'' network. It is then important
to understand what, in principle, is the appropriate way of describing
the dynamics in such a subnetwork. This is the aim of this paper, and
our main result is that such a description must in principle always
involve memory effects in addition to the well-studied extrinsic noise
caused by the presence of the bulk \cite{Swain2002, Paulsson2004}. 
We focus in our analysis on the
specific example of protein interaction networks with unary and binary
reactions, but expect that our qualitative conclusions are rather
general, as suggested by the generic nature of the intuitive
explanation of memory effects: the state of the subnetwork in the past
will influence the bulk, and this will feed back into the subnetwork
dynamics in the present (Fig.~\ref{fig:memorypic}).

We apply the method to investigate the dynamics of a subnetwork model
of epidermal growth factor signalling \cite{Normanno2006}. We show that the subnetwork dynamics, in the presence of Shc and Shc-interacting proteins, are more accurately
modelled by including memory terms originating from the Shc-centred bulk network in which the subnetwork is embedded. The models we use
obey conservation laws so that no increased gene expression or
destabilisation is incorporated. The analysis thus serves as a first
step towards quantitative modelling of experimentally tractable
perturbations and observable responses of both time courses and steady
state concentrations \cite{Rubin2014}, which may include signalling pathways with multiple ligands such as the ErbB signalling network \cite{Birtwistle2007}.

There is a substantial literature on methods of model reduction that
attempt to simplify an initial large model down to a subnetwork
description. The aim is to do this
whilst retaining the main features of the behaviour of the
original system \cite{Okino1998, Radulescu2012}. These methods are
often based on (a) sensitivity analysis, (b) timescale separation, (c) splitting the system
into modules or (d) lumping together components to obtain a smaller number
of parameters or variables. In most 
of these approaches,
it is assumed that the subnetwork can be freely chosen to make the
model reduction most effective. We consider the more difficult task of
finding a reduced description for a subnetwork that is fixed in
advance, e.g.\ because of its relevance to the overall biological
question being asked, or by experimental constraints on which 
molecular species can feasibly be monitored.

Sensitivity analysis tries to determine which 
molecular species are insignificant to the dynamic system of interest 
\cite{Huang2010}. A parameter is classified as insignificant if it has
a low sensitivity, in that its precise value does not have a large effect on
the concentrations of the rest of the species in the network.
Low sensitivity parameters are then eliminated or replaced by a
smaller number of effective species. 
However, sometimes it is necessary to keep a low sensitivity parameter to
ensure the results are biologically valid. 

Timescale separation techniques are used to focus on the species that
contribute most to the long-time dynamics of a system, by removing
molecular species whose dynamics takes place on much shorter
timescales. This is reasonable because biochemical processes occur on a range of
timescales; changes in gene expression levels, for example, may take place over hours whereas
protein signalling takes seconds.
Timescale separation approaches have been used by e.g.\ \citet{Gardiner1984}
and \citet{Thomas2012}, with the subnetwork then containing all the
slow molecular species and the bulk the fast ones. 
Thus, while these authors used projection techniques as we do,
memory effects did not arise: they become negligible if the bulk is
fast enough to respond effectively instantaneously -- on the timescale
of the subnetwork dynamics -- to the state of the subnetwork. Here we consider signalling networks where the timescales
of the dynamics of the subnetwork and the bulk are
comparable, so that timescale separation methods are not directly applicable. 

Another way to reduce the system is to split it into modules where
each module has a different function and a limited number of interactions with
the other modules \cite{Hartwell1999}. \citet{Conzelmann2004} apply
dimensional reduction to the modules so that the modules have reduced
complexity but show similar input and output behaviour. 

Lumping together variables with similar features also allows one to reduce
the size of a model \cite{Sunnaker2011,Conzelmann2004};
however, lumping components together may make it difficult to interpret
the results because the lumped variables may not retain their
original meaning.  Similarly 
\citet{Liebermeister2005} reduce the bulk surrounding a chosen
subnetwork, whilst the subnetwork is kept in its original form. As one
might expect, accounting for the bulk in this way, i.e.\ considering
the environment surrounding the subnetwork, yields a reduced model
that is more accurate than modelling just the isolated subnetwork. Our
work extends this result by 
showing that the inclusion of memory effects arising from the bulk gives a
significantly more
accurate description of the subnetwork dynamics.
\citet{Apri2012} remove or modify reactions and parameters based on
their effect on the output behaviour of the system. They consider which
parameters can be removed or lumped together to obtain output data
correct to within a certain tolerance. Although no detailed prior biological
knowledge of the system is needed, there must be some qualitative
understanding of the system dynamics to ensure no species which are
generally considered to be an important part of the network dynamics
are
removed. 

\begin{figure}[p!]
  \centering
  \subfloat[]{\label{fig:reactionfig}\includegraphics[scale=1.5]{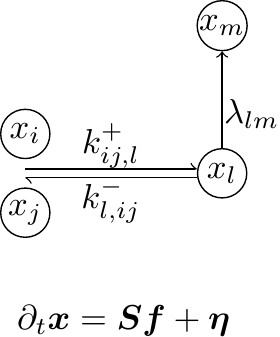}}\qquad
  \subfloat[]{\label{fig:subbulkpic}\includegraphics[scale=1]{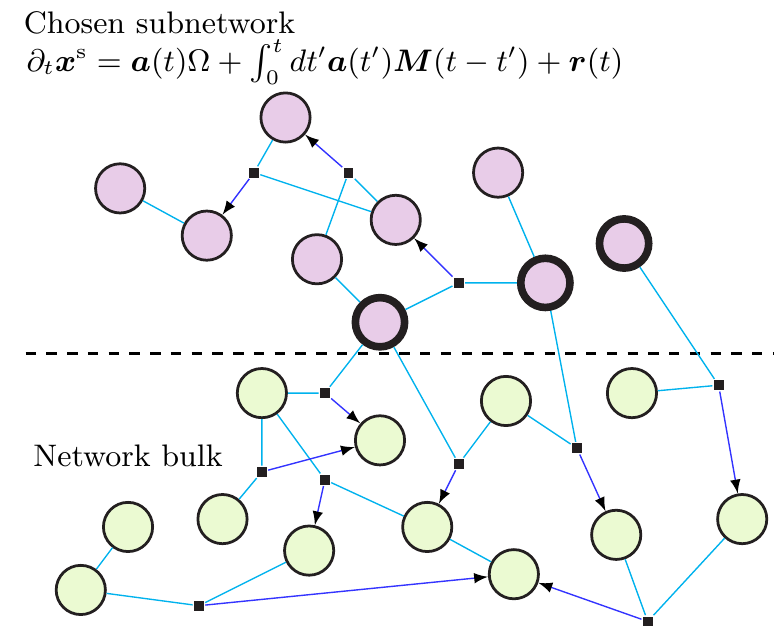}}\\
  \subfloat[]{\label{fig:amppics}\includegraphics[scale=0.65]{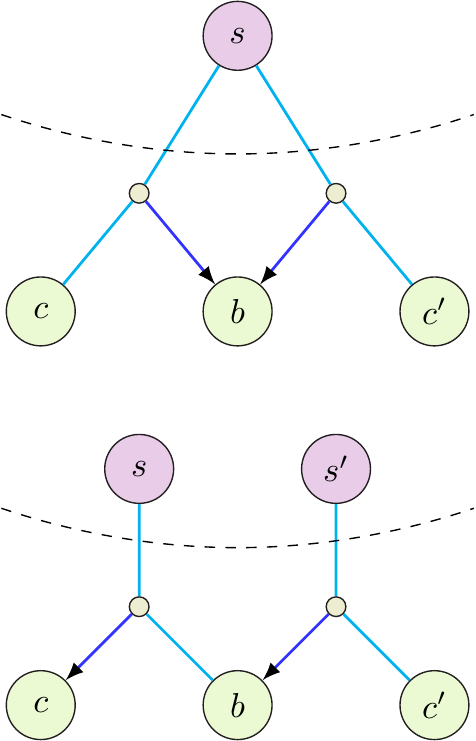}}\qquad
  \subfloat[]{\label{fig:boundarypic}\includegraphics[scale=0.6]{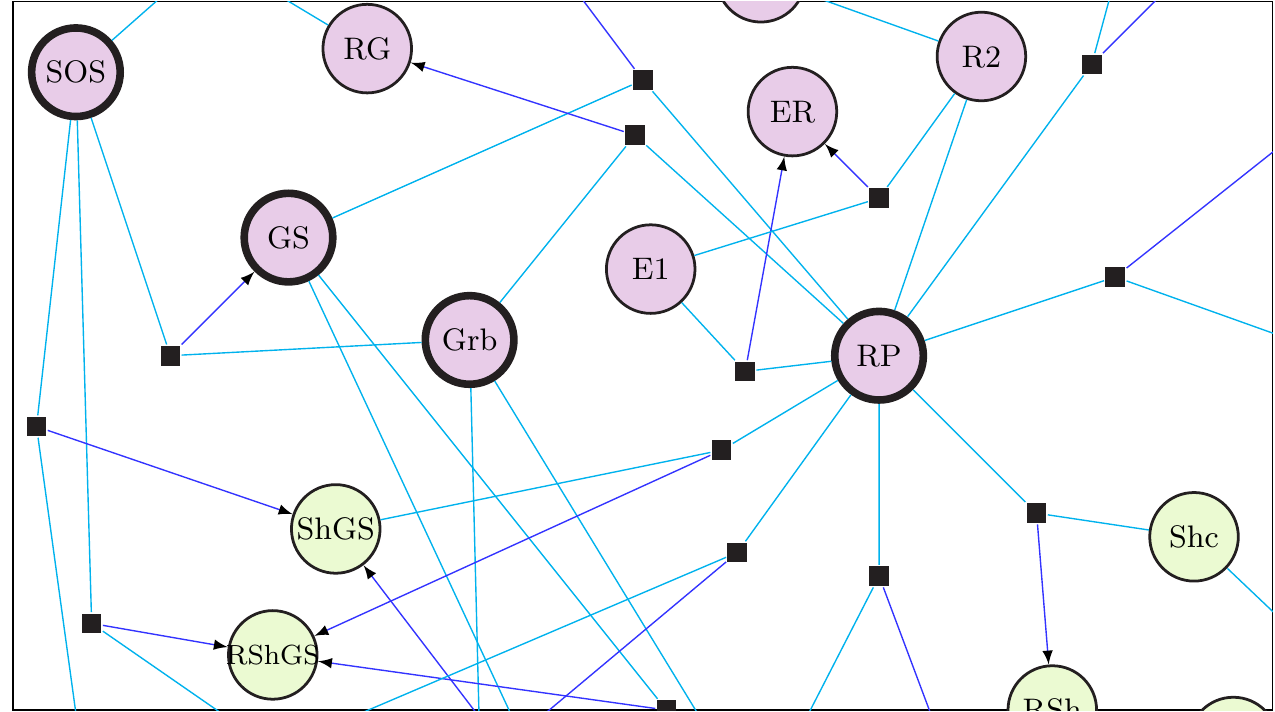}}
%
\caption{(a) Description of protein interaction networks: we use mass
  action kinetics ($x_i$, concentration of species $i$ = particle
  number per unit volume) with rates of binary complex formation and
  dissociation ($\kp{i}{j}{l}, \km{l}{i}{j}$) and unary transformation
  ($\lambda_{lm}$). Reaction equations can be written in terms of the
  stoichiometry matrix ${\bf S}$ and reaction flux vector ${\bf f}$
(Sec.~\ref{sec:reactioneqns}). Copy number fluctuations add intrinsic noise
${\bm{\eta}}$  with strength $\epsilon = 1/V$, the inverse volume of
the system (Sec.~\ref{sec:stochdynamics}).
 (b) 
Factor graph representation: a square denotes an interaction, an arrow
points to the resulting complex. The network is divided into the
subnetwork of interest (purple nodes, upper) and the ``bulk'' (green
nodes, lower).
The time evolution of concentrations $\mathbf{x}$ within the
subnetwork is described by projected equations
(Sec.~\ref{sec:projection_method}), where $\bm{a}(t)$ includes
$\bm{x}^\text{s}$ and its products (Sec.~\ref{sec:choiceofobs}). The rate matrix $\mathbf{\Omega}$
describes contributions that are local in time. Memory functions $\mathbf{M}(t)$
determine how strongly past values of $\mathbf{x}$ affect the present
rates of change $\partial_t x_i(t)$ (Secs.~\ref{sec:memory_overview},
\ref{sec:lindynamics}, \ref{sec:nonlindynamics}); the random force $\mathbf{r}(t)$
represents extrinsic noise. Memory terms feature the history of
species on the boundary of the subnetwork only (bold nodes,
Sec.~\ref{sec:boundarystructure}).
 (c) Exemplar subnetwork-bulk interactions for the calculation and
 properties of memory functions (Secs.~\ref{sec:amplitudes}). In a larger subnetwork, a memory function
 resulting from multiple reactions
 may be decomposed into ``source'' and ``receiver'' channels using these
 simple interaction structures (Sec.~\ref{sec:timescales_channels}).
 (d) Application to a model of the EGFR signalling pathway
 (Secs.~\ref{sec:EGFR_setup}, \ref{sec:memproperties}; see
 Fig.~\ref{fig:egfrnetwork} for full network). The bulk (green nodes) is chosen to be Shc and
 complexes containing Shc. The boundary species appearing in the
 memory terms are SOS, GS, Grb  
 and
 RP (bold). Quantitative comparisons shows that accurate modelling of
 subnetwork time courses requires the inclusion of memory terms
%
(Sec.~\ref{sec:quanttests}).
\label{fig:overview}
}
\end{figure}

Our approach starts from kinetic equations for the concentrations of a
set of molecular species in a large protein interaction network, allowing
for small amounts of intrinsic noise caused by fluctuations in the
copy number of each species as shown in Fig.~\ref{fig:reactionfig}.
We then use a projection operator formalism to obtain a set of dynamical
equations for selected variables from the network, which define the
chosen subnetwork. This approach
retains information from the remainder of the larger network, i.e.\
the bulk, and allows us to obtain a
reduced set of equations for the subnetwork (Fig.~\ref{fig:subbulkpic}). These 
%
%
projected equations contain extrinsic noise arising from the bulk dynamics as expected, 
but crucially the noise is accompanied
by memory terms (Fig.~\ref{fig:memorypic}). The memory terms are represented mathematically as
integrals over the past history of the subnetwork, modulated by {\em
  memory functions}. These are the focus of our analysis. 
%
%
In Section~\ref{sec:projection} we explain the projection approach and
how it can be applied to protein interaction networks. We also to
illustrate the method with a simple example that already captures some
general properties of memory functions (Fig.~\ref{fig:amppics}).  Next, in
Section~\ref{sec:genmem} we obtain closed-form expressions for memory
functions in protein interaction network dynamics and discuss and
illustrate some of their properties, e.g.\ the amplitudes and what
they us about reactions between the subnetwork and the bulk.
Finally in Section~\ref{sec:EGFR} we apply our approach to the EGFR
protein signalling network for short-term signalling from \citet{Kholodenko99} and study 
the memory functions for a chosen subnetwork (Fig.~\ref{fig:boundarypic}). We analyse
the dominant contributions to the memory functions, and show that our
projected equations with memory give a significantly more accurate
description of the subnetwork dynamics than can be obtained without memory.

This paper makes two main contributions. The first is a demonstration
of the conceptual and quantitative need to include memory terms in the
description of generic subnetwork dynamics. The second is of a
more technical nature, namely the derivation of closed-form memory functions for
the full nonlinear dynamics of protein interaction networks. To
reflect this contribution, and because the application of projection
methods to derive memory effects in biological networks is novel, we
describe the calculations in some detail. This is done in
Sec.~\ref{sec:lindynamics} for dynamics linearised around a fixed
point and then for the full nonlinear dynamics in
Sec.~\ref{sec:nonlindynamics}. Readers more interested in the conceptual
aspects and applications of our work might wish to skip these
sections. A graphical overview of the content of the paper is given in
Fig.~\ref{fig:overview}.


\section{Projection 
}
\label{sec:projection}
\subsection{Reaction equations}
\label{sec:reactioneqns}
We consider a protein interaction network described using mass action
kinetics.  The molecular reactions can be either binary or unary. In a
binary reaction two molecules react to form a molecule of a different
species (complex formation); the reverse process is the dissociation
of a complex 
into two molecules. In a unary reaction, one species transforms into
another via a conformational change like phosphorylation.  In our
setup we do not restrict the nature of the molecules that come
together in a binary reaction, and in particular we include the
possibility that a complex formed in some initial binary reaction may
react again with another molecule to form a higher order complex.  As
a convenient notational shorthand we nevertheless refer generically to
the two molecules that join together in a binary reaction as
``proteins'', and to the molecule that is formed as a complex.

The deterministic reaction equations for such a protein interaction
network containing $N$ molecular species can be written in the
form 
\begin{equation}
\label{deterministiceqns}
\begin{split}
\frac{\partial}{\partial t}\protein{i}&= \sum_{j(\neq i),l}
\left(\km{l}{i}{j}\protein{l}-\kp{i}{j}{l}
\protein{i}\protein{j}\right)+ \frac{1}{2}\sum_{j\neq l}\left(\kp{j}{l}{i}
\protein{j}\protein{l}- \km{i}{j}{l}\protein{i}\right)\\
&\quad + \sum_{l}
\left(2\km{l}{i}{i}\protein{l}-\kp{i}{i}{l}
\protein{i}\protein{i}\right) + \sum_{j}\left(\frac{1}{2}\kp{j}{j}{i}
\protein{j}\protein{j} -  \km{i}{j}{j}\protein{i}\right)\\
&\quad +\sum_j\left(\lam{j}{i}
\protein{j}-\lam{i}{j}\protein{i}\right)
\end{split}
\end{equation}
where $\protein{i}$ is the concentration of species $i$. In our notation we follow to a large extent the paper by \citet{Coolen2009}, which
presented an average-case analysis using generating functionals of the
dynamics in large protein interaction networks. We denote by
$\kp{i}{j}{l}$ the rate of formation of complex $l$ from proteins $i$
and $j$, and by $\km{l}{i}{j}$ the rate for the reverse process of
dissociation of complex $l$ into proteins $i$ and
$j$. 
To avoid ordering restrictions on the protein indices we set
$\kp{i}{j}{l} = \kp{j}{i}{l}$ and $\km{l}{i}{j} = \km{l}{j}{i}$. The
factor of 1/2 in the first line above is then needed to avoid double
counting of reactions of two different molecular species. The second
line relates to homodimer formation and dissociation, where two
proteins of the same species $i$ react. The extra factor of 2 arises
because dissociation of a homodimer $l$ creates two molecules of
species
$i$. 
The factor $1/2$ in the term describing formation of $i$ from two
molecules of species $j$ represents the reduction in number of
possible reaction pairs, compared to the case of formation of a
heterodimer where the two reacting species are different. The unit
prefactor of the $\kp{i}{i}{l}$ term arises as the combination of
these two effects. Finally, the last line of
\eqref{deterministiceqns} accounts for unary reactions, with
$\lam{i}{j}$ the rate of species $i$ changing into species $j$.

The reaction equations \eqref{deterministiceqns} can be written in
terms of a stoichiometry matrix and vector of reaction fluxes. Let the
number of
reactions with nonzero rates be $R$, where each forwards and backwards reaction is
counted separately. Then the stoichiometry matrix $\bm{S}$
is made up of integers $S_{i\mu}$, with $i = 1,2,\ldots,N$ and $\mu =
1,2,\ldots,R$. Each $S_{i\mu}$ records by how much the molecule count
of species $i$ changes in reaction $\mu$. Specifically, $S_{i\mu}$ is $-1$ if the
molecular species $\protein{i}$ is a reactant in reaction $\mu$ and
$+1$ if it is a reaction product. For homodimer reactions one
correspondingly has $S_{i\mu}=\pm 2$ when two molecules of species $i$
are used up or produced.

The vector of reaction fluxes $\bm{f}$ has entries
$f_\mu$ that give the reaction rate of reaction $\mu$ multiplied by the
concentrations of the proteins involved in that reaction. For example
the flux related to the formation of a complex $x_i$ from proteins $j$
and $l$ is $\kp{j}{l}{i}\protein{j}\protein{l}$ and the reverse reaction flux is
$\km{i}{j}{l}\protein{i}$. 
 For the conformational change of
protein $i$ to protein $j$ the forward reaction has reaction flux
$\lam{i}{j}\protein{i}$ and the reverse reaction flux is
$\lam{j}{i}\protein{j}$. 

With the stoichiometry matrix $\bm{S}$ and reaction flux vector
$\bm{f}$ defined as above, the reaction equations
\eqref{deterministiceqns} can be written in the compact form $\partial_t\protein{i} =
\sum_{\mu} S_{i\mu}f_{\mu}$. One benefit of this formulation is that
it shows transparently how conservation laws arise, where the sum of a
number of concentrations is constant in time. Quantitatively,
the number of conservation laws is given by the dimension of the left
nullspace of $\bm{S}$. 
If this nullspace is spanned by the (column) vectors 
$\bm{e}^{(a)}$, $a=1,2,\ldots$ then each such vector obeys
$\bm{e}^{(a)^{\T}}\bm{S}=0$. Accordingly the quantity
$\sum_i\protein{i}e^{(a)}_i$ is conserved:
$\partial_t\sum_i\protein{i}e^{(a)}_i = \bm{e}^{(a)^{\T}}\bm{S}\bm{f}= 0$.

\subsection{Stochastic Dynamics}
\label{sec:stochdynamics}
The deterministic reaction equations \eqref{deterministiceqns} apply
in the case where the number of molecules of each species, $x_iV$ in a
reaction compartment of volume $V$, is large enough so that
stochastic fluctuations around the mean value can be neglected. In
reality such copy number fluctuations are always present because the
number of molecules of any species is discrete, and when it changes over time
it does so due to elementary reactions that take place stochastically. The
relative size of the fluctuations in any $x_i$ will be of order
$1/\sqrt{x_i V}$, because any change in $x_i$ results from the cumulative
effect of many reactions and the number of reactions occurring within any fixed time interval
grows lineary with $V$.

We therefore next describe the stochastic extension of
\eqref{deterministiceqns} to the case of small copy number
fluctuations. The inverse volume of the system, $\epsilon=1/V$, will
be used to characterize the strength of this intrinsic noise.
We note that such a stochastic description is also important for our
use of the projection operator formalism \cite{Mori1965} to derive
subnetwork dynamical equations, as this approach starts from the time
evolution of a probability distribution over states of the network.

For small $\epsilon$, the appropriate stochastic version of
\eqref{deterministiceqns} is a Fokker-Planck equation for the time
evolution of the probability density $P(\bm{x},t)$. Truncating a
Kramers-Moyal expansion \cite{Gardiner09} after the first order in
$\epsilon$, this equation can be written in terms of 
the stoichiometry
matrix, $\bm{S}$, and reaction flux vector, $\bm{f}$, as
\begin{equation}
\label{eq:FP}
 \frac{\partial P(\bm{x},t)}{\partial t} = 
-\frac{\partial}{\partial \bm{x}}\left(\bm{Sf} P\right) 
  + \frac{\eps}{2}\frac{\partial^2}{\partial \bm{x}^2}
\left(\bm{BB}^{\T}P \right) =  \mathcal{L}^{\T}P(\bm{x},t)
\end{equation}
where
\begin{equation}
\bm{BB}^{\T} = \bm{S}\,\text{diag}\left(\bm{f}\right)\bm{S}^{\T}
\label{eq:BB_defn}
\end{equation}
and $\eps = {1}/{V}$ is the inverse reaction volume as before. This
formulation is useful for us as we can continue to describe each
species concentration with a single variable $x_i$, rather than having
to treat its mean time evolution and fluctuations separately as would be
done in a van Kampen system size expansion \cite{vanKampen07,Elf03}. 
Moreover, a recent
analysis \cite{Grima2011} shows that \eqref{eq:FP} is more accurate
than the van Kampen description, capturing the mean and variance of
the $x_i$ to higher order in $\epsilon$.

We will sometimes find it useful to switch from the above
Fokker-Planck description to the corresponding ``chemical Langevin
equation'' 
\cite{Gillespie2000}, which reads
\begin{equation}
  \frac{\partial}{\partial t}\bm{x} = \bm{Sf}\bm{x} + \bm{\eta}
\label{eq:CLE}
\end{equation}
The noise $\bm{\eta}$ 
is multiplicative as its statistics depend on $\bm{x}$; adopting the
Ito interpretation \cite{Gardiner09}, one has explicitly
$\langle\bm{\eta}(t)\bm{\eta}^{\T}(t')\rangle =
\eps\bm{BB}^{\T}\delta(t-t')$.

Returning to the Fokker-Planck equation \eqref{eq:FP}, the time evolution
it encodes can be thought of in terms of either an evolving
$P(\bm{x},t)$ or evolving observables $a(\bm{x},t)$ of the system; see e.g.\cite{Mori1965,Ritort03}. 
The time variation of $P(\bm{x},t)$ is the 
solution of \eqref{eq:FP}, which can be written formally as
$P(\bm{x},t) = e^{\mathcal{L}^{\T}t}P(\bm{x},0)$. Here
the operator exponential in $e^{\mathcal{L}^{\T}t}$ is defined as 
$e^{\mathcal{L}^{\T}t} =
\sum_{n=0}^\infty (\mathcal{L}^{\T}t)^n/n!$, requiring in
principle the application of 
successive powers of $\mathcal{L}^{\T}t$ to $P(\bm{x},0)$.


Now let $a(\bm{x})$ be an observable of the system, for example
one of the protein concentrations $\protein{i}$. Its time
average evolves in time as
\begin{equation}
\begin{split}
\langle a(t)\rangle &= \int d\bm{x}\ a(\bm{x}) P(\bm{x},t) =\int d\bm{x}\ a(\bm{x})
 e^{\mathcal{L}^{\T} t}P(\bm{x},0)
\end{split}
\label{eq:Pt_vs_at}
\end{equation}
Here we have introduced $\mathcal{L}$ as the adjoint operator to $\mathcal{L}^{\T}$, defined by
$\int d\bm{x}\, (\mathcal{L}a(\bm{x}))b(\bm{x}) = 
\int d\bm{x}\,a(\bm{x})\mathcal{L}^{\T}b(\bm{x})$. We have also defined 
\begin{equation}
a(\bm{x},t) = e^{\mathcal{L}t} a(\bm{x})
\label{eq:at}
\end{equation}
As the last equality of \eqref{eq:Pt_vs_at} shows, this is the average
value of $a$ at time $t$ conditional on the system initially being in state
$\bm{x}$. Its time evolution is given by \eqref{eq:at}, and
reads in differential form
\begin{equation}
\partial_t a(\bm{x},t) = \mathcal{L}a(\bm{x},t)
\label{eq:da_dt_is_La}
\end{equation}
with initial condition $a(\bm{x},0) = a(\bm{x})$.

Before we write down the adjoint Fokker-Planck operator,
we make a change of variables. 
For reasons explained further in Section~\ref{sec:projection} below, 
it will be useful to have variables with a mean value of zero in 
steady state. We
therefore define $\bm{x} = \bm{y} +
\bm{\delta x}$ where $\bm{y}$ is the mean steady state
value of $\bm{x}$, calculated as the fixed point of the mass-action
equations \eqref{deterministiceqns}, and $\bm{\delta x}$ is the
deviation away from this. Where the meaning is clear from the context,
we will then often use the shorthand ``concentration'' for the
concentration deviations from steady state, $\prot{i}$.
The time-evolving probability distribution is then $P(\bm{\delta
  x},t)$, and observables $a(\bm{\delta x})$ are likewise functions of
$\bm{\delta x}$. In terms of these variables the adjoint
Fokker-Planck operator $\mathcal{L}$ then writes 
\begin{equation}\label{eq:adjointFP}
\begin{split}
\mathcal{L} &= \sum_{i,j,l:i\neq j}\left[\km{l}{i}{j}\prot{l}
- \kp{i}{j}{l}(\y{j}\prot{i}+\y{i}\prot{j}+\prot{i}\prot{j})\right]
\frac{\partial}{\partial\prot{i}}\\
&\quad +\frac{1}{2} \sum_{i,j,l:j\neq l}\left[\kp{j}{l}{i}(\y{l}\prot{j}
+\y{j}\prot{l} + \prot{j}\prot{l}) - \km{i}{j}{l}\prot{i} \right]
\frac{\partial}{\partial\prot{i}}\\
&\quad + \sum_{i,j}\biggl\lbrace\left[2\km{j}{i}{i}\prot{j}
- \kp{i}{i}{j}(2\y{i}\prot{i}+\prot{i}\prot{i})\right]\\
&\quad +\left. \frac{1}{2}\left[\kp{j}{j}{i}(2\y{j}\prot{j}+\prot{j}\prot{j})
 - \km{i}{j}{j}\prot{i}\right]\right\rbrace
\frac{\partial}{\partial\prot{i}}\\
&\quad + \sum_{i,j}\left(\lam{j}{i}\prot{j}-\lam{i}{j}\prot{i}\right)\frac{\partial}{\partial\prot{i}}
 + \frac{\epsilon}{2} \sum_{i,j}
\left(\bm{BB}^{\T}\right)_{ij} \frac{\partial^2}{\partial\prot{i}\,
\partial\prot{j}} \end{split}
\end{equation}
All terms here except for the last describe deterministic
evolution. To write the reaction flux prefactors from
\eqref{deterministiceqns} we have replaced $x_i=y_i+\delta x_i$ and
exploited the fact that when $\bm{x}=\bm{y}$, i.e.\ $\bm{\delta x}=0$,
the deterministic drift terms must vanish. Note that $\mathcal{L}c=0$
for any constant $c$, so that from \eqref{eq:da_dt_is_La} the average of
such an ``observable'' is constant in time as it should be. Looking at
\eqref{eq:Pt_vs_at}, this property is equivalent to conservation of
probability in the original Fokker-Planck equation.

\subsection{Projection method}
\label{sec:projection_method}
We next summarise the salient features of the Zwanzig-Mori projection
method we use 
to derive equations describing the time evolution of the
concentrations in any chosen subnetwork of a larger protein
interaction network \cite{Mori1965,Hansen90,Ritort03}.
The approach allows one generally to derive such equations for the
conditional averages $a_i(\bm{\delta x},t)$ of any chosen set of
observables $\lbrace a_\alpha(\bm{\delta x})\rbrace$. One first defines a projection
operator $\mathcal{P}$ that projects any observable $b$
onto the space spanned by the chosen set of observables:
\begin{equation}\label{Poperator}
\left( \mathcal{P}b\right)(\bm{\delta x}) =
\sum_{\alpha,\beta}a_\beta(\bm{\delta x})\left(\bm{C}^{-1}\right)_{\beta\alpha}\left(a_\alpha,b\right)
\end{equation}
Here $\bm{C}$ is a correlation matrix with elements
\begin{equation}
C_{\alpha\beta}=\left(a_\alpha,a_\beta\right)
\label{eq:C_ij}
\end{equation}
defined in terms of an inner product $\left(a,b\right)$. The latter is just an
average over the steady state distribution $P_{\text{ss}}(\bm{\delta x})$
of $\bm{\delta x}$: 
\begin{equation}
  \left(a,b\right) = \langle ab \rangle_{\text{ss}} =
  \int d\bm{\delta x}\ a(\bm{\delta x})b(\bm{\delta
    x})P_{\text{ss}}(\bm{\delta x}).
\label{eq:ab}
\end{equation}
We see now explicitly that we need stochastic dynamics, i.e.\ nonzero $\epsilon$, to be able to deploy the projection formalism, even if we are interested in the limit of small $\eps$. If we were to set $\eps=0$ directly, the steady state distribution would become a Dirac delta function at the fixed point $\bm{\delta x}=0$, giving for the covariance matrix 
$C_{\alpha\beta}=a_\alpha(0)a_\beta(0)$. As the outer product of a vector -- with elements $a_\alpha(0)$ -- with itself this has rank one and so is not invertible except in the case of a single observable, making the projection operator \eqref{Poperator} ill-undefined. 

\

Once $\mathcal{P}$ is defined, the orthogonal projection operator
$\mathcal{Q}$ follows as $\mathcal{Q}=1-\mathcal{P}$. Then
$\mathcal{Q}b$ can be interpreted as the contribution to observable
$b$ that is uncorrelated in steady state with any of the chosen
observables $a_i$. In our case the latter will be a set of obervables
from the system such as protein 
and complex concentrations from the subnetwork, as discussed in more
detail below.

With the shorthand $a_\alpha(\bm{\delta x},t) = a_\alpha(t)$, the projected equations are written
\cite{Mori1965,Hansen90,Ritort03}
\begin{equation}\label{projection}
\frac{\partial}{\partial t}a_\alpha(t) = \sum_\beta a _\beta(t)\Omega_{\beta\alpha} +
\int_0^tdt'\sum_\beta a_\beta(t')M_{\beta\alpha}(t-t') + r_\alpha(t)
\end{equation} 
The first term on the r.h.s.\ is local in time. We will call the
coefficients 
\begin{equation}
\Omega_{\beta\alpha}=\sum_{\gamma}\left(\bm{C}^{-1}\right)_{\beta\gamma}\left(a_\gamma,\mathcal{L}a_\alpha\right)
\label{eq:Omega}
\end{equation}
the elements of the rate matrix $\bm{\Omega}$; in other contexts,
e.g.\ systems with inertial
dynamics, it is often referred to as the frequency matrix.
The second term represents the memory effects, as an integral over
past values of the observables weighted by a function of the time lag,
the {\em memory function}. The latter can be expressed as 
\begin{equation}
M_{\beta\alpha}(\dt) =
\sum_\gamma\left(\bm{C}^{-1}\right)_{\beta\gamma}
\left(a_\gamma,\mathcal{LQ}e^{\mathcal{QLQ}\dt} \mathcal{QL}a_\alpha\right)
\label{eq:Mij}
\end{equation}
where $\dt = t-t'$.
The memory function $M_{\beta\alpha}$ determines how strongly the past values of
observable $a_\beta$ affect the present rate of change of $a_\alpha$; sometimes
it will be useful to think of the $M_{\beta\alpha}(\dt)$ as the elements of a
memory matrix $\bm{M}(\dt)$ whose size is, as for the rate matrix, the number of
observables $a_\alpha$. The third term in \eqref{projection}, finally, is called the random
force and is written
\begin{equation}
r_\alpha(t)=e^{\mathcal{QLQ}t}\mathcal{QL}a_\alpha.
\label{eq:random_force}
\end{equation}
The name comes from the fact that the value of $r_\alpha(t)$ at any time
$t$ is uncorrelated with the 
initial values of the observables $a_\beta(0)\equiv a_\beta$; 
mathematically this property is expressed as 
$\left(a_\alpha,r_\beta(t)\right)=0$. Note that this notion of randomness does not imply that the random force resembles white noise as in e.g.\ Langevin equations. This is natural given that it appears in the time evolution of the $a_\alpha(t)$, which are conditional averages over dynamical fluctuations. In fact we will see later in \eqref{eq:random_force_linearised} that the random force encodes primarily the initial conditions of the bulk variables.

The projected equations \eqref{projection} are exact as written, and
have several remarkable features. Firstly, they emphasize that memory
terms must arise generically once we go from a description of the full
system, in terms of $\bm{\delta x}$, to one in terms of a reduced
number of observables. Secondly, they provide an ``almost'' closed set
of equations for the chosen observables, with all non-autonomous
effects collected in the random force term. Specifically, while the
time evolution of each $a_\alpha(\bdx,t)$ depends in principle on all
details of the initial system state $\bdx$, the projected equations
\eqref{projection} with the random force term omitted can be solved
knowing only the initial values of the chosen observables, $a_\alpha(0)$.

To make use of the projected equations, we must be able to calculate 
the rate matrix and the memory functions, and say something about the
statistics of the random force. Calculations of the rate matrix
\eqref{eq:Omega} and
memory functions \eqref{eq:Mij} are discussed in more detail in
Section~\ref{sec:genmem}. Here we note only two useful identifies,
which follow from the definitions of these quantities and that of the
projection operator \eqref{Poperator}:
\begin{equation}\label{eq:projoperators}
  \begin{split}
   \sum_\beta a_\beta\Omega_{\beta\alpha}  &= \mathcal{PL}a_\alpha\\
   \sum_\beta a_\beta M_{\beta\alpha}(\dt) &= \mathcal{PLQ}e^{\mathcal{QLQ}\dt}\mathcal{QL}a_\alpha\\
  \end{split}
\end{equation}
To find $\Omega_{\beta\alpha}$ and $M_{\beta\alpha}(\dt)$ we can then first evaluate the
r.h.s.\ of these identities, and identify the coefficients of the
different $a_\beta$.

As regards the statistics of the random force, there is a simple scenario
where all correlation functions $\langle r_\gamma(t')r_\alpha(t)\rangle$ (for
$t\geq t'$) can be deduced from the memory functions. This is the case
where the operator $\mathcal{L}$ is self-adjoint with regards to the
product $(a,b)$, such that $(a,\mathcal{L}b) = (\mathcal{L}a,b)$ for
any observables $a$ and $b$. Using that
$\mathcal{Q}$ automatically has the same property one then finds \cite{Mori1965,Hansen90} 
\begin{equation}
\sum_{j} C_{\gamma\beta}M_{\beta\alpha}(t-t') =
\left(\mathcal{QL}a_\gamma,e^{\mathcal{QLQ}(t-t')}\mathcal{QL}a_\alpha\right) =
\left(e^{\mathcal{QLQ}t'}\mathcal{QL}a_\gamma,e^{\mathcal{QLQ}t}\mathcal{QL}a_\alpha\right) =
\langle r_\gamma(t')r_\alpha(t)\rangle
\label{eq:memory_vs_random_force_correlators}
\end{equation}
showing that random force correlators are indeed determined by the
memory functions. The self-adjointness of $\mathcal{L}$ required here
normally holds in physical systems: these obey detailed balance,
meaning that in the steady state there are no unbalanced probability
fluxes. Protein interaction networks do not in general have this
property \footnote{Note also that even if detailed balance holds for a system described fully in terms of discrete numbers of molecules, it may be lost when going to our 
Kramers-Moyal expansion truncated at second order.},
so that random force statistics have to be calculated separately. We
therefore leave this matter as a point of investigation for a separate
publication, and note here only that the random force has the
biological meaning of extrinsic noise acting on the subnetwork,
arising from it being embedded in the bulk network.

What specific projected equations one obtains from the framework
summarised above is of course largely
dependent on the choice of observables $a_\alpha$. This is discussed in
more detail in Sec.~\ref{sec:choiceofobs}. Here we just note that one useful
convention is to employ observables with vanishing steady state
average, $\langle a_\alpha\rangle \equiv (a_\alpha,1)=0$, which can always be
achieved by subtracting any nonzero average from $a_\alpha$. This
convention has two benefits: first, it guarantees that the matrix $\bm{C}$
defined in \eqref{eq:C_ij} really is a correlation matrix for
fluctuations around the steady state. Second, the projection operator
then obeys $\mathcal{P}c=0$, hence $\mathcal{Q}c=c$ and
$\mathcal{QLQ}c=\mathcal{QL}c=0$. The operator $\mathcal{QLQ}$ thus
inherits from $\mathcal{L}$ the property that its application to any
constant gives zero. As argued above for $\mathcal{L}$, this is
equivalent to saying that the adjoint operator $(\mathcal{QLQ})^{\T}$
conserves probability in the time evolution it generates.
One can therefore think of the time-dependencies
in the memory function and random force as resulting from a
``projected evolution'' of the system with this operator. In
applications to physical systems, this is often used to argue that as
a first approximation $\mathcal{QLQ}$ can be replaced by $\mathcal{L}$ \cite{Goetze1991,Goetze1995}, 
though this is not a path we follow here as we want to
retain a quantitatively accurate projected description.



In order to evaluate the rate matrix \eqref{eq:Omega} and memory
functions \eqref{eq:Mij} we have to calculate the various observable
products $(a,b)$ that occur, and from \eqref{eq:ab} these are defined
in terms of the steady state distribution of $\bdx$. In our case the
latter is a vector of concentrations, shifted to zero mean. Our
general strategy will be to consider suitably large reaction volumes
so that the noise strength $\epsilon=1/V$ is small. More specifically
we require that for typical concentrations of any species, the
absolute number of molecules be large, say $Vy_i \gg 1$ for all $i$ if
we take the steady state concentrations as typical. \footnote{In the EGFR network discussed in Sec.~\ref{sec:EGFR}, steady
state concentrations range from 0.05 to 1000nMol \cite{Kholodenko99}. If we estimate
cells to have a of diameter 20$\mu$m and hence a volume of order (20$\mu$m)$^3$,
this gives absolute steady state molecule numbers $Vy_i$ in the range
240 to 4.8$\cdot 10^6$ and the criterion $Vy_i\gg 1$ is well satisfied. In
separate large-scale studies in specific human cell lines \cite{Beck2011,Nagaraj2011}, protein abundances of up to 2$\cdot 10^7$ molecules per cell have been
reported, with a median number across species of 1.8$\cdot 10^4$. The
distribution of number of molecules is broad, but almost all (97.7\%) species have more than 100 molecules per cell,
so that a small noise approximation should again be justified.}
The steady state fluctuations $\bdx$
will then be small, and we can find their distribution as the steady
state to an approximate Fokker-Planck operator, obtained from
$\mathcal{L}^{\T}$ by linearizing around $\bdx=0$. We emphasise that
this simplification is used only for the steady state, and does not
restrict the deviations from the steady state $\bdx$ that can be
considered in the projected equations, e.g.\ while the system evolves from some
non-steady initial state.

In the linearised version of $\mathcal{L}$, the diffusion matrix
$\bm{BB}^{\T}$ is evaluated at $\bdx=0$, i.e.\ at the steady state
concentrations. The deterministic drift is linearised in $\bm{\delta x}$ so
that it can be written in terms of a drift matrix $\bm{A}$ and a
vector $\bm{\delta x}$ as $\bm{Sf} = \bm{A\delta x}$.  The steady
state $P_{\text{ss}}(\bdx)$ of such a Fokker-Planck operator is a
Gaussian distribution for the
$\bm{\delta x}$ with zero mean and a covariance matrix $\bm{\Sigma}$
that is a solution of the Lyapunov equation \cite{Klipp2009}
\begin{equation}\label{eq:lyapunov}
\bm{A \Sigma} + \bm{\Sigma A}^{\rm T} + \epsilon\bm{BB}^{\T} =0
\end{equation}
Once $\bm{\Sigma}$ is known, the inner products \eqref{eq:ab} in the
projection can then be evaluated as Gaussian averages.

One proviso with this approach to finding $\bm{\Sigma}$, and hence
$P_{\rm ss}(\bdx)$, is that the solution to the Lyapunov
equation is not unique. This is because of 
the conservation laws: each fixed value of the conserved quantities
leads to a different steady state distribution, and the generic solution
for $\bm{\Sigma}$ represents a superposition of these
distributions. 
In simple networks that we have analysed -- and more generally in any network with the detailed balance property discussed above -- one particularly simple solution of
this type is the one where each molecular species
has independent Poisson fluctuations at steady state. For small
$\epsilon$ this product of Poisson distributions becomes a Gaussian with a
diagonal covariance matrix $\bm{\Sigma}$. Because under Poisson
statistics the variance of the number of molecules for each species,
$Vx_i$, equals its mean $Vy_i$, one has $\langle (V\prot{i})^2\rangle
= Vy_i$ and hence $\Sigma_{ii}=\langle 
(\prot{i})^2\rangle = y_i/V = \epsilon y_i$. For brevity we
will call such a covariance matrix ``Poissonian''.

We will see below that having a steady state distribution with
Poissonian covariance matrix has a number of benefits. 
The main one is that the rate matrix terms in the
projected equations will reproduce precisely those terms from the
original evolution equations for the full network that describe
reactions within the chosen subnetwork. The memory terms can then be
interpreted directly as arising from the presence of the bulk. In
view of this, we will use the Poissonian choice of covariance matrix
throughout. This means that, depending on the network under study,
one $P_\text{ss}(\bdx)$ will only be an approximation of the true steady
state distribution. However, this is not a serious obstacle: if one
looks at the derivation \cite{Mori1965} of the projected equations
\eqref{projection},
one sees that in principle any distribution can
be used to define the projection operator. (The exception is the
detailed balance property discussed around
\eqref{eq:memory_vs_random_force_correlators}, but we do not rely on this
in our analysis.) The price we pay is that the random force is then
``random'', i.e.\ uncorrelated with the initial values of our chosen
observables, under the Poisson distribution we have chosen, while
under the true steady state distribution it will generally have
nonzero correlations. This is a proviso that has to be born in mind,
but it is easily outweighed by the fact that the projected equations will
be simpler to interpret.

In summary, while the Poissonian covariance matrix assumption does represent a valid steady state in simple networks, more generally it should be viewed as an auxiliary construct that produces the simplest form of the projected equations for the subnetwork dynamics.


\subsection{Choice of subnetwork observables}
\label{sec:choiceofobs}

\begin{figure}[!ht]
\centering
  \includegraphics[scale=0.8]{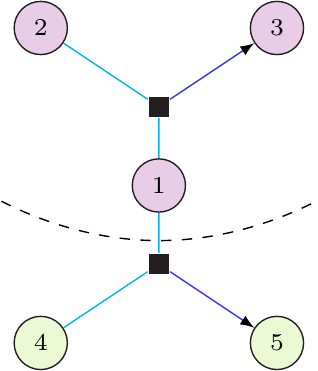}  
\caption{Sketch of a simple model protein interaction network. Protein 1
  reacts with protein 2 to form complex 3, and in 
  reverse 3 can dissociate into 1 and 2. 
  There is an analogous reaction where 1 reacts with 4 to form
  5, and the reverse dissociation. We choose the subnetwork to be
  species 1, 2 and 3 and the bulk to be 4 and 5, as indicated by the
  dashed line. 
}
\label{fig:3p2c}
\end{figure}


To calculate the projected equations we need to choose the set of
observables $\lbrace a_i(\bm{\delta x})\rbrace$ that we will project
on. We assume that a set of molecular species has been chosen
as the subnetwork of interest, e.g.\
because of relevance to some biological function or experimental
accessibility. As before we will call
the rest of the species in the network the bulk. However, this
division still leaves an element of choice in which subnetwork
observables to use in the projection. To illustrate the issues, we
consider a small example, represented graphically in 
Figure~\ref{fig:3p2c}, with two complex formation and dissociation
reactions as indicated below:
\begin{equation}
  \label{eq:3p2c}
  \begin{split}
    1 + 2 &\xrightleftharpoons[\km{3}{1}{2}]{\kp{1}{2}{3}} 3 \\
   1 + 4 &\xrightleftharpoons[\km{5}{1}{4}]{\kp{1}{4}{5}} 5
  \end{split}
\end{equation}
The chemical Langevin equations for this network are
\begin{subequations}\label{eq:eqns3p2c}
\begin{equation}\label{eq:subeqns3p2c}
\begin{split}
   \frac{\partial}{\partial t}\prot{1} &= \km{3}{1}{2}\prot{3} - \kp{1}{2}{3}
 \left(\y{2}\prot{1}+\y{1}\prot{2}+\prot{1}\prot{2}\right)\\
&\quad +\km{5}{1}{4}\prot{5}
  - \kp{1}{4}{5}\left(\y{4}\prot{1}+\y{1}\prot{4}+\prot{1}\prot{4}\right)
  + \noise{1}\\
   \frac{\partial}{\partial t}\prot{2} &= \km{3}{1}{2}\prot{3} - \kp{1}{2}{3}
 \left(\y{2}\prot{1}+\y{1}\prot{2}+\prot{1}\prot{2}\right) + \noise{2}\\
   \frac{\partial}{\partial t}\prot{3} &=  \kp{1}{2}{3}
 \left(\y{2}\prot{1}+\y{1}\prot{2}+\prot{1}\prot{2}\right) -
 \km{3}{1}{2}\prot{3} + \noise{3}\\
\end{split}
\end{equation}
\begin{equation}\label{eq:bulkeqns3p2c}
\begin{split}
   \frac{\partial}{\partial t}\prot{4} &= \km{5}{1}{4}\prot{5} - \kp{1}{4}{5}
 \left(\y{4}\prot{1}+\y{1}\prot{4}+\prot{1}\prot{4}\right) + \noise{4}\\
   \frac{\partial}{\partial t}\prot{5} &=  \kp{1}{4}{5}
 \left(\y{4}\prot{1}+\y{1}\prot{4}+\prot{1}\prot{4}\right) -\km{5}{1}{4}\prot{5}
  + \noise{5}
\end{split}
\end{equation}
\end{subequations}
Here the terms
$\noise{i}$ are the contributions from the (intrinsic) noise. As in
the general form of the adjoint Fokker-Planck operator
\eqref{eq:adjointFP}, we have written concentration products
$\protein{i}\protein{j}$ from the original mass action form
\eqref{deterministiceqns} in terms of $\prot{i}$ and
$\prot{j}$ and removed constant terms that cancel in steady
state, giving $\y{j}\prot{i}+\y{i}\prot{j}+\prot{i}\prot{j}$.  

We assume that the subnetwork of interest in this example consists of
species 1, 2 and 3, and want to select observables for the projection
method accordingly. The goal is to keep the number of variables small,
for computational and conceptual expediency, while retaining an
explicit description of the subnetwork reaction \eqref{eq:subeqns3p2c} in its
original form 

As a first choice one could consider projecting onto only the \emph{protein concentrations}
in the subnetwork, $(\prot{1},\prot{2})$. Explicitly, this means we
use only two observables, $a_1(\bdx)=\prot{1}$ and 
$a_2(\bdx)=\prot{2}$ where $\bdx=(\prot{1},\ldots,\prot{5})^\T$. When we
write down the projected equations \eqref{projection}, we should in
principle write $a_1(t)$ and $a_2(t)$ and bear in mind that these are the conditional
averages -- over the stochastic noise from copy number fluctuations --
of $\prot{1}$ and $\prot{2}$. However, as we are interested throughout
in the {\em limit of small} $\epsilon$, where the effect of averaging
over the noise becomes negligible, we write directly $\prot{1}$ and $\prot{2}$.

Deferring for now a discussion of how rate matrix and memory functions are
calculated in practice (see Secs.~\ref{sec:lindynamics} and
\ref{sec:nonlindynamics}), we state directly the projected equation for
$\prot{1}$ that results from the above choice of subnetwork observables:
\begin{equation}
 \begin{split}
  \frac{\partial}{\partial t}\prot{1} &= -\kp{1}{2}{3}\left(\y{2}\prot{1} + \y{1}\prot{2}\right) - \kp{1}{4}{5}\y{4}\prot{1}\\
 &\quad + \int_0^t dt' \left[\kp{1}{4}{5}\y{4}(\km{5}{1}{4} + \kp{1}{4}{5}\y{1})
e^{-(\km{5}{1}{4}+\kp{1}{4}{5}\y{1})(t-t')} 
+ \km{3}{1}{2}\kp{1}{2}{3}\y{2}e^{-\km{3}{1}{2}(t-t')}\right]\prot{1}(t') \\
&\quad + \int_0^t dt'\left[\km{3}{1}{2}\kp{1}{2}{3}\y{1}e^{-\km{3}{1}{2}(t-t')}\right]\prot{2}(t')
+ r_1(t)\\ 
\end{split}
\label{eq:dx1_projected}
\end{equation}
The terms from the rate matrix, which are the local-in-time
contributions in the first line, are linear in 
$\prot{1}$ and $\prot{2}$ as expected from \eqref{projection}. They
therefore do not capture all terms from the complex formation
and dissociation reactions within the subnetwork, as written in the
first line of \eqref{eq:subeqns3p2c} 

To include nonlinear terms, one could consider adding a third
observable, $a_3=\prot{1}\prot{2} - \langle \prot{1}\prot{2}\rangle$,
giving a projection onto \emph{protein concentrations and products of protein
  concentrations}. We have written the subtraction of
$\langle\prot{1}\prot{2}\rangle$ here for clarity to emphasize that
also nonlinear observables must have zero mean, though for our chosen
Poissonian steady state this average vanishes. The projected equation
for $\prot{1}$ that results is similar to \eqref{eq:dx1_projected},
but now explicitly includes the $\prot{1}\prot{2}$ term from the first
line of \eqref{eq:subeqns3p2c}. However, the complex dissociation term
$\km{3}{1}{2}\prot{3}$ still does not feature because the
complex, species 3, remains ``eliminated'' from the subnetwork
description. Its contribution is retained indirectly through the
memory function, but not in a very transparent way.

The best option is therefore to project onto the \emph{protein
  concentrations, products of protein concentrations and complex
  concentrations} from the subnetwork. With the vector of observables now
$(a_1,\ldots,a_4)=\left(\prot{1},\prot{2},\prot{3},\prot{1}\prot{2} - \langle
  \prot{1}\prot{2}\rangle\right)$, the projected 
equation for $\prot{1}$ becomes
\begin{equation}\label{mem3p2c}
 \begin{split}
  \frac{\partial}{\partial t}\prot{1} &= \km{3}{1}{2}\prot{3} 
-\kp{1}{2}{3}\left(\y{2}\prot{1} + \y{1}\prot{2}  + \prot{1}\prot{2}\right)
  - \kp{1}{4}{5}\y{4}\prot{1}\\
  &\quad + \int_0^t dt'\left[ M_{1,1}(t-t')\prot{1}(t')  
 + M_{12,1}(t-t')\prot{1}(t')\prot{2}(t')\right] + r_1(t)
\end{split}
\end{equation}
All contributions relating to the subnetwork reaction $1 + 2
\xrightleftharpoons[]{} 3$ now appear directly via the local-in-time
rate matrix terms: compare the first line of \eqref{mem3p2c} to
\eqref{eq:subeqns3p2c}. 
 The bulk, which here comprises just
species 4 and 5, contributes an additional local-in-time term because
of the reaction of $4$ with $1$. The other bulk effects are captured
in the memory terms: as expected from our general discussion, feedback
effects from the subnetwork into the bulk and back at a later time
lead to the evolution of $\prot{1}$ being coupled linearly to its own
history, via a ``self memory'' term; there is also memory term that is
nonlinear in concentration fluctuations.  The linear self-memory
function can be written explicitly as $M_{1,1}(t-t') =
\kp{1}{4}{5}\y{4}(\km{5}{1}{4} + \kp{1}{4}{5}\y{1})
e^{-(\km{5}{1}{4}+\kp{1}{4}{5}\y{1})(t-t')}$; we omit the full
expression for $M_{12,1}(t-t')$ for the sake of brevity. As expected,
the reaction rates $\kp{1}{4}{5}$ and $\km{5}{1}{4}$ relating to
the bulk protein and complex that are being projected out from the
description appear in the memory functions.

The upshot of the discussion so far is that we should project onto the
concentrations of all molecular species from the subnetwork -- both
proteins and complexes -- and all products of these
concentrations. This gives projected 
equations where (a) all reactions taking place within the subnetwork
are represented in their original form, as if the subnetwork was
isolated, and (b) memory terms arise only from the presence of the
bulk. One final choice left open here is which concentration products
to include, only those occurring in the subnetwork reactions like
$\prot{1}\prot{2}$ above, or all possible concentration products
(e.g.\ $\prot{1}\prot{2}$, $\prot{1}^2$, $\prot{2}^2$, with nonzero
averages subtracted as necessary). We will see
below that the latter choice has advantages in a general treatment, in
that it leads to smaller random force contributions.

\subsection{Memory functions: initial orientation}
\label{sec:memory_overview}

We conclude this section by using the simple example above to provide
some initial insights into the properties and intuitive meaning of
memory functions.

\begin{figure}[ht!]
   \centering
   \includegraphics[scale=0.8]{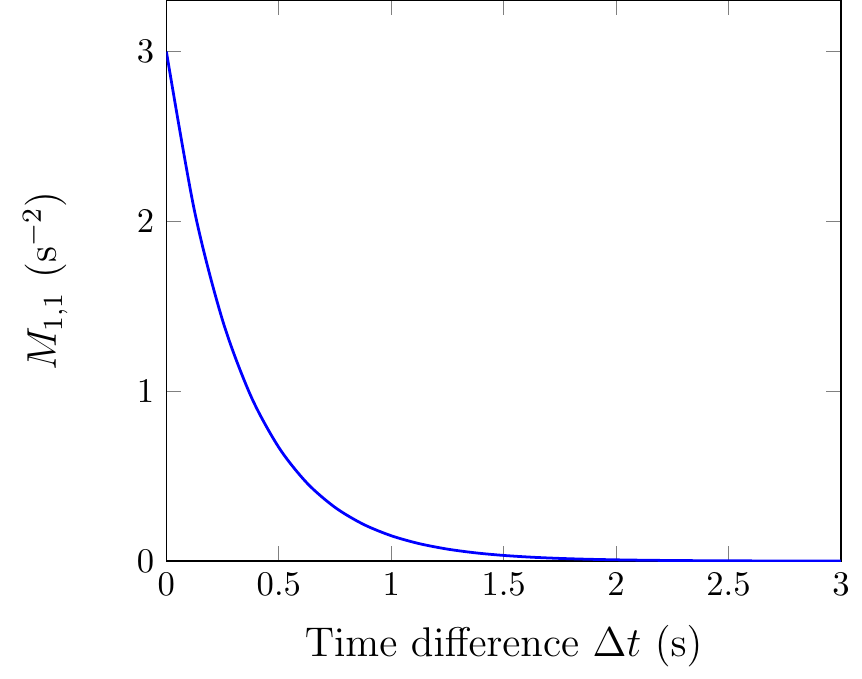}
   \caption{The coefficient of the self memory function for $\prot{1}$
     with rate constants $\kp{1}{2}{3} = \kp{1}{4}{5} = 1$,
     $\km{3}{1}{2}=\km{5}{1}{4}=2$, and steady state values $\y{1} =
     \y{2} = \y{3} = 1$. 
   }
   \label{fig:mem3p2c}
 \end{figure}

We focus initially on the self-memory function
$M_{1,1}(t-t')$. Figure~\ref{fig:mem3p2c} shows a sketch of this
function, for a 
simple choice of reaction rates in appropriate dimensionless
units. The self-memory function is positive, and decays exponentially
with the time difference to the presence. The sign implies that having
a higher concentration of species $1$ at some previous time $t'$
($\prot{1}(t')>0$) will lead to
more $1$ being created at time $t$. To see why this is so, note that
if more 1 is present at $t'$, then more of species $5$ will be created
from the reaction with 4; this will then dissociate back into $1$
at a later time, increasing the concentration of $1$. This effect
weakens as the concentration of 5 reverts to its steady state with
time, explaining the decay of the memory function with the time
difference $t-t'$. 

Looking at the self-memory function more quantitatively, one
recognises that it reflects conservation laws of the proteins and
complexes in the bulk, as it should. For the above example there is
one such conservation law: the total concentration of species 4 and 5
is conserved, implying that deviations away from steady state are
equal and opposite: $\prot{4} =
-\prot{5}$. Therefore the equation \eqref{eq:bulkeqns3p2c} 
 for the complex
$\prot{5}$ can be rewritten as
\begin{equation}
  \frac{\partial}{\partial t}\prot{5} = \kp{1}{4}{5}\left(\y{4}\prot{1} 
    - \prot{1}\prot{5}  \right) - \left(\kp{1}{4}{5}\y{1} 
    + \km{5}{1}{4}\right)\prot{5}
\end{equation}
If we now drop the $\prot{1}\prot{5}$ term, which would contribute
to the random force and to nonlinear memory terms 
 and then integrate this equation we obtain $\prot{5}(t) = 
\int_0^t dt' \kp{1}{4}{5}\y{4}
e^{-(\km{5}{1}{4}+\kp{1}{4}{5}\y{1})(t-t')}\prot{1}(t')$ (up to an
initial condition-dependent term which would give another contribution to the
random force). Inserting into the equation for $\prot{1}$ in
\eqref{eq:subeqns3p2c} and using $\prot{4}=-\prot{5}$ then gives
the linear memory term in \eqref{mem3p2c}, showing that this accounts
for the bulk conservation law as it should. 

If we next look at the general structure of the memory terms in the
projected equation for \eqref{mem3p2c}, we notice that in the
linear memory terms only the history of $\prot{1}$ features, not that
of $\prot{2}$. The same is true in the projected equation of motion
for $\prot{2}$, which we have not given explicitly. As
explained in more detail in Section~\ref{sec:memproperties1}, 
this is a general property of linear memory terms:
the only variables
that appear in these are the concentrations of ``boundary
species''. Here a boundary species is one that has a direct reaction
with a bulk species. In our example above, 1 is the only boundary
species, while 2 and 3 are in the interior of the subnetwork. The
intuitive reason why their histories do not appear in linear memory
terms is that their effects on the bulk can only be ``transmitted''
indirectly via the time course of the concentration of species 1,
rather than directly.

 \begin{figure}[ht!]
   \centering
\includegraphics[scale=0.8]{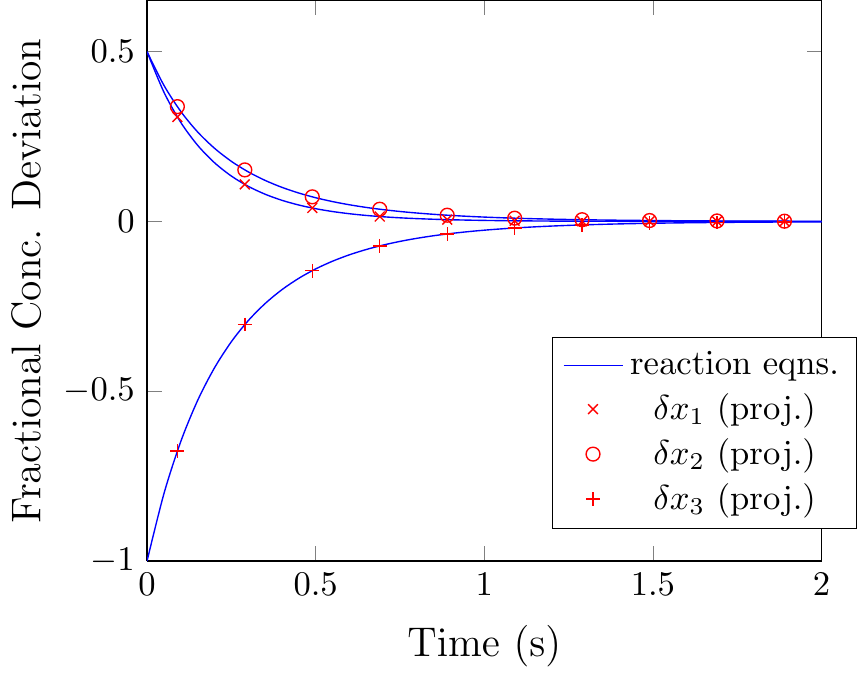}
\caption{Concentration time courses for example
  \eqref{eq:3p2c}, comparing the solution of the full reaction equations
  \eqref{eq:eqns3p2c} (solid lines) and the solution of the subnetwork
  projected equations \eqref{mem3p2c} (symbols). Note the
  excellent agreement even though random force terms were omitted from
  the projected equations. The $y$-axis shows fractional concentration
  deviations from the steady state, $\prot{i}/y_i$, so that a value of
  $-1$ corresponds to concentration $x_i=0$.
  Rate constants $\kp{1}{2}{3} = \kp{1}{4}{5} = 1$ and
  $\km{3}{1}{2}=\km{5}{1}{4}=2$; initial conditions
  $\prot{1}(0)/\y{1}=\prot{2}(0)/\y{2}=1/2$, $\prot{3}(0)/\y{3}=-1$, $\prot{4}(0)=\prot{5}(0)=0$.
}
\label{fig:3p2csoln}
 \end{figure}

 Finally we demonstrate the quantitative accuracy of the projected
 equations, i.e.\ \eqref{mem3p2c} and the analogous equations for
 $\prot{2}$ and $\prot{3}$. We know that the equations are exact in
 the small noise limit $\epsilon\to0$ that we have already taken, but
 the random force terms cannot be expressed in closed form, as
 discussed in more detail below. Our interest is therefore in
 assessing how accurate the projected subnetwork description is {\em
   when the random force terms are omitted but memory terms are
   retained}. Fig.~\ref{fig:3p2csoln} compares the solution of the
 resulting approximate projected equations with the solution of the
 full set of reaction equations \eqref{eq:eqns3p2c}. The two sets of
 time courses are visually indistinguishable, confirming that the
 projected subnetwork equations give a highly accurate description of
 the dynamics. The initial conditions were chosen so that
 concentrations of bulk species were at their steady state
 values. This is the regime where we expect the omitted random force
 terms to be smallest, as discussed in Sec.~\ref{sec:quanttests}
 below. We will also 
 compare to alternative reduced descriptions
 of subnetwork dynamics.


 \section{Memory functions: explicit expressions and general
   properties}
\label{sec:genmem}

In this section we give explicit expressions for memory functions

describing the dynamics of protein interaction subnetworks. We study
their general properties, in particular with a view to how they encode
subnetwork-bulk interactions. In Sec.~\ref{sec:lindynamics} we study
first a simplified scenario, where the dynamical equations of the
original large network are linearised around the steady
state. Applying the projection method to obtain a description of the
subnetwork dynamics, the memory functions can be found explicitly; we
validate the approach by comparing with the simpler approach of
integrating out the bulk degrees of freedom directly. In
Sec.~\ref{sec:nonlindynamics} we then demonstrate that, more
surprisingly, we can obtain the memory functions explicitly even
for the full nonlinear dynamics. Here as throughout we focus on the
small noise, large reaction volume limit $\epsilon\to 0$.  Finally, in
Sec.~\ref{sec:memproperties1} we discuss some generic properties of
memory functions.

\subsection{Linearised dynamics}
\label{sec:lindynamics}
To get some insight into the general form of the
projected equations we first consider a simplified problem, starting from a
linearised description for the full network.
The linearised reaction equations including copy number noise are 
\begin{equation}
  \partial_t\bm{\delta x} = \bm{A\,\delta x} + \bm{\eta}
\label{eq:linearised_CLE}
\end{equation}
where $\bm{A}$ is as defined just before 
\eqref{eq:lyapunov} and the covariance
matrix $\epsilon \bm{B}\bm{B}^\T$ of the noise $\eta$, which normally
is $\bdx$-dependent, is evaluated at steady state ($\bdx=0$). 
The
corresponding adjoint Fokker-Planck operator is
\begin{equation}
\mathcal{L} =  \sum_{ij} \prot{j}A_{ji} \frac{\partial}{\partial
  \prot{i}} + \frac{\epsilon}{2} \sum_{ij} (\bm{B}\bm{B}^\T)_{ij}
\frac{\partial^2}{\partial \prot{i}\partial \prot{j}}
\label{eq:L_linearised}
\end{equation}

In
Section~\ref{sec:choiceofobs} we showed that in general, the most 
appropriate choice of
subnetwork observables $\lbrace a_i\rbrace$ consists of the subnetwork
concentrations and all their products. Now that we are considering
linearised dynamics, we will 
only want to project onto the concentrations themselves, omitting the
products. The
linearised projected equations can then be written in the general form
\begin{equation}
 \frac{\partial}{\partial t}\prot{i}(t) = \sum_{j=1}^{\Ns} \prot{j}(t)\Omega_{ji} +
\int_0^tdt' \sum_{j=1}^{\Ns}\prot{j}(t')M_{ji}(t-t') + r_i(t)
\label{eq:projected_lin_general_form}
\end{equation}
and our aim will be to find explicit expressions for the rate matrix
entries $\Omega_{ji}$ and the memory functions $M_{ji}(t-t')$. Note
that, as it should be for a description of the subnetwork dynamics,
the sums over $j$ above runs only over subnetwork concentrations. We
assume here that these concentrations make up the first entries of
the vector $\bdx$, i.e.\ $x_j$ with $j=1\ldots \Ns$ where $\Ns$ is the
number of subnetwork species. We will denote the subnetwork part of
$\bdx$ by $\bdxs$, and the remaining bulk part by $\bdxb$, so
that $\bdx^{\T}=(\bdxs{}^\T, \bdxb{}^\T)$. 

To find the rate matrix and memory functions from the general
expressions \eqref{eq:Omega} and
\eqref{eq:Mij}, or equivalently \eqref{eq:projoperators}, we need to
be able to find the action of the operators 
$\mathcal{L}$, $\mathcal{P}$ and $\mathcal{Q}$ on the observables
$a_i=\prot{i}$ ($i=1,\ldots,\Ns$) and evaluate products of the form
$(a,b)$. Starting with the latter, we choose for the (approximate)
steady-state distribution a Gaussian over $\bdx$ with mean zero and
Poissonian covariance matrix $\bm{\Sigma}$. The elements of this
matrix then give the products
$(\prot{i},\prot{j})=\Sigma_{ij}$. More specifically, if we partition
the covariance matrix depending on
whether the relevant molecular species are in the subnetwork or bulk,
as done for the vector $\bdx$, it can be written in the form
\begin{equation}
   \bm{\Sigma} =
  \begin{pmatrix}
    \bm{\Sigma}^{\superss} & \bm{0}\\
    \bm{0} & \bm{\Sigma}^{\superbb}
  \end{pmatrix}
\end{equation}
The Poissonian form for $\bm{\Sigma}$ forces zeros
on the off-diagonal blocks as we have written. It also implies that
$\bm{\Sigma}^{\superss}$ are $\bm{\Sigma}^\superbb$ are
diagonal, although we will not need this property in the following.

We can now write down the action of the projection operator
\eqref{Poperator} on a
generic observable. For linear observables $\prot{i}$, which will be sufficient
for our purposes, we obtain
\begin{equation}
\mathcal{P}\prot{i} = \sum_{j,k\leq \Ns}\left(\prot{i},\prot{j}\right)
\left(\bm{\Sigma}^{\superss}\right)^{-1}{}_{jk}\prot{k}
\end{equation}
Here we have used that the observable correlation matrix $\bm{C}$,
i.e.\ covariance of the subnetwork concentrations, is just the top
left block $\bm{\Sigma}^\superss$ of $\bm{\Sigma}$. For
$i=1,\ldots,\Ns$ the first product is simply $\Sigma^{\superss}_{ij}$ so that
$\mathcal{P}\prot{i}=\prot{i}$. Conversely for $i=\Ns+1,\ldots,N$ the
product vanishes because of the block structure of $\bm{\Sigma}$, and
$\mathcal{L}\prot{i}=0$. If we collect these results, and the
corresponding ones for the orthogonal projector $\mathcal{Q}$, in the form
\begin{equation}\label{defn:operatorsPQ}
\begin{split}
  \mathcal{P}\prot{i}&= \sum_{j=1}^N\prot{j}P_{ji}\\
  \mathcal{Q}\prot{i}&= \sum_{j=1}^N\prot{j}Q_{ji}
\end{split}
\end{equation}
then the coefficients $P_{ji}$ and $Q_{ji}$ can be arranged into
matrices with the simple block form
%
%
\begin{equation}\bm{P} = 
  \begin{pmatrix}
    \mathbb{1} & \bm{0}\\
    \bm{0} & \bm{0}
  \end{pmatrix},
\quad
\bm{Q} = 
  \begin{pmatrix}
    \bm{0} & \bm{0}\\
    \bm{0} & \mathbb{1}
  \end{pmatrix}
\label{eq:PandQ_blocks} 
\end{equation}
These results are intuitively obvious: when we project onto the
subnetwork, the only observables that survive are those from
the subnetwork. Similarly when applying the orthogonal projection
operator $\mathcal{Q}$, only bulk observables remain. 

Finally we can also write the adjoint Fokker-Planck operator in a
similar matrix form. Looking at \eqref{eq:L_linearised},
$\mathcal{L}\prot{j} = \sum_i A_{ji}\prot{i}$, so if we define
\begin{equation}\label{defn:operatorsL}
\begin{split}
  \mathcal{L}\prot{i} &= \sum_{j=1}^N\prot{j}L_{ji}
\end{split}
\end{equation}
then $\bm{L}=\bm{A}^\T$ is the transpose of the dynamical matrix. 
This makes sense as the equation of motion \eqref{eq:da_dt_is_La} for
the conditionally averaged concentrations,
\begin{equation}\label{deterministiceqnslin}
  \frac{\partial}{\partial t}\prot{i} = 
\mathcal{L}\prot{i} = \sum_j\prot{j}L_{ji}
\end{equation}
has to agree with the noise-averaged
rate equation \eqref{eq:linearised_CLE},  
$\partial_t\prot{i} = \sum_jA_{ij}\prot{j} =
\sum_j\prot{j}(\bm{A}^{\T})_{ji}$. We can partition the matrix
$\bm{L}$ representing the adjoint Fokker-Planck operator $\mathcal{L}$
into subnetwork and bulk contributions again, according to
\begin{equation}\label{lmatlin}
  \bm{L} =
  \begin{pmatrix}
    \bm{L}^{\superss} & \bm{L}^{\supersb}\\
    \bm{L}^{\superbs} & \bm{L}^{\superbb}
  \end{pmatrix}
\end{equation}
From the definition \eqref{defn:operatorsL} one then sees that 
${L}^{\superbs}$, for example, contains the coefficients of
bulk concentrations in the equations of motion for subnetwork concentrations.

%
Note that the
matrix representations \eqref{defn:operatorsPQ} and
\eqref{defn:operatorsL} have been defined so that the vector $\bdx$
sits on the left, e.g.\ $\mathcal{P}\bdx^\T = \bdx^\T \bm{P}$. This
has the advantage of maintaining the ordering of the matrices when
operators are composed, for example $\mathcal{PL}\prot{i} =
\mathcal{P}\sum_j \prot{j} L_{ji} = \sum_{jk} \prot{k}P_{kj}L_{ji}$,
or in vector form $\mathcal{PL}\bdx^\T = \bdx^\T \bm{PL}$.

This identity can now be employed directly to get the rate matrix terms in
the projected equations \eqref{projection}. We use
\eqref{eq:projoperators}, i.e.\
$\sum{j=1}^{\Ns} \prot{j}\Omega_{ji} = \mathcal{PL}\prot{i}$. This has to hold
for all $i=1,\ldots,\Ns$, so one reads off that 
$\bm{\Omega}$ is the top left block of 
$\bm{PL}$, which because of the simple form of \eqref{eq:PandQ_blocks}
is simply the top left block of $\bm{L}$ in \eqref{lmatlin}, i.e.
\begin{equation}
\bm{\Omega} = \lblockb{s}{s}
\label{eq:Omega_lin}
\end{equation}
Similarly, the memory function obeys the identity
\eqref{eq:projoperators}:
\begin{equation}
\sum_{j=1}^{\Ns} \prot{j}M_{ji}(t) =
\mathcal{PLQ}e^{\mathcal{QLQ}t}\mathcal{QL}\prot{i}
\end{equation}
Exploiting the
correspondence between operators and matrices again, the r.h.s.\ is
the $i$-th entry of the vector $\bdx^\T \bm{PLQ}e^{\bm{QLQ}t}\bm{QL}$. 
Comparing with the l.h.s.\ shows that the matrix of memory functions
$\bm{M}(t)$ is the top left block of $\bm{PLQ}e^{\bm{QLQ}t}\bm{QL}$,
where $e^{\bm{QLQ}t}$ is a matrix exponential. Inserting the block
decomposition \eqref{lmatlin} of $\bm{L}$ and exploiting again
\eqref{eq:PandQ_blocks} shows that this can be written
explicitly as
\begin{equation}
\bm{M}(t) = \lblockb{s}{b}e^{\lblockb{b}{b}t}\lblockb{b}{s}
\label{eq:M_lin}
\end{equation}
With this and \eqref{eq:Omega_lin} we have obtained the desired explicit
expressions for the rate and memory matrices of the projected
equations \eqref{eq:projected_lin_general_form}. We note as an aside
that, for the linearised scenario we are considering here, an expression for the
random force can also be found. The definition 
\eqref{eq:random_force} becomes
$r_i(t)=e^{\mathcal{QLQ}t}\mathcal{QL}\prot{i}$, which is the $i$-th
entry of the vector $\bm{\delta x}^{\T}e^{\bm{QLQ}t}\bm{QL}$. If we
gather the required entries for $i=1,\ldots,\Ns$ into a vector $\bm{r}(t)$, this
can be simplified to
\begin{equation}
\bm{r}^\T(t) = \bdxb{}^\T e^{\lblockb{b}{b}}\lblockb{b}{s}.
\label{eq:random_force_linearised}
\end{equation}

We have gone through the application of the projection approach to the
linearised dynamics to illustrate the steps involved in calculating
the rate matrix and memory functions. For this relatively simple
setup one can obtain the projected equations also more directly, by
integrating out the bulk degrees of freedom. One starts
from the equations of motion for the (conditionally averaged)
concentrations, which read $\partial_t \bdx^\T = \bdx^\T \bm{L}$
or after decomposing into subnetwork and bulk terms 
\begin{subequations}
\begin{align}
\partial_t\bdxs{}^\T &= \bdxs{}^\T \lblockb{s}{s} + \bdxb{}^\T \lblockb{b}{s}\label{eq:linearised_split_sub}
\\
\partial_t\bdxb{}^\T &= \bdxs{}^\T \lblockb{s}{b} + \bdxb{}^\T \lblockb{b}{b}\label{eq:linearised_split_bulk}
\end{align}
\end{subequations}
The solution for the bulk concentrations reads 
\begin{equation}
\bdxb{}^\T(t) = \bdxb{}^\T(0) e^{\lblockb{b}{b}t} + \int_0^tdt'\,\bdxs{}^\T(t')\lblockb{s}{b}e^{\lblockb{b}{b}(t-t')}
\end{equation}
and substituting into the first line of \eqref{eq:linearised_split_sub} 
gives for the subnetwork concentrations
\begin{equation}
\partial_t \bdxs{}^\T(t) = \bdxs{}^\T \lblockb{s}{s} 
+ \int_0^tdt'\,\bdxs{}^\T(t')\lblockb{s}{b}e^{\lblockb{b}{b}(t-t')}\lblockb{b}{s}
+ \bdxb{}^\T(0)e^{\lblockb{b}{b}t}\lblockb{b}{s}
\end{equation}
which is exactly \eqref{eq:projected_lin_general_form} with the rate
matrix \eqref{eq:Omega_lin}, memory matrix \eqref{eq:M_lin} and random
force \eqref{eq:random_force_linearised}. This derivation shows
explicitly how memory arises from bulk degrees of freedom being
influenced by past behaviour of the subnetwork, and then feeding this
influence back to the subnetwork at a later time. One also sees either
from this or from \eqref{eq:random_force_linearised} that the random
force terms account for the effects of potentially unknown bulk
initial conditions $\bdxb(0)$. When $\bdxb(0)=0$, i.e.\ when the bulk
is initially in steady state, then the random force vanishes. The
solution of the projected equations for the subnetwork with the random
force omitted will then match exactly the solution of the original
linearised reaction equations \eqref{deterministiceqnslin}. This
motivates the good agreement we observed between the two sets of
solution in the simple example of Sec.~\ref{sec:choiceofobs}, cf.\
Fig.~\ref{fig:3p2c}, although there we were dealing with the full
nonlinear reaction equations. This is the topic we consider next.


\subsection{Nonlinear dynamics}
\label{sec:nonlindynamics}

The projection approach as examplified for linearised dynamics in the

previous section may seem formal and somewhat indirect, given
that bulk degrees of freedom can be eliminated directly. The method
comes into its own, however, when we consider the full nonlinear reaction
equations \eqref{deterministiceqns}, where a direct elimination
approach is not feasible. We show in this section that, non-trivially 
for a nonlinear case, explicit expressions for the rate matrix and
memory functions in the projected equations can be found. We will
appeal to the small noise limit $\epsilon\to 0$ as before, and will
need to examine carefully what terms survive in this limit. Note that
this was not necessary for the linearised dynamics, where the noise
drops out from the equations for the conditionally averaged
concentrations, whatever the value of $\epsilon$. Guided by the
discussion of the linear case, we will again aim to find a suitable
matrix representation for the operators involved.

Regarding the choice of observables $\lbrace a_\alpha\rbrace$ 
for the projection we follow the discussion in
Section~\ref{sec:choiceofobs} and include the concentrations of the
subnetwork species and all products of these concentrations,
shifted to zero mean as necessary. The list of observables is then
$\bm{a}^\T = (\prot{1},\ldots,\prot{\Ns},\prot{1}^2-\langle \prot{1}^2\rangle,
\prot{1}\prot{2}-\langle \prot{1}\prot{2}\rangle, \ldots,
\prot{\Ns}^2-\langle \prot{\Ns}^2\rangle)$, containing in total
$\Ns+\Ns(\Ns+1)/2 = \Ns(\Ns+3)/2$ distinct quantities. For the steady state
distribution we take again a zero mean Gaussian for $\bdx$, with a
Poissonian covariance matrix. The steady state averages $\langle
\prot{i}\prot{j} \rangle$ are then $\order(\epsilon)$ and can be neglected
against terms of order unity. Applying this simplification, the
nonlinear projected equations for the subnetwork concentrations then
follow from the general result \eqref{projection} as
\begin{equation}\label{eq:proj_dimensional}
\begin{split}
  \frac{\partial}{\partial t}\prot{i} &= \sum_{j=1}^{\Ns} \prot{j}\Omega_{ji}^\superss + 
\sum_{1\leq j\leq k\leq \Ns}\prot{j}\prot{k}\Omega_{(jk)i}^\supersss\\
&\quad + 
  \int_0^tdt'\left(\sum_{j=1}^{\Ns}\prot{j}(t')M_{ji}^\superss(t-t')
    +\sum_{1 \leq j\leq k \leq \Ns}\prot{j}(t')\prot{k}(t')M_{(jk)i}^\supersss(t-t')\right) + r_i(t)
\end{split}
\end{equation} 
Here we have
split sums over 
observables into linear and nonlinear observables to show the structure more
clearly. Accordingly there are now linear and nonlinear rate matrix
and memory contributions. The bracket on the subscript in the nonlinear
terms $\Omega_{(jk)i}^\supersss$ and
$M_{(jk)i}^\supersss$ indicates the effect of a concentration
product $\prot{j}\prot{k}$ on the time evolution of $\prot{i}$. As
before we have not distinguished in our notation between the original
concentration variables $\prot{i}$ or $\prot{i}\prot{j}$ and their averages conditional on a
given initial set of concentrations across the network, because the
two become identical for $\epsilon\to 0$. Finally, all indices relate
only to subnetwork variables and so lie in the range $1,\ldots,\Ns$.

Our goal in this section is to find explicit expressions for the
linear and nonlinear rate matrix and memory functions in
\eqref{eq:proj_dimensional}.
To establish whether we can achieve this using matrix representations
of the relevant operators, we first look at the terms we obtain by
applying the 
operators $\mathcal{L,P} $ and $\mathcal{Q}$ to concentrations and
products of concentrations.
%
%
The adjoint Fokker-Planck operator $\mathcal{L}$
from \eqref{eq:adjointFP} contains single derivatives for the
deterministic drift terms, multiplied by terms of order $\delta x$ and
$\delta x^2$, and second derivatives for the diffusion terms from copy
number noise. The latter come with a factor $\epsilon$ and are
multiplied by elements of the matrix $\bm{BB}^\T$. From
\eqref{eq:BB_defn} these get their $\delta x$-dependence from the
reaction fluxes $\bm{f}$ and thus contain terms of up to quadratic order
in $\delta x$. Applying then $\mathcal{L}$ to any linear concentration
observable gives
 \begin{equation}
   \begin{split}
   \mathcal{L}\prot{i} &= \dxany + \dxany^2
   \end{split}
\label{eq:Lproperty1a}
 \end{equation}
because the diffusion piece does not contribute. The symbolic shorthand on the
r.h.s.\ indicates a linear combination of terms of the form $\prot{i}$
and $\prot{j}\prot{k}$. 
The analogous 
statement for a product of concentrations is
 \begin{equation}
   \begin{split}
   \mathcal{L}\prot{i}\prot{j} &= \dxany^2 + \dxany^3
   +\order(\epsilon)
   \end{split}
\label{eq:Lproperty1b}
 \end{equation}
because in the deterministic piece of $\mathcal{L}$ the first
derivative now leaves one power of $\delta x$. The terms generated by
the diffusion part are of order $\epsilon$, $\epsilon\,\delta x$ and
$\epsilon\,\delta x^2$, and we denote such terms summarily by
$\order(\eps)$. To summarize the last two relations, define 
$\bm{z}$ as a vector containing all concentrations $\prot{i}$ from the entire
network as well as all concentration products $\prot{j}\prot{k}$: 
$\bm{z}^\T = (\prot{1},\ldots,\prot{N},\prot{1}^2-\langle \prot{1}^2\rangle,
\prot{1}\prot{2}-\langle \prot{1}\prot{2}\rangle, \ldots,
\prot{N}^2-\langle \prot{N}^2\rangle)$. Note that this is different
from the vector $\bm{a}$, which only contains the elements of $\bm{z}$
that relate exclusively to the subnetwork. We can now write
\eqref{eq:Lproperty1a} and \eqref{eq:Lproperty1b} together in the form
\begin{equation}
  \mathcal{L}z_\alpha = \sum_\beta z_\beta L_{\beta \alpha} + \dxany^3 + \order(\eps)
\label{eq:Lproperty1}
\end{equation}
where $\alpha$ and $\beta$ lie in the range $1,\ldots,N(N+3)/2$ and
$L_{\beta\alpha}$ is a suitably defined matrix. Finally we have for
the action of $\mathcal{L}$ on an $n$-th order product of
concentrations
\begin{equation}
\begin{split}
  \mathcal{L}\dxany^n &= \dxany^n + \dxany^{n+1} + \order(\eps),\quad n\geq 3
\end{split}
\label{eq:Lproperty2}
\end{equation}
where the first two terms on the r.h.s.\ again come from the
deterministic drift.

The projection operators $\mathcal{P}$ and $\mathcal{Q}$ have similar
properties as we show next. 
From the definition
\eqref{Poperator}, we need the correlations $(a_\alpha,a_\beta)$ to
get the projector. Fortunately, these are diagonal for our choice of a
Poissonian covariance matrix $\bm{\Sigma}$. Firstly, because odd
moments of a zero mean Gaussian vanish, there are no correlations
between linear and quadratic observables. Correlations among linear
observables are diagonal as before,
$(\prot{i},\prot{j})=\Sigma_{ij}=\epsilon y_i \delta_{ij}$. The
correlation among quadratic variables can be worked out using Wick's
theorem \cite{Wick1950}
\begin{equation}
\begin{split}
(\prot{i}\prot{j}-\langle \prot{i}\prot{j}\rangle,
\prot{k}\prot{l}-\langle \prot{k}\prot{l}\rangle)
&= \langle \prot{i}\prot{j}\prot{k}\prot{l}\rangle - 
\langle \prot{i}\prot{j}\rangle \langle \prot{k}\prot{l}\rangle\\
&= \langle \prot{i}\prot{k}\rangle \langle \prot{j}\prot{l}\rangle
+ \langle \prot{i}\prot{l}\rangle \langle \prot{j}\prot{k}\rangle
\end{split}
\label{eq:Wick}
\end{equation}
The surviving first term is nonzero only if $i=k$ and $j=l$, and
similarly for the second one. Taking the indices as ordered ($i\leq j$ and
$k\leq l$) then shows that the only nonzero correlations are the
diagonal ones:
\begin{equation}
(\prot{i}\prot{j}-\langle \prot{i}\prot{j}\rangle,
\prot{i}\prot{j}-\langle \prot{i}\prot{j}\rangle) = 
(1+\delta_{ij})\langle \prot{i}^2 \rangle \langle \prot{j}^2\rangle
= (1+\delta_{ij}) \epsilon^2 y_i y_j
\end{equation}
where the $\delta_{ij}$ accounts for the fact that the last term in
\eqref{eq:Wick} contributes only when $i=j$.

The projection 
operator now becomes, if we collect the above results for the
observable correlation matrix 
$C_{\alpha\beta} = (a_\alpha,a_\beta)$ and split the sum
over observables in the general definition
\eqref{Poperator} into linear and nonlinear terms, 
\begin{equation}
\mathcal{P}b = \sum_{i=1}^{\Ns} \prot{i} \frac{\langle
  \prot{i}b\rangle}{\epsilon y_i} 
+ \sum_{i=1}^{\Ns} (\prot{i}^2 - \epsilon y_i) \frac{\langle
  (\prot{i}^2 - \epsilon y_i)b\rangle}{2\epsilon^2 y_iy_j} 
+ \sum_{1\leq i<j\leq \Ns} \prot{i}\prot{j} \frac{\langle
  \prot{i}\prot{j} b\rangle}{\epsilon^2 y_i y_j}
\label{eq:nonlinear_P_explicit}
\end{equation}
For linear observables it now follows that
$\mathcal{P}\prot{i}=\prot{i}$ for subnetwork concentrations
($i=1,\ldots,\Ns$) and $=0$ for bulk concentrations
($i>\Ns$). Similarly, $\mathcal{P}\prot{i}\prot{j}=\prot{i}\prot{j}$
when both indices are in the subnetwork range, and $=0$
otherwise. The only exception is the case of two equal indices
($i=j$) in the subnetwork, where $\mathcal{P}\prot{i}^2 =
\prot{i}^2-\eps y_i$. Using again the vector $\bm{z}$ collecting all linear and
quadratic observables from the network this means there is a matrix
$\bm{P}$, given explicitly in \eqref{eq:PLmatrices} below, such that
\begin{equation}\label{eq:nonlinP1}
\begin{split}
  \mathcal{P}z_\alpha &= \sum_\beta z_\beta P_{\beta\alpha} + \order(\eps)
\end{split}
\end{equation}
We also need to know the action of $\mathcal{P}$ on higher order observables
$b=\dxany^n$ with $n\geq 4$. If $n$ is odd, then only the linear terms
in \eqref{eq:nonlinear_P_explicit} contribute, with $\langle\prot{i}b\rangle$
proportional to $\epsilon^{(n+1)/2}$ from the scaling of the
covariances. Thus $\mathcal{P}\dxany^n$ is of order $\epsilon^{(n-1)/2}
\dxany$, which is $\order(\epsilon)$ as $n\geq 3$. For even $n$, we
get only the quadratic terms from \eqref{eq:nonlinear_P_explicit};
the products with $b$ are proportional to $\epsilon^{n/2+1}$ and so
$\mathcal{P}\dxany^n$ is order $\epsilon^{n/2-1}[\dxany^2+\order(\eps)]$, hence
again $\order(\epsilon)$ as the smallest even value of $n$ is $4$.
Thus
\begin{equation}\label{eq:nonlinP2}
\begin{split}
  \mathcal{P}\dxany^n &= \order(\eps),\quad n\geq 3
\end{split}
\end{equation}
The analogous properties of the orthogonal projector $\mathcal{Q}$ follow
directly from the definition $\mathcal{Q}b=b-\mathcal{P}b$: its action
on linear or quadratic observables, can be represented by a matrix,
\begin{equation}\label{eq:nonlinQ1}
\begin{split}
  \mathcal{Q}z_\alpha &= \sum_\beta z_\beta Q_{\beta\alpha}
\end{split}
\end{equation}
while higher order terms remain of higher order:
\begin{equation}\label{eq:nonlinQ2}
\begin{split}
  \mathcal{Q}\dxany^n &= \dxany^n + \order(\eps),\quad n\geq 3\\
\end{split}
\end{equation}

We can now summarise the matrix representations for the nonlinear
case. These matrices are defined by the action of the operators on
linear or quadratic observables, up to terms that are at least cubic
in concentration or proportional to $\epsilon$:
\begin{equation}
\begin{split}
  \mathcal{L}z_\alpha &= \sum_\beta z_\beta L_{\beta \alpha} +
  \dxany^3 + \order(\eps)\\
  \mathcal{P}z_\alpha &= \sum_\beta z_\beta P_{\beta \alpha} +
  \dxany^3 + \order(\eps)\\
  \mathcal{Q}z_\alpha &= \sum_\beta z_\beta Q_{\beta \alpha} +
  \dxany^3 + \order(\eps)
\end{split}
\label{defn:nonlinoperators}
\end{equation}
On the other hand for higher order observables, only terms of the same
or higher order are created, or ones proportional to $\epsilon$:
\begin{equation}
\begin{split}
  \mathcal{L}\dxany^n &= \order(\dx^n) + \order(\epsilon),\quad n\geq 3
\end{split}
\label{eq:LPQproperty2}
\end{equation}
and similarly for $\mathcal{P}$ and $\mathcal{Q}$.  Terms of order
$\epsilon$ also remain of order $\epsilon$ or higher when one of the
three operators is applied. It then follows that, as in the linear
case, the product (composition) of any two operators has the same
properties, and its matrix representation is the product of the
matrices for the two operators. To see this consider e.g.\
\begin{equation}
\mathcal{LQ}z_\alpha = \mathcal{L}\biggl(
\sum_\beta z_\beta Q_{\beta\alpha} + \order(\dx^3)+\order(\epsilon)\biggr)
= \sum_{\beta,\gamma} z_\gamma L_{\gamma\beta}Q_{\beta\alpha} +
\order(\dx^3)+\order(\epsilon)
\end{equation}
This is the key result, which extends by induction to products over
any number of operators.

It will be useful later to have the explicit forms of the nonlinear
matrix representations:
\begin{equation}\label{eq:PLmatrices}
  \bm{P} =
  \begin{pmatrix}
    \mathbb{1} & \bm{0} & \bm{0} & \bm{0} & \bm{0}\\
    \bm{0} & \bm{0} & \bm{0} & \bm{0} & \bm{0}\\
    \bm{0} & \bm{0} & \mathbb{1} & \bm{0} & \bm{0}\\
    \bm{0} & \bm{0} & \bm{0} & \bm{0} & \bm{0}\\
    \bm{0} & \bm{0} & \bm{0} & \bm{0} & \bm{0}
  \end{pmatrix},
\quad
\bm{L} =
\begin{pmatrix}
  \bm{L}^\superss & \bm{L}^{\rm{s},\rm{b}} & \bm{0} &  \bm{0} & \bm{0} \\
  \bm{L}^{\rm{b},\rm{s}} & \bm{L}^\superbb & \bm{0} &  \bm{0} & \bm{0}\\
  \bm{L}^\supersss & \bm{L}^{\rm{ss},\rm{b}} & \bm{L}^{\rm{ss},\rm{ss}} 
  & \bm{L}^{\rm{ss},\rm{sb}} & \bm{L}^{\rm{ss},\rm{bb}}\\
  \bm{L}^{\rm{sb},\rm{s}} & \bm{L}^{\rm{sb},\rm{b}} & \bm{L}^{\rm{sb},\rm{ss}}
  & \bm{L}^{\rm{sb},\rm{sb}} & \bm{L}^{\rm{sb},\rm{bb}}\\
  \bm{L}^{\rm{bb},\rm{s}} & \bm{L}^{\rm{bb},\rm{b}} & \bm{L}^{\rm{bb},\rm{ss}}
  & \bm{L}^{\rm{bb},\rm{sb}} & \bm{L}^{\rm{bb},\rm{bb}}  
\end{pmatrix}
\end{equation}
Here the five rows and columns of the block structure relate to:
linear subnetwork observables (s), linear bulk observables (b),
product of subnetwork concentrations (ss), mixed subnetwork-bulk
products of concentrations (sb) and products of bulk concentrations
(bb). The fact that the top right blocks of $\bm{L}$ vanish comes from
the statement \eqref{eq:Lproperty1b}: application of
$\mathcal{L}$ to quadratic observables does not give linear terms. The
matrix representation $\bm{Q}$ of $\mathcal{Q}$ is analogous to that
of $\bm{P}$, with the roles of zero and identity matrices in the
diagonal blocks interchanged.

We can proceed at this point to find the rate matrix for the nonlinear
projected equations. Using \eqref{eq:projoperators}, we need to apply 
first the adjoint Fokker-Planck operator $\mathcal{L}$ to an
observable, and then the projector $\mathcal{P}$. The matrix
representation of this product of operators is $\bm{PL}$, thus
\begin{equation}
\mathcal{PL}z_\alpha = (\bm{z}^\T\bm{PL})_\alpha + \order(\epsilon)
\end{equation}
where there are no $\order(\delta x^3)$ terms because the projector
satisfies not just \eqref{defn:nonlinoperators} but in fact the
stronger \eqref{eq:nonlinP2}. The $\order(\epsilon)$ term can
furthermore be dropped when $\epsilon\to 0$. We now only need to
choose for $\alpha$ 
the indices that give us the subnetwork entries from $\bm{z}$, and can
then read from \eqref{eq:projoperators} the rate matrix entries. The
relevant indices are those in the first and third block columns of the matrices
in \eqref{eq:PLmatrices}. Writing out those columns of $\bm{PL}$ shows
that the linear rate matrix, whose elements are written
$\Omega_{ji}^\superss$ in the projected equations
\eqref{eq:proj_dimensional}, is simply the block $\bm{L}^\superss$ of
$\bm{L}$:
\begin{equation}
\bm{\Omega}^{\rm{s,s}} = \bm{L}^\superss
\label{eq:Omega_linear}
\end{equation}
As one might have expected, this is the same result as for the
linearised dynamics discussed in Sec.~\ref{sec:lindynamics}. Similarly the
nonlinear rate matrix is the block
\begin{equation}
\bm{\Omega}^{\rm{ss,s}} = \bm{L}^\supersss
\label{eq:Omega_quadratic}
\end{equation}

The same logic can now be applied to the calculation of the linear and
nonlinear memory functions, starting from \eqref{eq:projoperators}. 
The required operator involves an operator exponential, which can be
expressed as a series:
\begin{equation}
\mathcal{PLQ}e^{\mathcal{QLQ}t}\mathcal{QL}
= 
\sum_{n=0}^\infty
\frac{t^n}{n!}\mathcal{PLQ}(\mathcal{QLQ})^n\mathcal{QL}
\end{equation}
Every term in the sum is a product of operators, whose matrix
representation will be the product of the matrices for the individual
operators. Adding the series back up gives a matrix exponential, so
that
\begin{equation}
\mathcal{PLQ}e^{\mathcal{QLQ}t}\mathcal{QL}z_\alpha
= 
(\bm{z}^\T\bm{PLQ}e^{\bm{QLQ}t}\bm{QL})_\alpha +
\order(\epsilon)
\end{equation}
As before $\order(\delta x^3)$ terms are absent because the leftmost
operator is the projector $\mathcal{P}$, and we can drop the
$\order(\epsilon)$ terms when $\epsilon\to 0$. The remainder of the
reasoning is as for the rate matrix: the linear memory functions 
$M_{ji}^\superss(t)$ are the elements of the top left (s,s)
block of $\bm{PLQ}e^{\bm{QLQ}t}\bm{QL}$, while the nonlinear memory
functions $M_{(jk)i}^\supersss(t)$ are those of the (ss,s) block.

Also for the memory functions one can show that the linear
contributions are the same as for the linearised dynamics. To see
this, one can write the building blocks of
$\bm{PLQ}e^{\bm{QLQ}t}\bm{QL}$ in block form:
\begin{equation}\label{eq:more_matrices}
\begin{split}
  \bm{PL} &=
\begin{pmatrix}
  \bm{L}^\superss & \bm{L}^{\rm{s},\rm{b}} & \bm{0} &  \bm{0} & \bm{0} \\
  \bm{0} & \bm{0} & \bm{0} & \bm{0} & \bm{0} \\
  \bm{L}^\supersss & \bm{L}^{\rm{ss},\rm{b}} & \bm{L}^{\rm{ss},\rm{ss}} 
  & \bm{L}^{\rm{ss},\rm{sb}} & \bm{L}^{\rm{ss},\rm{bb}}\\
  \bm{0} & \bm{0} & \bm{0} & \bm{0} & \bm{0} \\
  \bm{0} & \bm{0} & \bm{0} & \bm{0} & \bm{0}
\end{pmatrix}
\\
\bm{QLQ} &=
\begin{pmatrix}
  \bm{0} & \bm{0} & \bm{0} & \bm{0} & \bm{0} \\
  \bm{0} & \bm{L}^\superbb & \bm{0} &  \bm{0} & \bm{0}\\
  \bm{0} & \bm{0} & \bm{0} & \bm{0} & \bm{0} \\
  \bm{0} & \bm{L}^{\rm{sb},\rm{b}} & \bm{0}
  & \bm{L}^{\rm{sb},\rm{sb}} & \bm{L}^{\rm{sb},\rm{bb}}\\
  \bm{0} & \bm{L}^{\rm{bb},\rm{b}} & \bm{0}
  & \bm{L}^{\rm{bb},\rm{sb}} & \bm{L}^{\rm{bb},\rm{bb}}  
\end{pmatrix}
\\
\bm{QL} &=
\begin{pmatrix}
  \bm{0} & \bm{0} & \bm{0} & \bm{0} & \bm{0} \\
  \bm{L}^{\rm{b},\rm{s}} & \bm{L}^\superbb & \bm{0} &  \bm{0} & \bm{0}\\
  \bm{0} & \bm{0} & \bm{0} & \bm{0} & \bm{0} \\
  \bm{L}^{\rm{sb},\rm{s}} & \bm{L}^{\rm{sb},\rm{b}} & \bm{L}^{\rm{sb},\rm{ss}}
  & \bm{L}^{\rm{sb},\rm{sb}} & \bm{L}^{\rm{sb},\rm{bb}}\\
  \bm{L}^{\rm{bb},\rm{s}} & \bm{L}^{\rm{bb},\rm{b}} & \bm{L}^{\rm{bb},\rm{ss}}
  & \bm{L}^{\rm{bb},\rm{sb}} & \bm{L}^{\rm{bb},\rm{bb}}  
\end{pmatrix}
\end{split}
\end{equation}
If we momentarily think of the sb and bb columns and rows as one,
denoted ``$*$b''below,
then $\bm{QLQ}$ has a lower triangular block structure. It follows that
$e^{\bm{QLQ}t}$ has the same structure, with the diagonal blocks the
exponentials of those of $\bm{QLQ}$. In particular the (b,b) block of 
$e^{\bm{QLQ}t}$ is $e^{\bm{L}^\superbb t}$. Multiplying by
$\bm{PL}$ and $\bm{QL}$ from the left and right, one then finds a matrix with
$(s,s)$ block equal to
\begin{equation}
\bm{M}^{\rm{s,s}}(\dt) = \lblockb{s}{b}e^{\lblockb{b}{b}\dt}\lblockb{b}{s}
\label{eq:M_linear}
\end{equation}
in agreement with the results from the linearised dynamics.

The nonlinear memory matrix can be obtained similarly with a bit of
algebra. Writing $\bm{E}(\dt) = e^{\bm{QLQ}t}$, the result has the
form
\begin{equation}
\bm{M}^{\rm ss,s}(\dt) = \lblockb{ss}{b} \bm{E}^{\rm b,b}(\dt)\lblockb{b}{s} +
\lblockb{ss}{*b} \bm{E}^{\rm *b,b}(\dt)\lblockb{b}{s} +
\lblockb{ss}{*b} \bm{E}^{\rm *b,*b}(\dt)\lblockb{*b}{s}
\label{eq:M_quadratic_general}
\end{equation}

We comment finally on the random force terms $r_i(t)$ in the nonlinear
projected equations \eqref{eq:proj_dimensional} for the subnetwork dynamics.
From \eqref{eq:random_force} we have $r_i(t) =
e^{\mathcal{QLQ}t}\mathcal{QL}\prot{i}$. Given that the $\prot{i}$
make up the first $\Ns$ components of the vector $\bm{z}$, we apply
the same argument as for the memory function:
\begin{equation}
e^{\mathcal{QLQ}t}\mathcal{QL}z_i = 
(\bm{z}^\T e^{\bm{QLQ}t}\bm{QL})_i + \order(\dx^3) + \order(\epsilon)
\end{equation}
For $\epsilon\to0$ the last term can again the dropped, but the
$\order(\dx^3)$ terms remain as we do not have a projection operator 
$\mathcal{P}$ applied last that would remove them. The random force
can therefore not be calculated in closed form from the matrix
representations we have introduced.

The linear and quadratic contributions to the random force
$r_i(t)$ are known explicitly, and given by the $i$-th entry of
$\bm{z}^\T e^{\bm{QLQ}t}\bm{QL}$. We look briefly at which
concentrations $\prot{i}$ enter here. Expanding out the matrix
exponential, one sees that all terms in $e^{\bm{QLQ}t}\bm{QL}$ contain
$\bm{Q}$ as the leftmost factor. From \eqref{eq:PLmatrices} together
with $\bm{Q}=\mathbb{1}-\bm{P}$, only the block rows labelled b, sb
and bb of this matrix are nonzero. Thus $\bm{z}^\T
e^{\bm{QLQ}t}\bm{QL}$ involves only linear terms $\prot{i}$ from the
bulk, and quadratic terms $\prot{j}\prot{k}$ with one or both factors
in the bulk. All linear and quadratic terms in the random force
therefore vanish when the bulk initial conditions are at steady state,
$\prot{i}=0$ for $i>\Ns$. Only third order and higher terms remain, so
we expect the random force to be small or negligible in this
case, consistent with the results of Fig.~\ref{fig:3p2csoln} above.

We can now also see why it is useful to include {\em all} products of
subnetwork concentrations among our set of observables for the
projection. These products are then removed from the random force by
the orthogonal projector $\bm{Q}$, ensuring it vanishes to linear and
quadratic order for a bulk initially in steady state. If we choose only to
include {\em some} subnetwork concentration products, e.g.\ those that
appear in the original reaction equations \eqref{deterministiceqns},
then the remaining ones can and generically will appear in the random
force, giving non-vanishing quadratic random force terms even for an
initial bulk steady state. As we normally want to use the projected
equations in the approximated form where random force terms are
omitted, including all subnetwork products in the projection is
preferable as it will lead to smaller random force contributions.



\subsection{Properties of memory functions}
\label{sec:memproperties1}

We next discuss some generic properties of memory functions, based on
the explicit expressions for linear and nonlinear memory functions
obtained in the previous section. Considering which memory functions
are nonzero shows that the nonzero memory functions relate to
molecular species on the {\em boundary} of the subnetwork to the bulk
(Sec.~\ref{sec:boundarystructure}). For the nonzero memory functions we then discuss
what sets their amplitude, i.e.\ the value at zero time difference (Sec.~\ref{sec:amplitudes})
and the timescale of their decay as this time difference increases
(Sec.~\ref{sec:timescales_channels}).

\subsubsection{Boundary structure}
\label{sec:boundarystructure}

Before discussing which memory functions can be nonzero, we need to
agree some conventions for how molecular species can be divided
between a subnetwork and bulk. We will assume that a subnetwork
complex can only be 
created by two subnetwork proteins, whereas a bulk complex can be
created by either two bulk proteins or a bulk and a subnetwork
protein.  This is a reasonable biological
assumption: we are generally interested in subnetworks that are small, e.g.\ to
aid interpretability of the dynamics, and contain molecular species
that are well understood in the sense that they do not form
``unknown'' complexes that we would assign to the bulk. Similarly
for complexes retained in the subnetwork description we can expect
that it is known how they are formed, and that the constituent proteins
are included in the subnetwork.


To see the consequences for the (nonlinear) matrix representation of
the adjoint Fokker-Planck operator in \eqref{eq:PLmatrices}, recall
that the equation of motion for linear and quadratic observables is,
from \eqref{eq:da_dt_is_La} and \eqref{eq:Lproperty1}, $\partial_t
z_\alpha = \sum_\beta z_\beta L_{\beta\alpha} + \dxany^3 +
\order(\eps)$. Thus the second index in $L_{\beta\alpha}$ determines
which equation of motion we are considering, while the
first index labels the variables featuring in this equation. Our
assumptions above then mean that some of the blocks of $\bm{L}$ are
zero. This applies to $\lblockb{ss}{b}$, which encodes
contributions quadratic in subnetwork concentrations to the equation
of motion for a bulk concentration. Looking back at the equations of
motion \eqref{deterministiceqns}, such contributions could arise
only from a bulk complex being formed from two subnetwork proteins,
which we have excluded. Similarly, as subnetwork complexes can only be
formed from subnetwork proteins, $\lblockb{bb}{s}$ must vanish; in
$\lblockb{sb}{s}$ only elements where the first and second index refer
to the same subnetwork species can be nonzero, which then correspond
to formation rates for bulk complexes from a subnetwork and a bulk
protein.
For the equations of
motion of quadratic observables, it is easiest to note that
\begin{equation}
\frac{\partial}{\partial t} \prot{i}\prot{j} = 
\frac{\partial \prot{i}}{\partial t} \prot{j} + 
\prot{i} \frac{\partial\prot{j}}{\partial t} 
= \sum_k \prot{k}\prot{j} L_{ki} +
\sum_k \prot{i} \prot{k} L_{kj} + \dxany^3 + \order(\eps)
\label{eq:Lquadratic_to_linear}
\end{equation}
where the $L_{ij}$ are the linear-linear entries of $\bm{L}$.  The
quadratic-quadratic blocks of $\bm{L}$, such as $\lblock{ss}{ss}$, can
then be obtained directly from the linear-linear ones. Because all terms on the
r.h.s.\ of \eqref{eq:Lquadratic_to_linear} contain a factor of either
$\prot{i}$ or $\prot{j}$, one then sees that the blocks $\lblockb{ss}{bb}$
and $\lblockb{bb}{ss}$ are generically zero. 

We summarize the discussion of the block structure of $\bm{L}$
briefly. Eq.~(\ref{eq:Lquadratic_to_linear}) implies generically that
all linear-quadratic blocks vanish as already shown in
\eqref{eq:PLmatrices}, and that
$\lblockb{ss}{bb}=\lblockb{bb}{ss}=0$. Our assumptions on the
subnetwork-bulk division imply further that $\lblockb{ss}{b}=0$ (bulk
complex never formed from two subnetwork proteins) and
$\lblockb{bb}{s}=0$ (subnetwork complex always formed from two
subnetwork proteins). Most entries of $\lblockb{sb}{s}$ are also
zero except for those of the form $L_{(sb),s}$, where the same
subnetwork species $s$ appears in the quadratic first index and the
linear second index.
Here and in the
following we use indices $b,c,c'$ etc for bulk species and
$s,\sp,\spp, u$ etc
for subnetwork species to make the distinction obvious from the notation.
These constraints then simplify the expression
\eqref{eq:M_quadratic_general} for the nonlinear memory matrix
considerably:
\begin{equation}
\bm{M}^{\rm ss,s}(t) = 
\lblockb{ss}{sb} [\bm{E}^{\rm sb,b}(t)\lblockb{b}{s} +
\bm{E}^{\rm sb,sb}(t)\lblockb{sb}{s}]
\label{eq:M_quadratic}
\end{equation}

Before looking at the consequences for the memory terms in the
nonlinear projected equations of motion, we comment briefly on the
local-in-time terms from the linear and nonlinear rate matrices, as
shown in the first line of \eqref{eq:proj_dimensional}. The discussion
above implies that reactions within the subnetwork contribute terms to
the equations of motion for subnetwork concentrations only via
$\lblockb{s}{s}$ and $\lblockb{ss}{s}$. As these just give the rate
matrices, cf.\ \eqref{eq:Omega_linear} and \eqref{eq:Omega_quadratic},
one deduces that all subnetwork reactions are captured, in their
original form, in local-in-time terms. This was one of the desiderata
for our projected description of the subnetwork dynamics.

More importantly, for the memory functions we can now deduce which can
be nonzero in a given 
subnetwork. Let us term ``boundary species'' the molecular species
from the subnetwork that interact directly with any bulk species, and
``interior'' species the others. 
Given our assumptions above, the
interaction of a boundary species with the bulk could be either a
unary reaction, where a subnetwork species $s$ is transformed into a
bulk species $b$ (by phosphorylation, say). This would give nonzero
entries in the blocks $\lblockb{b}{s}$ and $\lblockb{s}{b}$,
specifically $L_{bs}$ and $L_{sb}$, while such entries will be zero for
interior species $i$. More commonly, a boundary subnetwork protein $s$ and a
bulk protein $b$ can form a bulk complex $c$, contributing in addition
to $\lblockb{sb}{s}$, via the element
$L_{(sb),s}$, and to $\lblockb{sb}{b}$ via $L_{(sb),b}$ and $L_{(sb),c}$. 

The first
statement we can deduce about memory functions is that {\em memory
terms appear only in the equations of motion for boundary
species}. Mathematically, $M^{\superss}_{\sp s}(t)=
M^{\supersss}_{(\sp\spp),s}(t)=0$ when $s$ is an interior species. This follows
directly from \eqref{eq:M_linear} and \eqref{eq:M_quadratic} because, 
from the discussion above, the $s$-th columns of $\lblockb{b}{s}$ and
$\lblockb{*b}{s}$ vanish for an interior species $s$.

Turning now to boundary species $s$, we can further narrow down what
memory functions can be nonzero. For the linear memory $M^{\rm
  s,s}_{\sp s}(t)$ to be nonzero, the index of the species $\sp$
influencing the evolution of $s$ must be such that
$L_{\sp b}$ is nonzero for some bulk species $b$, giving a nonzero entry in
the $\sp$-th row of \eqref{eq:M_linear}. As we saw above, this is
possible only if $\sp$ is itself a boundary species, taking part in a
unary or binary reaction with the bulk. The conclusion is that linear
memory functions are nonzero {\em only for boundary species
influencing other boundary species}.


There are similar restrictions on the entries of the nonlinear memory matrix
$\bm{M}^{\rm ss,s}(t)$ for the time evolution of a
boundary species $s$. Looking at \eqref{eq:M_quadratic}, this
matrix is proportional to $\lblockb{ss}{sb}$, so that $M^{\rm
  ss,s}_{(\sp\spp),s}(t)$ can be nonzero only if there is a nonzero element
of the matrix $\bm{L}$ of the form $L_{(\sp\spp),(ub)}$ with $b$ a bulk
index. As $\sp$ and $\spp$ are both in the subnetwork, then also $u$ must
be in the subnetwork (as $\lblockb{ss}{bb}=0$). Our question then
becomes: which subnetwork products $\prot{\sp}\prot{\spp}$ appear in the
equation of motion for a subnetwork-bulk product $\prot{u}\prot{b}$?
Looking at \eqref{eq:Lquadratic_to_linear}, one has
\begin{equation}
\partial_t (\prot{u} \prot{b}) = \prot{u}\partial_t \prot{b} +
\ldots = \sum_{u'} L_{u'b} \prot{u}\prot{u'} + \ldots
\end{equation}
where $u'$ is a subnetwork index and the dots indicate terms that are
irrelevant here because they do not involve the product of two
subnetwork concentrations. One reads off that 
\begin{equation}
L_{(\sp \spp),(ub)}=\delta_{\sp u}L_{\spp b} + \delta_{\spp u}L_{\sp b}
\label{eq:Lss_sb_general}
\end{equation}
when $\sp<\spp$, while for $\sp=\spp$
\begin{equation}
L_{(\sp\sp),(ub)} = \delta_{\sp u}L_{\sp b}
\label{eq:Lss_sb_general_self}
\end{equation}
Therefore one of $\sp$ and $\spp$ must equal $u$, and the other index
must be a reaction partner of the bulk species $b$, hence a boundary
species. As $u$ is an arbitrary subnetwork species, this means that
{\em the only concentration products affecting the evolution of a
  boundary species via memory terms are products involving at least
  one boundary species}.


\subsubsection{Memory function amplitudes}
\label{sec:amplitudes}
We next want to analyse the amplitudes of the memory functions 
at zero time difference, to see what they can tell us
about the structure of the bulk and its interactions with the
subnetwork.

\

\noindent\textit{Linear amplitudes: Self-memory}

The linear memory matrix at zero time difference is, from
\eqref{eq:M_linear}, simply $\lblockb{s}{b}\lblockb{b}{s}$ because the
exponential $\bm{E}^{\rm b,b}(t)=e^{\lblockb{b}{b}t}$ reduces to the
identity matrix for $t=0$. To calculate these amplitudes we can look at the
possible structure of interactions between the subnetwork and bulk and
then identify the terms in $\lblockb{s}{b}\lblockb{b}{s}$ that
relate to these interactions. Initially we consider self memory
functions $M^{\rm s,s}_{s,s}(0)$, which give the coefficient of
$\prot{s}(t'=t)$ in the memory term of the equation for
$\prot{s}(t)$. As discussed above, only
boundary species in the subnetwork will have a nonzero self memory function.

\begin{figure}[ht!]
  \centering
  \subfloat[]{\label{fig:selfa2}\includegraphics[scale=0.75]{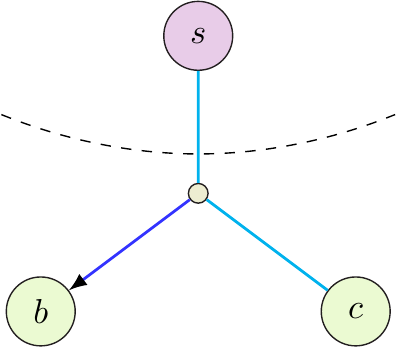}}\qquad
  \subfloat[]{\label{fig:selfa1}\includegraphics[scale=0.75]{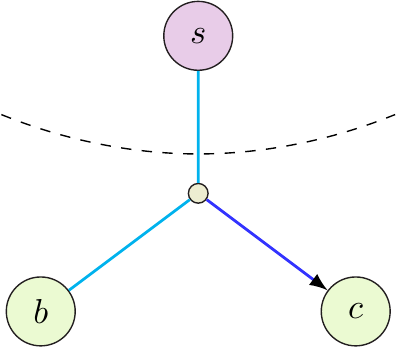}}\\
  \subfloat[]{\label{fig:selfa4}\includegraphics[scale=0.75]{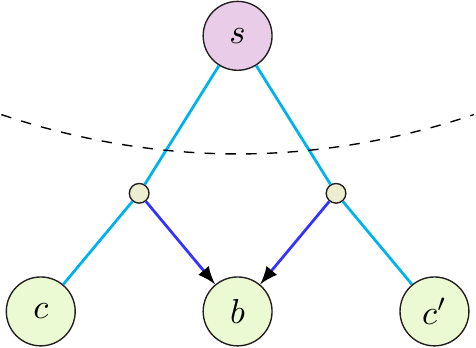}}\qquad
  \subfloat[]{\label{fig:selfa5}\includegraphics[scale=0.75]{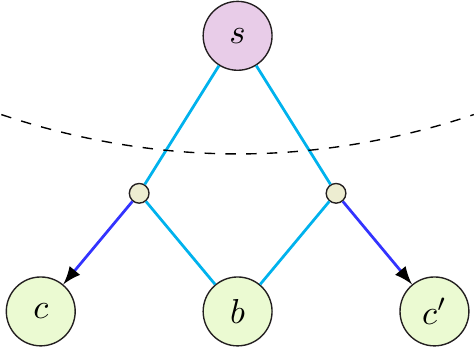}}\qquad
  \subfloat[]{\label{fig:selfa3}\includegraphics[scale=0.75]{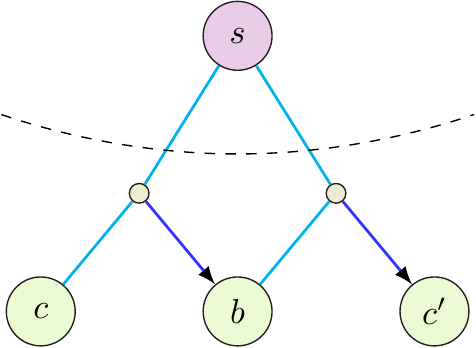}}
  \caption{Sketch of interaction patterns for self-memory terms. The
    subnetwork species $s$ reacts with (a) $c$ in the bulk to make
    $b$; (b) $b$ in the bulk to make $c$; (c) $c$ in the bulk
    to make $b$ and $c'$ in the bulk, also producing $b$; (d) $b$ in
    the bulk to make $c$ and $c'$; (e) $c$ in the bulk to make
    $b$, which reacts again with $s$ to produce $c'$.
  }
\end{figure}

Consider the self-memory for the subnetwork species $s$. By
considering all possible reactions between subnetwork and bulk one
finds for the relevant matrix elements of $\bm{L}$
\begin{equation}
\begin{split}
L_{sb}&=\sum_c(\kp{s}{c}{b}\y{c}-\kp{s}{b}{c}\y{b})+\lam{s}{b}\\
L_{bs}&=\sum_c(\km{b}{s}{c}-\kp{s}{b}{c}\y{s})+\lam{b}{s}
\end{split}
\label{eq:L_sb_explicit}
\end{equation}
and hence for the self-memory amplitude
\begin{equation}
M^{\rm s,s}_{ss}(0) = \sum_b L_{s,b}L_{b,s} 
= \sum_{b}\left[\sum_c(\kp{s}{c}{b}\y{c}-\kp{s}{b}{c}\y{b})+\lam{s}{b}\right]
\left[\sum_{c'}(\km{b}{s}{c'}-\kp{s}{b}{c'}\y{s})+\lam{b}{s}\right]
\label{eq:lin_self_mem}
\end{equation}
The first bracket here is a coefficient in the equation of motion for
species $b$, so encodes the initial effect of a deviation $\prot{s}$ of
the concentration of $s$ on the concentration $\prot{b}$, while the
second bracket gives the 
subsequent (after an infinitesimal time difference $t-t'$) feedback effect
from $b$ back to $s$. The different combinations of terms then
correspond to different interaction patterns. 

The product of the first terms in each bracket gives a contribution to
the self-memory amplitude of
\begin{equation}\label{selfa2}
\sum_{b}\left(\sum_{c'}\kp{s}{c'}{b}\y{c}\right)
\left(\sum_{c}\km{b}{s}{c}\right)
\end{equation}
This is a contribution only from reactions where bulk species $b$ is a
complex formed from subnetwork protein $s$ and another bulk protein
$c$ or $c'$. For the simplest case where there is only one such
reaction involving $b$ and $s$, this is sketched in
Fig.~\ref{fig:selfa2}. Intuitively, an increase in the concentration
of $s$ means that more $b$ will be formed
from the reaction with $c$. The bulk complex $b$ will then dissociate
again into $s$, thus increasing the rate of change of the
concentration of $s$. This produces a positive self-memory amplitude.

The product of the second terms in each bracket in
\eqref{eq:lin_self_mem} also give a positive contribution to the
self-memory amplitude, but from a combination of two negative effects:
\begin{equation}\label{selfa1}
   \sum_{b}\left(-\sum_{c'} \kp{b}{s}{c'}\y{b}\right)\left(-\sum_{c}
\kp{s}{c}{l}\y{s}\right) 
\end{equation}
For the simplest case ($c=c'$) this is sketched in 
Fig.~\ref{fig:selfa1}. 
Here the bulk species $b$ that transmits
the instantaneous memory forms a complex $c'$ together with the
subnetwork species $s$. An increase in the concentration of $s$ then
means that more of $b$ will be used in the 
formation of $c$: the concentration of $b$ is reduced. There will then
be less $b$ to react with $s$, 
giving overall a positive effect on the rate of change of $s$.

If the subnetwork species $s$ only takes part in one complex formation
reaction with the bulk, then the two terms
\eqref{selfa2} and \eqref{selfa1} give the total self-memory
amplitude, which will be positive. The same is true if $s$ reacts with
the bulk in several ways but none of these reactions share bulk
species.

If $s$ is involved in several overlapping complex formation reactions
with the bulk then one still gets the positive self-memory amplitude
contributions \eqref{selfa2} and \eqref{selfa1}, but now $c$ and $c'$
can be different as sketched in Figs.~\ref{fig:selfa4} and \ref{fig:selfa5}. 
In addition, however,
there can be negative memory contributions where a positive initial
effect from $s$ on $b$ combines with a negative subsequent effect,
cf.\ Fig.~\ref{fig:selfa3}, or vice versa. These are the cross terms from
\eqref{eq:lin_self_mem}, reading explicitly:
\begin{equation}
  \sum_{b}\left(\sum_{c'}\kp{s}{c'}{b}\y{l}\right)\left(-\sum_{c}
    \kp{s}{b}{c}\y{s}\right) +
  \sum_{b}\left(-\sum_{c'}\kp{b}{s}{c'}\y{b}\right)
  \left(\sum_c\km{b}{s}{c}\right)
\label{eq:negative_amplitude}
\end{equation}
The first product accounts for the fact that an increase in $s$ means that there will be more of it
to react with $c'$ to form $b$; $b$ then reacts with $s$ to make $c$,
having a negative effect on
the concentration of $s$. The second product corresponds to 
$s$ reacting with $b$ to form $c'$ (negative effect) and then $b$ dissociating into $s$ and $c$
(positive effect). Because such negative self-memory contributions
rely on a single bulk species being both formed in one complex
reaction and acting as reaction partner in a further complex
formation, they 
necessarily involve ternary subnetwork-subnetwork-bulk complexes. On this
basis one would expect that {\em positive} linear self-memory
amplitudes are the norm while negative ones, where contributions of the
type \eqref{eq:negative_amplitude} would have to dominate, the exception.

We have not yet discussed the unary reaction contributions to the self-memory amplitude. Where such reactions do not overlap with other subnetwork-bulk reactions, they make a positive contribution of $\lam{b}{s}\lam{s}{b}$ to the amplitude \eqref{eq:lin_self_mem}. Negative contributions would result only from overlap, where a unary reaction partner $b$ of a subnetwork species $s$ is also a reaction partner in a complex formation reaction with $s$.

\

\noindent\textit{Linear amplitudes: Cross-memory}
\begin{figure}[ht!]
  \centering
  \subfloat[]{\label{fig:crossb}\includegraphics{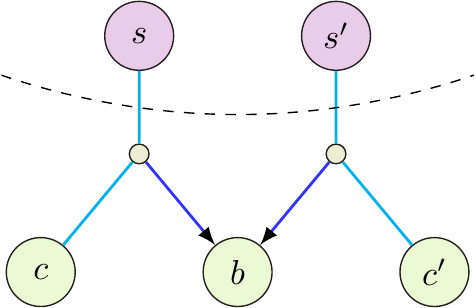}}\qquad
  \subfloat[]{\label{fig:crossa}\includegraphics{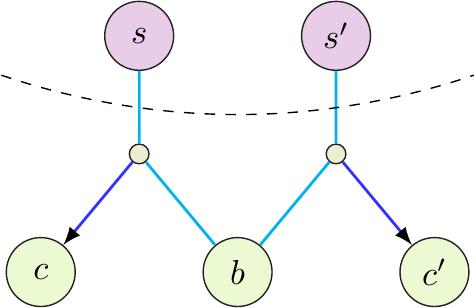}}\\
  \subfloat[]{\label{fig:crossc}\includegraphics{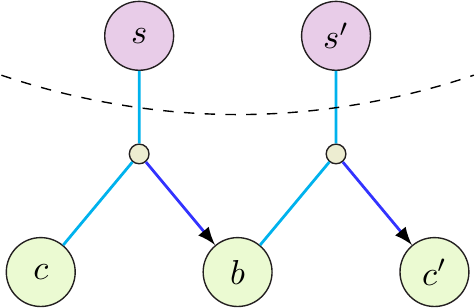}}\qquad
  \subfloat[]{\label{fig:crossd}\includegraphics{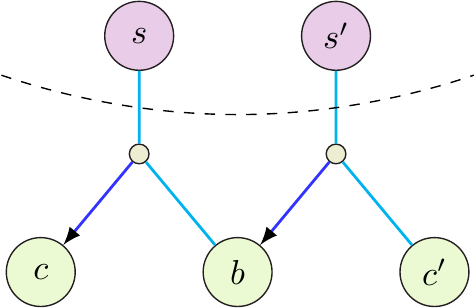}}
  \caption{Interaction patterns for cross-memory amplitudes. See text for discussion.}
  \label{fig:crossterms}
\end{figure}

All other linear memory function amplitudes $M^{\rm s,s}_{\sp s}(0)$, where the concentration of $\sp$ influences the rate of change of that of $s$, can be calculated similarly. Explicitly, each cross-memory amplitude is given by the analogue of \eqref{eq:lin_self_mem},
\begin{equation}
M^{\rm s,s}_{\sp s}(0) = \sum_b L_{\sp b}L_{bs} 
= \sum_{b}\left[\sum_{c'}(\kp{\sp}{c'}{b}\y{c'}-\kp{\sp}{b}{c'}\y{b})+\lam{\sp}{b}\right]
\left[\sum_{c}(\km{b}{s}{c}-\kp{s}{b}{c}\y{s})+\lam{b}{s}\right]
\label{eq:lin_cross_mem}
\end{equation}
Leaving aside unary reactions, there are four possible cases as shown
in Figure~\ref{fig:crossterms}. A positive contribution to the
amplitude is obtained when the subnetwork species $s$ and $s'$ have a
shared reactant or a shared product $b$ as in Figures~\ref{fig:crossb}
and \ref{fig:crossa}, 
and a negative amplitude contribution is obtained when the shared
species $b$ is a reactant in one reaction and a product in the other
reaction as in Figures~\ref{fig:crossc} and \ref{fig:crossd}. As
before this relies on the existence of ternary
subnetwork-subnetwork-bulk complexes and so positive contributions
would generically be expected to dominate. For example in EGFR we only
have two negative cross-memory amplitudes, in the cross terms between
Grb2 and SOS as shown in
Sec.~\ref{sec:memproperties}. 

\

\noindent\textit{Nonlinear amplitudes: Self-memory}

We can look similarly at the amplitude of nonlinear memory functions
$M^{\rm ss,s}_{(\sp\spp),s}(0)$. These are the elements of the matrix
$\bm{M}^{\rm ss,s}(0)$, which from \eqref{eq:M_quadratic} is given by
$\lblockb{ss}{sb}\lblockb{sb}{s}$ because $\bm{E}(0)$ is the identity
matrix. Recall now
that the only nonzero elements of $\lblockb{sb}{s}$ are of the form
$L_{(sb),s}$, so that
\begin{equation}
M^{\rm ss,s}_{(\sp \spp),s}(0) = \sum_b L_{(\sp\spp),(sb)}L_{(sb),s} 
\label{eq:M_nonlinear_amplitude}
\end{equation}
The first factor in turn was determined above
in (\ref{eq:Lss_sb_general},\ref{eq:Lss_sb_general_self}), while the
second one is given explicitly by 
\begin{equation}
L_{(sb),s} =
-\sum_{c}\kp{s}{b}{c}
\label{eq:L_sb_s}
\end{equation}
as can be read off from the equation of
motion for $\prot{s}$.

We look at the simpler case $\sp=\spp$ first, which gives the quadratic
self-memory amplitude. Inserting \eqref{eq:Lss_sb_general_self} into
\eqref{eq:M_nonlinear_amplitude} 
and using the explicit form of $L_{(sb),s}$ and $L_{sb}$ (cf.\
\eqref{eq:L_sb_explicit}) yields the constraint $\sp=s$, so that the
only nonzero amplitude for this case is
\begin{equation}
M^{\rm ss,s}_{(ss),s}(0) =
\sum_b\left[\sum_{c'}(\kp{s}{c'}{b}\y{c'}-\kp{s}{b}{c'}\y{b})+\lam{s}{b}\right]
\left(-\sum_{c}\kp{s}{b}{c}\right)
\label{eq:selfprod}
\end{equation}
Pairing the first and second term in the first bracket with the second
factor corresponds to the interaction patterns shown above in
Figs.~\ref{fig:selfa5} and~\ref{fig:selfa3}.
The sign of each contribution to 
the quadratic self-memory amplitude is the same as the
corresponding contribution to the linear self-memory. 

Explicitly, the second pairing from above gives the amplitude of the
self product of \ref{fig:selfa5}.
This
is
\begin{equation}\label{selfproda} 
  \sum_{b}\left(-\sum_{c'}\kp{s}{b}{c'}\y{b}\right)
  \left(-\sum_{c}\kp{s}{b}{c}\right) 
\end{equation}
where an increase in $s$ means that it will be used up in the reaction
with $b$ to form $c'$. There will then be less $s$ to react with $b$ to
form $c$ subsequently, having a positive effect on the rate of change of $s$. 
Note that such a contribution to the quadratic
self-memory amplitude will be present for any
boundary species, with the restriction $c=c'$ if it participates only
in non-overlapping bulk reactions. 

The first combination from \eqref{eq:selfprod} corresponds to
Fig.~\ref{fig:selfa3} with $c$ and $c'$ swapped 
and gives a contribution
\begin{equation}
  \sum_{b}\left(\sum_{c'}\kp{s}{c'}{b}\y{c'}\right)
  \left(-\sum_{c}\kp{s}{b}{c}\right) 
\end{equation}
Here, an increase in $s$ allows more $b$ to be formed from $s$ and
$c'$. The $b$ then reacts with $s$, having a negative effect on the rate
of change of $s$.

\

\noindent\textit{Nonlinear amplitudes: Cross-memory}

One can discuss nonlinear cross-memory amplitudes, where $\sp<\spp$,
in a similar 
fashion. Starting from \eqref{eq:Lss_sb_general} one finds
\begin{equation}
M^{\rm ss,s}_{(\sp\spp),s}(0) = \delta_{\spp s}
\sum_b\left[\sum_{c'}(\kp{\sp}{c'}{b}\y{c'}-\kp{\sp}{b}{c'}\y{b})
+\lam{\sp}{b}\right]
\left(-\sum_{c}\kp{s}{b}{c}\right) + (\sp \leftrightarrow \spp)
\label{eq:crossprod}
\end{equation}
The shorthand $(\sp \leftrightarrow \spp)$ indicates that the analogous term
with $\sp$ and $\spp$ swapped has to be added, to account for the two
alternative cases $\sp<\spp=s$ and $s=\sp<\spp$. The delta function prefactor
indicates that we get nonzero amplitudes only for concentration
products where one factor equals 
$\prot{s}$, the concentration of the species being influenced; the
other factor has to relate to a boundary 
species. More generic products, constrained only by the fact
that one factor has to relate to a boundary species, can thus contribute
to memory terms only at nonzero time difference.

As before one can combine each of the first two terms in the first
factor in \eqref{eq:crossprod} with the second factor and then sees
that these correspond to the interaction patterns sketched in
Fig.~\ref{fig:crossc} and \ref{fig:crossd}. 
The signs of the amplitude contributions are again
the same as for the linear cross-memory. 

\subsubsection{Memory function timescales}
\label{sec:timescales_channels}

So far we have discussed the amplitude of the memory functions. For
overall effect of the memory terms equally important is the timescale
on which the memory functions decay.  For a generic memory function
$M(t)$ we will identify this timescale by dividing its time integral
by the amplitude: $\timescale{}=[1/M(0)] \int_0^\infty dt\,M(t)$. If $M(t)$
decays as a single exponential, $M(t)=M(0)e^{-t/\tau}$, this would
give back the decay time $\tau$. 

Applying this definition to the linear self-memory $M^{\rm
  s,s}_{ss}(t)$ and using \eqref{eq:M_linear} gives an explicit
expression for the timescale of the form 
%
\begin{equation}
\label{eq:timescale}
\timescale{s} =
\frac{[\lblockb{s}{b}(\lblockb{b}{b})^{-1}\lblockb{b}{s}]_{ss}}
{[\lblockb{s}{b}\lblockb{b}{s}]_{ss}}
\end{equation}
Qualitatively, one sees from this that each $\timescale{s}$ is a
weighted average of elements of $(\lblockb{b}{b})^{-1}$. As the
elements or $\lblockb{b}{b}$ are all proportional to reaction rates
within the bulk, this shows that generally the memory function
timescales will scale with the inverse of these rates: memory functions
decay more quickly the faster the dynamics in the bulk. One can check
this explicitly by scaling up all bulk reaction rates by a certain
factor; the timescales $\timescale{s}$ will then decrease by the same
factor. One proviso here is that the steady state concentrations
$\y{i}$ must be maintained because contributions of $\lblockb{b}{b}$
from complex formation reactions are of the form $\kp{}{}{}\y{}$. In
the example \eqref{eq:3p2c}, one would need to keep the ratio of
$\kp{1}{4}{5}$ and $\km{5}{1}{4}$ constant while scaling up both rates.


In practice the reaction rates have to be treated as given so the
scaling argument does not necessarily help to understand the order of
magnitude of the memory timescales $\timescale{s}$. One concept that
can provide quantitative insight is to think of the memory functions
as decomposed according to source and receiver ``channels''. The
(linear) memory matrix \eqref{eq:M_linear} is proportional to
$\lblockb{s}{b}$ and $\lblock{b}{s}$. Both of these can be seen as
a superposition of contributions from individual reactions between
boundary species and bulk. If we identify each such contribution as a
source channel (for $\lblockb{s}{b}$) or a receiver channel (for
$\lblockb{b}{s}$), then the memory function is a sum of
all possible contributions of source and receiver channels. Here the
source tells us by which boundary reaction a concentration
fluctuation in the past initially propagates into the bulk, and the
receiver channel defines by which route it feeds back into the
subnetwork. We will explore this decomposition of memory functions in
the concrete example that is the subject of the next
section.

Each combination of source and receiver will have a specific
``propagation timescale'' in the bulk, which will consist of
combinations of entries of $(\lblockb{b}{b})^{-1}$. The overall memory
timescale $\timescale{s}$ from \eqref{eq:timescale} can then be viewed
as a weighted average of these propagation timescales, but bearing in
mind that the weights can be both positive and negative.

The memory functions are dominated by the reactions that have larger
contributions in the blocks $\lblockb{s}{b}$ and $\lblockb{b}{s}$. For
example, if we consider a reaction of the form $s + b \rightarrow b'$ then this will dominate if either the steady state concentration
$\y{b}$ or the complex formation rate $\kp{s}{b}{b'}$ are large enough for the product $\kp{s}{b}{b'}\y{b}$ to be significantly larger than for any competing reaction.

\

A very similar decomposition can be performed for nonlinear memory
functions, using \eqref{eq:M_quadratic}. If we write $\bm{QLQ}^{-1}$ in block form as
\begin{equation}
  \label{eq:qlqi}
  \bm{QLQ}^{-1} =
  \left(\begin{array}{ccc}
    (\lblockb{b}{b})^{-1}& \bm{0}&\bm{0}\\
\bm{I}^{\rm{sb},\rm{b}}&\bm{I}^{\rm{sb},\rm{sb}}&\bm{I}^{\rm{sb},\rm{bb}}\\
\bm{I}^{\rm{bb},\rm{b}}&\bm{I}^{\rm{bb},\rm{sb}}&\bm{I}^{\rm{bb},\rm{bb}}
  \end{array}\right)
\end{equation}
then the timescale for e.g.\ the nonlinear self-memory functions is
\begin{equation}
  \label{eq:timescalenonlin}
   \timescale{ss}= \frac{\left[\lblockb{ss}{sb}\left(\bm{I}^{\rm{sb},\rm{b}}\lblockb{b}{s}+ \bm{I}^{\rm{sb},\rm{sb}}\lblockb{sb}{s}\right)\right]_{(ss),s}}{\left[\lblockb{ss}{sb}\lblockb{sb}{s}\right]_{(ss),s}}
\end{equation}

The source channels are
identified by grouping the contributions to $\lblockb{ss}{sb}$, which
are given explicitly in \eqref{eq:Lss_sb_general} and
\eqref{eq:Lss_sb_general_self}, according to reactions between
boundary and bulk species. The terms in $\lblockb{b}{s}$ and
$\lblockb{sb}{s}$, cf.\ \eqref{eq:L_sb_s}, can be grouped analogously into
receiver channels.


\section{Application to EGFR}
\label{sec:EGFR}
\begin{figure}[!ht]
\centering
  \includegraphics[scale=0.6]{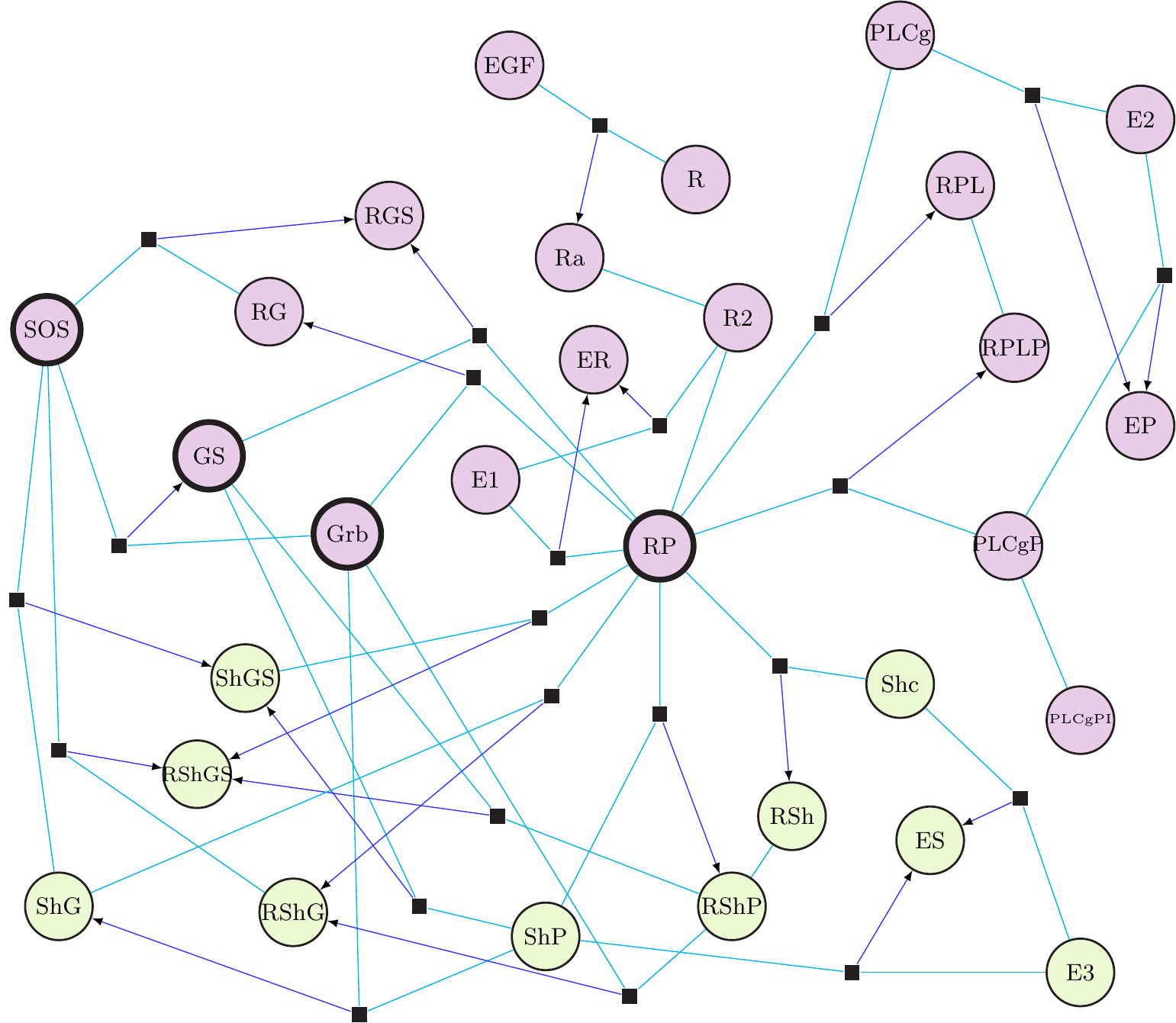}
  \caption{Factor graph of EGFR network as described in
    \citet{Kholodenko99} adapted to include enzyme reactions. Three added enzyme reactions with enzymes denoted E1-3 and enzyme-substrate complexes ``enzyme-R'' (denoted ER), ``enzyme-PLC$\gamma$'' (denoted EP), ``enzyme-Shc'' (denoted ES) are also shown to capture Michaelis-Menten contributions to the dynamics. 
    The purple nodes denote subnetwork species while the bulk species
    are shown in green. See \ref{appendix1} for a full list of abbreviations for each network component.
  }
\label{fig:egfrnetwork} 
\end{figure}

We apply the projection method to a model for the signalling network
of epidermal growth factor receptor (EGFR) developed by
\citet{Kholodenko99}, see Figure~\ref{fig:egfrnetwork}. 

\subsection{Setup of EGFR model for application of projection technique}
\label{sec:EGFR_setup}

We use the mass action rate parameters from \citep{Kholodenko99}. Most
of these fit directly into our setup of unary and binary
reactions. The network also involves three Michaelis-Menten reactions
transforming a ``substrate'' into a ``product''. We incorporate these
by adding to the description one enzyme and one enzyme-substrate
complex species per reaction, with the rates for formation and
dissociation of the complex large enough in order to force the two
added species to be in equilibrium at any time with the prevailing
substrate and product concentrations
\cite{Murray2001}. 
Initial
conditions for the added species are then also derived from those of
the substrate according to this equilibrium criterion. We choose the
above method of incorporating Michaelis-Menten reactions as it
allows direct application of the framework developed so far. We will
report separately on an alternative approach where the enzyme
reactions do not need to be represented explicitly.

To apply the projection method we need to first select a subnetwork
and bulk from the EGFR network. We have chosen the bulk to be the
protein Src homology and collagen domain protein (Shc) and any complexes that include Shc, consistent with our
convention that if a protein is
in the bulk, any complexes containing that protein will also be in
the bulk. Shc and its complexes interact directly with four subnetwork
species, which therefore form the boundary of our subnetwork; they
are phosphorylated EGFR (denoted RP), growth factor receptor-binding protein 2 (Grb2, Son of Sevenless homolog protein (SOS), and protein complex Grb2-SOS (denoted GS; see \ref{appendix1} for a full list of abbreviations for network components). 
We then apply the projection method to obtain
a set of equations for the subnetwork species. For the interior
species these will have the original mass action form, while the
boundary species will acquire additional memory (and random force) terms.

To avoid having to carry around concentration units in the following,
we will switch to dimensionless concentrations defined as
\begin{equation}
  \label{eq:nondimdx}
\protnondim{i} = \prot{i}/\y{i} 
\end{equation}
Intuitively, the $\protnondim{i}$ are fractional
concentration deviations from the steady state. The lowest value is
$-1$, corresponding to a concentration of zero ($100\%$ below steady
state), while e.g.\ $\protnondim{i}=2$ indicates that the concentration of
$i$ is three times that in steady state.

The projected equations
\eqref{projection} written in terms of the dimensionless
$\protnondim{i}$ take the form 
\begin{equation}\label{eq:projnondim}
\begin{split}
  \frac{\partial}{\partial t}\protnondim{i} &=
  \sum_{j=1}^{\Ns}\protnondim{j}\tilde{\Omega}_{ji}^\superss +
  \sum_{1\leq j\leq k\leq\Ns}\protnondim{j}\protnondim{k}\tilde{\Omega}_{(jk)i}^\supersss\\
  &\quad +
 \int_0^tdt'\left( \sum_{j=1}^{\Ns}\protnondim{j}(t')\tilde{M}_{ji}^\superss(t-t')
    +\sum_{1\leq j\leq k\leq\Ns}\protnondim{j}(t')\protnondim{k}(t')\tilde{M}_{(jk)i}^\supersss(t-t')\right)
  + \tilde{r}_i(t)
\end{split}
\end{equation} 
where
\begin{equation}
\begin{split}
&\tilde{\Omega}_{ji}^\superss =
\y{j}\Omega_{ji}^\superss\y{i}^{-1}, \qquad
\tilde{\Omega}_{(jk)i}^\supersss =
\y{j}\y{k}\Omega_{(jk)i}^\supersss\y{i}^{-1},\\
\tilde{M}_{ji}^\superss(t) =
\y{j}M_{ji}^\superss(t)\y{i}^{-1},& \qquad
\tilde{M}_{(jk)i}^\supersss(t) =
\y{j}\y{k}M_{(jk)i}^\supersss(t)\y{i}^{-1}, \qquad
\tilde{r}_i(t) = r_i(t)\y{i}^{-1}
\end{split}
\end{equation}
The rescaled rate matrix entries and random forces have dimensions of
rate, i.e.\ inverse time, while the rescaled memory functions have dimensions of
rate squared. The rate matrices and memory functions are calculated
by first constructing the matrix $\bm{L}$ for the network, then
using \eqref{eq:Omega_linear}, \eqref{eq:Omega_quadratic},
\eqref{eq:M_linear} and \eqref{eq:M_quadratic}, and finally switching
to dimensionless concentrations as explained above.

We will first discuss qualitative features of the memory functions
themselves.  Quantitative tests of the projected equations are
presented next; as before, we will drop the random force terms so that
the equations are a closed system determining the timecourses of the
subnetwork concentrations for any given initial condition.  To solve
this numerically, we implement a solver for systems of
integro-differential equations \cite{Day1967}. 
The enzyme reactions are much faster than the
remainder of the kinetics and this causes the system of equations to become what is known as ``stiff''.
This can be handled by converting enzyme
reaction terms in the subnetwork back into Michaelis-Menten form
before using the numerical solver, or by transforming the projected
equations into an enlarged set of differential equations
\cite{Rubin2014a}
that can then be integrated using standard methods for stiff systems
\cite{Hairer1996}. 

\subsection{Memory function properties}
\label{sec:memproperties}
We will now look at how general properties of the memory functions described
in Section~\ref{sec:memproperties1} manifest themselves in the EGFR network.
We first analyse the amplitudes of the memory functions as in
Section~\ref{sec:amplitudes}, to see how these reflect the structure of the
network.

\begin{figure}[ht!]
  \centering
 \includegraphics[scale=0.8]{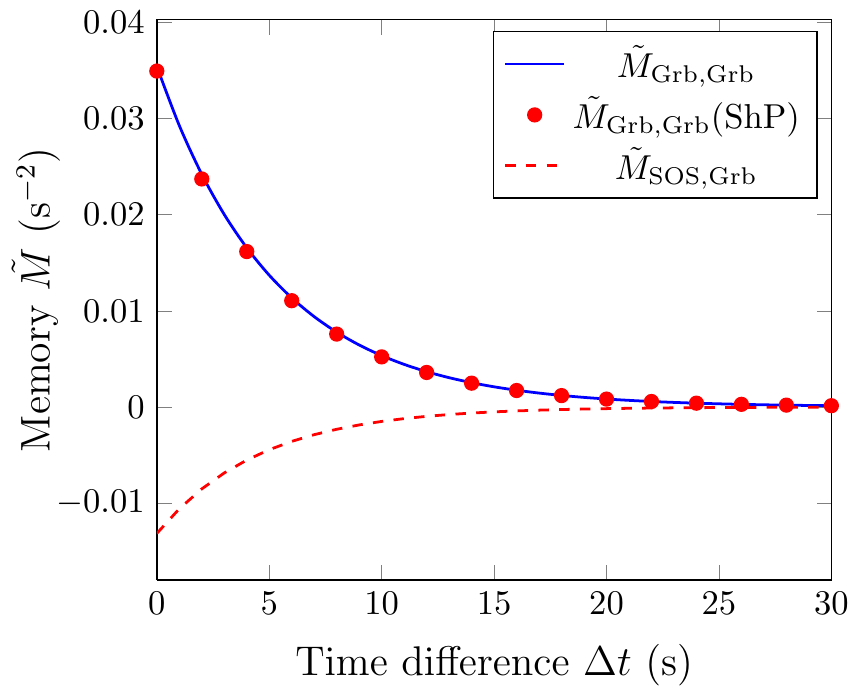}
  \caption{Memory functions in EGFR network: coefficients of
    Grb2 and SOS in the linear memory term for Grb2. The self memory 
function of Grb2 is compared to the contribution coming only from the reaction with phosphorylated Shc (ShP) as source and receiver channel. Contributions from other channels are no larger than $8\cdot10^{-4}$s$^{-2}$.
}
  \label{fig:grbmem}
\end{figure}

Figure~\ref{fig:grbmem} shows two of the linear memory functions in
the equation of motion for the concentration of Grb2, one the
self-memory and the other the cross-memory that determines the
influence of past concentration values of SOS. The amplitudes are
given by the intercepts with the $y$-axis ($\dt=0$): one sees that the
self-amplitude of Grb2 is positive. To understand why, we note that
Grb2 has two different reactions with bulk species: 
\begin{equation}
  \begin{split}
    \text{Grb2} + \text{ShP} &\leftrightharpoons \text{ShG}\\
    \text{Grb2} + \text{RShP} &\leftrightharpoons \text{RShG}
  \end{split}
\label{eq:Grb_reactions}
\end{equation}
As these do not overlap, each reaction gives separate contributions to
the self-memory amplitude of the
form shown in Figs.~\ref{fig:selfa2} and \ref{fig:selfa1}, which are always
positive.

The amplitude of the cross-memory of $\protnondim{\text{GRB}}$ to
$\protnondim{\text{SOS}}$, on the other hand, is negative as
Figure~\ref{fig:grbmem} shows. To rationalise this, note that there
are two bulk species that are shared between the bulk reactions of
SOS and Grb2, namely ShG (ShP-Grb2) and RShGS (RShG-SOS/ShGS-RP). The reaction patterns involving
these species both have the structure of Fig.~\ref{fig:crossd}, and
hence both give negative contributions.

Next we look at the time-dependence of the memory functions, and in
particular the channel decomposition described in
Sec.~\ref{sec:timescales_channels}. For the self-memory of Grb2, there
are two source and receiver channels, namely the two bulk reactions
\eqref{eq:Grb_reactions}. The memory function can be decomposed into
four pieces according to the combination of these four channels, e.g.\
``out via ShP (source) and in via RShP (receiver)''. It turns out in
this case that the channel via phosphorylated Shc (ShP) dominates entirely. This is shown
in Fig.~\ref{fig:grbmem}, which compares the total memory function
with its ``out and in via ShP'' contribution.


Note that the rates for both reactions \eqref{eq:Grb_reactions}
between Grb2 and the bulk are the same; 
however the steady state concentration of phosphorylated Shc (ShP) is much larger than the
steady state value of RShP and therefore the reaction between Grb2 and phosphorylated Shc (ShP) to make ShP-Grb2 (ShG) is the one that dominates the self memory function of
Grb2. Therefore it may be useful to study interactions between Shc and
Grb2 to help understand why this reaction dominates the memory.


\

\begin{figure}[!ht]
  \centering
 \includegraphics[scale=0.8]{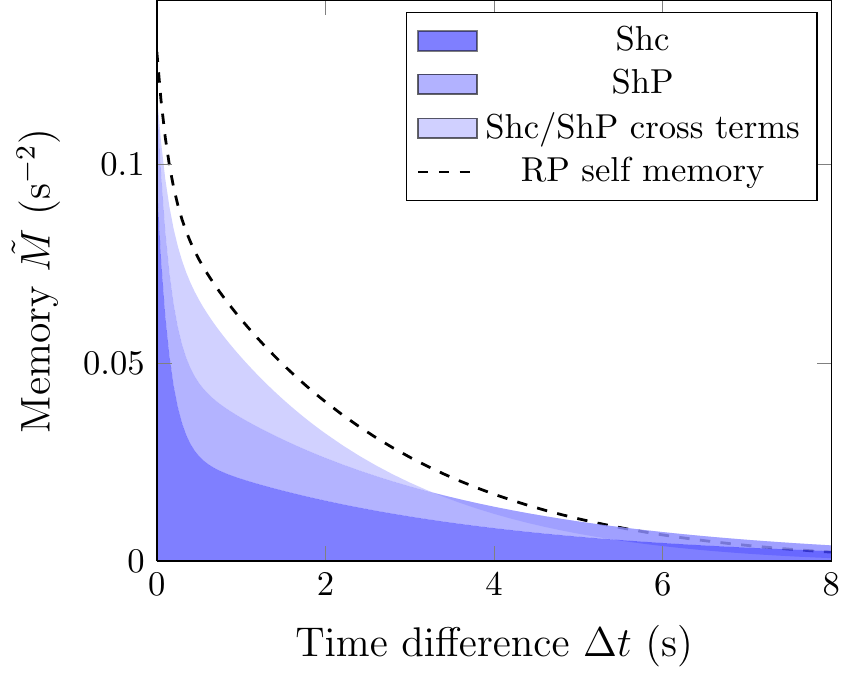}
 \caption{Comparison of self memory function of phosphorylated EGFR (RP) with dominant terms
   from the channel decomposition: in and out via Shc; in and out via phosphorylated Shc (ShP); and in via Shc and out via phosphorylated Shc (ShP) and vice versa, shown
   together. 
 }
  \label{fig:rpaccumulatemem}
\end{figure}

For the self-memory of phosphorylated EGFR (RP), which has four interactions with the bulk,
the channel decomposition is richer because there are now 16
combinations of the four source and receiver channels.
None of the $\kp{s}{b}{c}\y{b}$ 
values
for the interactions with the bulk proteins is large enough to be
entirely dominant and accordingly there are several
channel combinations that give significant contributions to the
memory. Fig.~\ref{fig:rpaccumulatemem} shows the four that are
largest: in and out via Shc; in and out via ShP; and in via Shc and
out via ShP and vice versa. The combination of these gives a good
account of the overall shape of the memory function, indicating that
two channels (Shc and phosphorylated Shc (ShP)) are dominant over the other two (ShP-Grb2 (ShG) and ShGS (Shc-Grb2-SOS)
). Looking at the figure closely one sees that the
cross-channel contribution between Shc and ShP is positive for short
time differences but becomes negative (and small) for longer time
differences. In contrast the two other terms, for which the source and receiver
channel is the same, are always positive. 

\begin{figure}[!ht]
  \centering
 \includegraphics[scale=0.8]{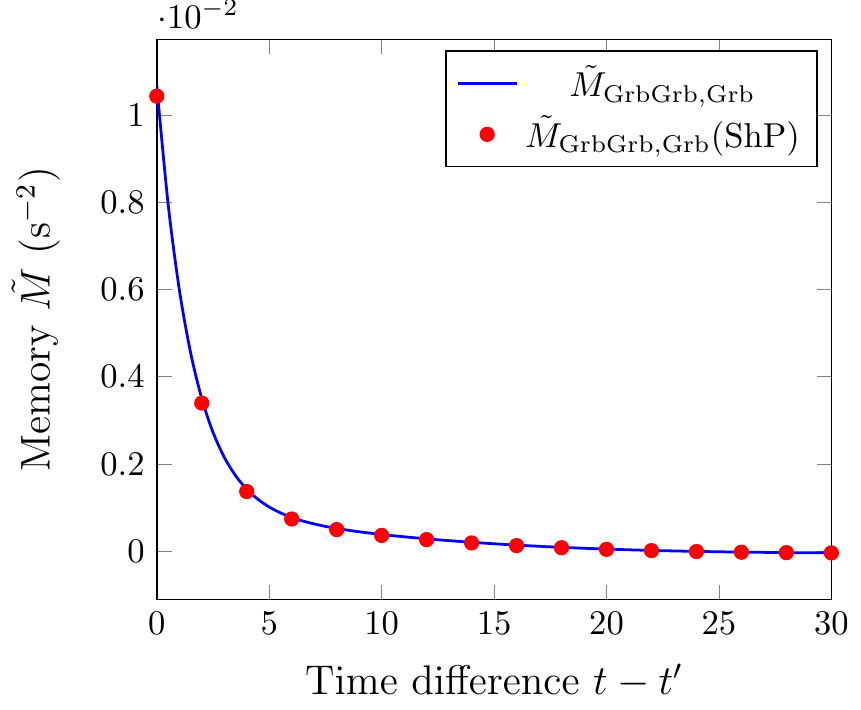}
 \caption{Comparison of nonlinear self memory function of Grb2 with
   the contribution from the channel combination ``in and out via phosphorylated Shc (ShP)''.}
  \label{fig:grbnonlinselfmem}
\end{figure}

An analogous channel decomposition can be performed for (linear) cross memory
and nonlinear memory functions as explained in
Sec.~\ref{sec:timescales_channels}. 
As for the linear self memory functions above, we find that
often 
only a few channels provide the dominant contribution. This occurs for
all the memory functions of EGFR. Fig.~\ref{fig:grbnonlinselfmem}
shows that the nonlinear self memory of Grb2 is dominated by the
reaction Grb2 + ShP $\rightarrow$ ShG acting as source and receiver,
i.e.\ by the channel combination ``in and out via phosphorylated Shc (ShP)''. This is not
unexpected as the same combination dominates the linear self memory
(see Fig.~\ref{fig:grbmem}).

The channel decomposition can also be used to analyse self memory function
timescales $\timescale{s}$ as defined in \eqref{eq:timescale}.
In particular, if there is a single channel that dominates
the memory function then the memory function contribution from this
channel will have a similar timescale to the full memory function.
%
%
For example, we find that the self memory of
Grb2 has a timescale $\tau_{\text{Grb}} = 5.31\,{\rm s}$. 
The contribution from the phosphorylated Shc (ShP) channel as source and
receiver, shown in Fig.~\ref{fig:grbmem}, has a timescale that is
very close to this, namely $5.25\,{\rm s}$.

The memory function of phosphorylated EGFR (RP) has a faster timescale, $\tau_{\text{RP}} =
1.68\,{\rm s}$. The contributions to the memory function which come
from the ``in and out via Shc'' and ``in and out via phosphorylated Shc (ShP)'' channel
combinations, on the other hand, have timescales of $1.08\,{\rm s}$
and $2.86\,{\rm s}$,  respectively. 
The timescales of the
contributions from the dominant channels are therefore sufficient to give
an order-of-magnitude estimate of the overall memory function
timescale. 








The dominance of certain channels encourages us to look at how the
system behaves if species or reactions that do not 
appear to contribute to the behaviour of the system are removed. The reaction
between Grb2 and RShP does not make a significant contribution to the
memory functions of Grb2. Excluding this
reaction does not cause many changes to most of the memory functions
involving Grb2, but some memory functions including
$\tilde{M}_{\text{RP},\text{Grb}}$ have large differences. The change
in $\tilde{M}_{\text{RP},\text{Grb}}$ occurs because removing the reaction between Grb2 and RShP
 means that phosphorylated EGFR (RP) and Grb2 only share interactions through phosphorylated Shc (ShP) and ShP-Grb2 (ShG). On the other hand, the channel decomposition shows that ``in via Shc and out via RP-Shc-Grb2 (RShG)'' and ``in via phosphorylated Shc (ShP) and out via RP-Shc-Grb2 (RShG)'' are
the dominant reactions in this memory function and therefore
removing this connection
between phosphorylated EGFR (RP) and Grb2 will have a large effect on the memory behaviour. Therefore because all the channels are connected one cannot necessarily remove channels that look weak in one memory function, as this will generally have an effect on the other species.

\begin{figure}[!ht]
  \centering
  \includegraphics[scale=0.8]{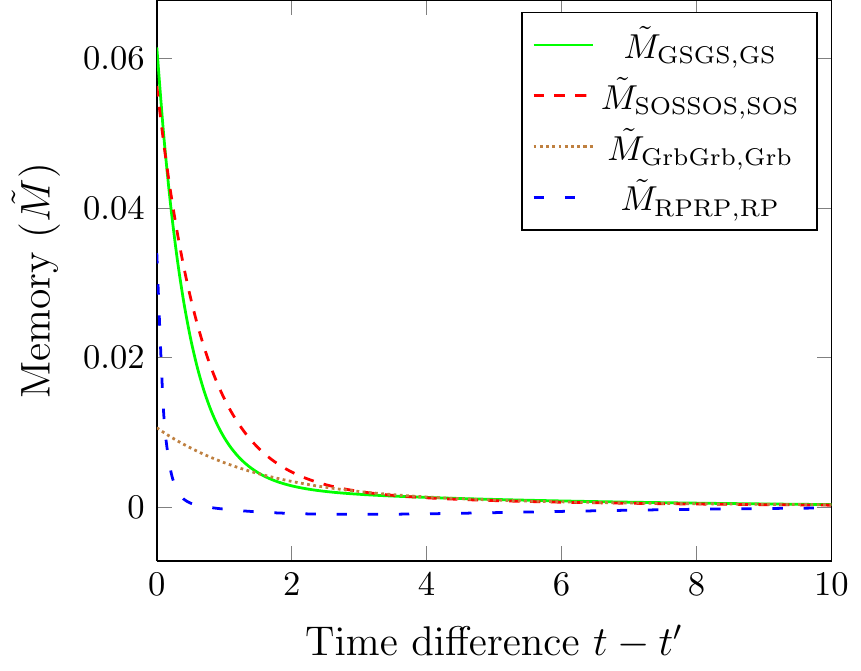}
  \caption{The nonlinear self memory functions of Grb2-SOS (GS), SOS, Grb2 and phosphorylated EGFR (RP).}
  \label{fig:nonlinselfmem}
\end{figure}

One benefit of our analysis is that we can characterise explicitly also the nonlinear memory functions, and in particular assess the relative size of their contribution compared to the linear memory terms.  
Figure \ref{fig:nonlinselfmem} shows the nonlinear self memory functions of the boundary species and
Table \ref{tb:ampsandtimes} lists the amplitudes and timescales of the linear and nonlinear self memory functions. It is easy to see that the nonlinear self memory amplitudes are all smaller than their respective linear amplitudes. Similarly the nonlinear self memory functions decay faster than their respective linear contributions as shown by their shorter timescales. This suggests that, where it is desirable to capture nonlinear memory terms only approximately, relatively simple approximations like short-timescale exponentials could be considered. The nonlinear self memory of RP is a special case: the memory function changes sign (see Fig.~\ref{fig:nonlinselfmem}) and the positive and negative contributions to the integral defining the timescale (see Sec.~\ref{sec:timescales_channels}) cancel almost exactly, giving a notional timescale that is much shorter than for the other boundary species.

There are at least two ways one could use information from the memory functions to estimate the values of the dimensionless concentrations where nonlinearities become important. Concentrating on the self memory functions as above, the linear and nonlinear instantaneous (small $\dt$) contributions become comparable when $\tilde{M}^\superss(0)\delta\tilde{x}_s=\tilde{M}^\supersss(0)\delta\tilde{x}_s^2$, leading to the estimate $\delta\tilde{x}_s^{\rm c,1}=\tilde{M}^\superss(0)/\tilde{M}^\supersss(0)$ for the size of the linear regime. More relevant for the long-time dynamics is to consider the total memory terms assuming constant $\delta\tilde{x}_s(t)$. 
Then $\delta\tilde{x}_s^{\rm c,2}=\int_0^{\infty} dt\,\tilde{M}^\superss(t)/\int_0^\infty dt\,\tilde{M}^\supersss(t)$ would delimit the extent of the linear regime, i.e.\ the ratio of amplitude times timescale for the linear and nonlinear self memory.

The two estimates $\delta\tilde{x}_s^{\rm c,1}$ and $\delta\tilde{x}_s^{\rm c,2}$ defined above are shown in Table~\ref{tb:ampsandtimes} alongside the memory amplitudes and timescales. 
We see that e.g.\ for Grb2-SOS (GS) $\delta\tilde{x}^{\rm c,2} = 29.92$ whereas for Grb2 $\delta\tilde{x}^{\rm c,2} = 9.37$; this suggests that for Grb2 nonlinear memory functions have a larger effect. To confirm this, we have run numerical experiments on time courses starting in steady state except for a perturbation in one of the four boundary species (see Table~\ref{tb:ampsandtimes}). We then compared the time courses for this species as predicted from the projected equations with and without the nonlinear memory terms, respectively, measuring the deviation between them as in Eq.~(\ref{eq:time_error}) below.  We find that these deviations are ordered among the four species in inverse proportion to their $\delta\tilde{x}^{\rm c,2}$, i.e.\ the larger this measure of the size of the linear regime, the smaller the nonlinear memory effects. Among the three species other than phosphorylated EGFR (RP) we also find quantitatively that deviations observed for an initial perturbation $\delta\tilde{x}_s(0)$ chosen as some constant fraction (say 1\%) of $\delta\tilde{x}^{\rm c,2}_s$ are of similar magnitude. This indicates that values of $\delta\tilde{x}^{\rm c,2}$ can give not just qualitative but also quantitative information. Our tests show it to be superior to $\delta\tilde{x}^{\rm c,1}$ in this regard.
For phosphorylated EGFR (RP), where because of the small notional nonlinear self memory timescale the value of $\delta\tilde{x}_s^{\rm c,2}$ is unrealistically large, we find that it still retains qualitative significance: the deviations that we measure due to the omission of the nonlinear memory terms are the smallest (by a factor of 100 compared to the next largest) among the four species tested.



\

\begin{table}[ht]
\centering
\begin{tabular}{|c|l|l|l|l|l|l|}
  \hline
&\multicolumn{2}{|c|}{linear}&\multicolumn{2}{|c|}{nonlinear}&\multirow{2}{*}{$\delta\tilde{x}_s^{\rm c,1}$}&\multirow{2}{*}{$\delta\tilde{x}_s^{\rm c,2}$}\\
\cline{2-5}
&amp.&$\tau$&amp.&$\tau$&&\\
\hline
RP&0.13&1.68&0.034&$-2.16\cdot10^{-6}$&3.82&2.97$\cdot10^{6}$\\
Grb2&0.036&5.31&0.01&2.04&3.6&9.37\\
SOS&0.12&4.61&0.056&0.82&2.14&12.05\\
GS&0.23&5.23&0.06&0.67&3.83&29.92\\
\hline
\end{tabular}
\caption{The amplitudes (in s$^{-2}$) and timescales (in s) of linear and nonlinear self memory functions. Also shown are the estimates for the size of the linear regime resulting from instantaneous and long-time memory contributions, $\delta\tilde{x}_s^{\rm c,1}$ and  $\delta\tilde{x}_s^{\rm c,2}$, respectively.
}
\label{tb:ampsandtimes}
\end{table}

\subsection{Changing the subnetwork}
\label{sec:subchange}

\begin{figure}[!ht]
  \centering
 \includegraphics[scale=0.8]{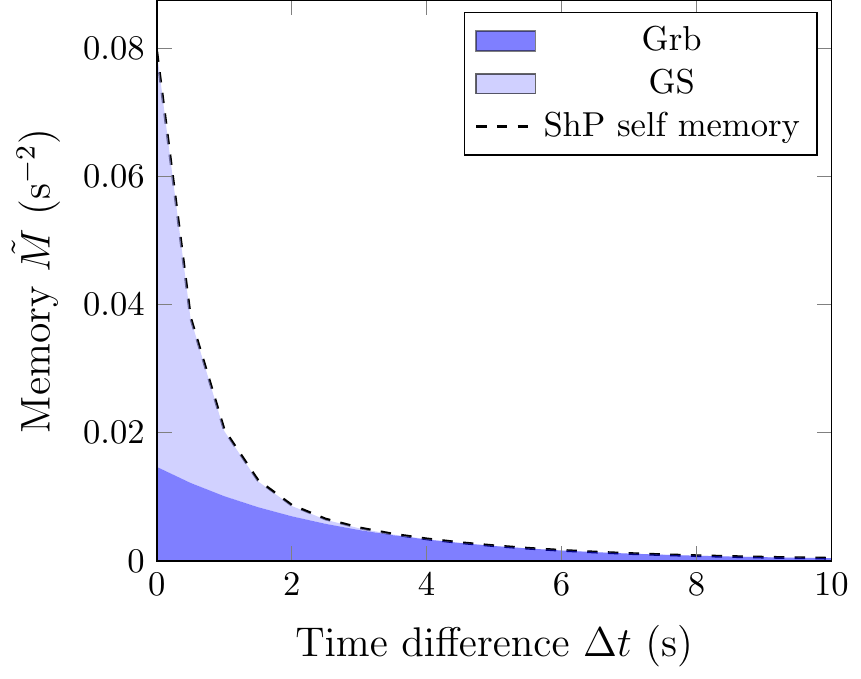}
 \caption{Comparison of self memory function of phosphorylated Shc (ShP) with dominant terms
   from the channel decomposition: in and out via Grb2; and in and out via
   Grb2-SOS (GS), shown
   together. 
 }
  \label{fig:shpaccumulatemem}
\end{figure}

If we take the EGFR network and choose a different subnetwork then we
will have a different set of boundary nodes and memory
functions. Although the memory functions will be quantitatively
different, their behaviour will still adhere to the general principles derived above. Let us take the EGFR network and change the subnetwork so that the bulk consists of Grb2 and all complexes that include Grb2. The boundary species of the subnetwork are now phosphorylated EGFR (RP), SOS, RShP and phosphorylated Shc (ShP). Figure~\ref{fig:shpaccumulatemem} shows the linear self memory function of phosphorylated Shc (ShP) in the equation for ShP. ShP has two reactions with bulk species 
\begin{equation}
\begin{split}
  \text{ShP} + \text{Grb2} &\leftrightharpoons \text{ShG}\\
  \text{ShP} + \text{GS} &\leftrightharpoons \text{ShGS}
\end{split}
\end{equation}
As with the self memory function of Grb2 in Sec.~\ref{sec:memproperties} each reaction gives a separate positive contribution to
the self-memory amplitude.

\

\begin{figure}[ht!]
  \centering
 \includegraphics[scale=0.8]{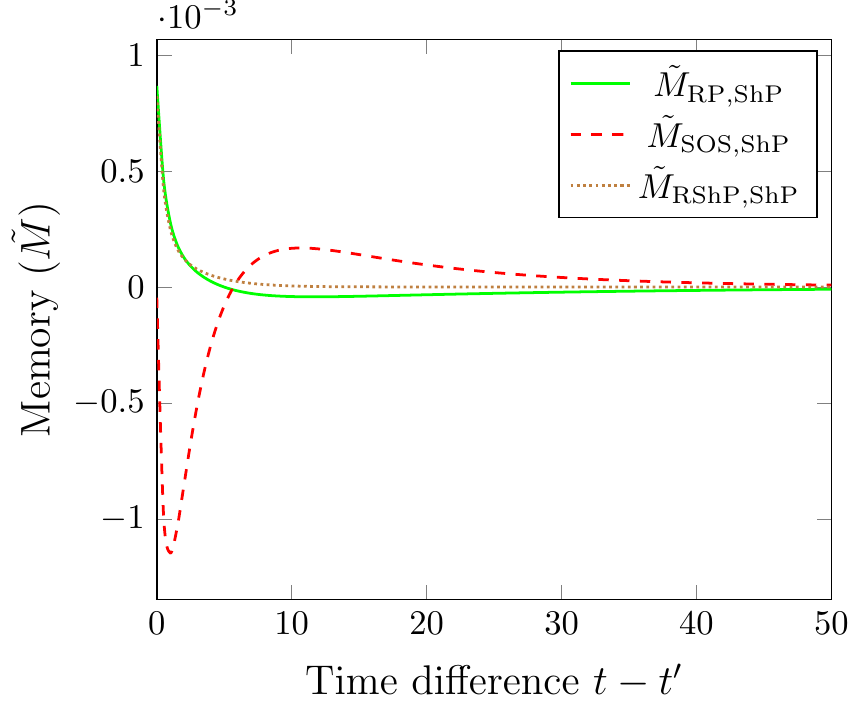}
  \caption{Memory functions in EGFR network: coefficients of
    phosphorylated EGFR (RP), SOS and RShP in the linear memory term for phosphorylated Shc (ShP). 
}
  \label{fig:shpcrossmem}
\end{figure}

Next we consider the cross-species effects in the memory term for the evolution of ShP. The amplitude of the memory to past values of $\protnondim{\text{SOS}}$ is negative, whereas the amplitudes for memory to $\protnondim{\text{RP}}$ and $\protnondim{\text{RShP}}$ is positive, as shown in Fig.~\ref{fig:shpcrossmem}. This last amplitude is made up of two reaction patterns with the structure of Fig.~\ref{fig:crossa} and hence has to be positive as we find. The cross memory functions of SOS and phosphorylated EGFR (RP) consist of a mixture of different reaction patterns from Fig.~\ref{fig:crossterms}, and the sign of their amplitude is therefore determined by the relative sizes of the contributions of different signs. The amplitude of the cross memory function of SOS is negative, whereas the amplitude of the cross memory function of phosphorylated EGFR (RP) is positive but approximately ten times larger in size than the amplitude of the memory function for SOS.

Also for the current changed subnetwork one can decompose the self memory functions into channels to understand the relative important of the latter. Fig.~\ref{fig:shpaccumulatemem} shows that the two channels ``out and in via Grb2'' and ``out and in via Grb2-SOS (GS)'' both give similar contributions to the memory function of ShP. This is in line with the earlier analysis in Sec.~\ref{sec:memproperties}: the $\kp{s}{b}{c}\y{b}$ values for both interactions are of the same order and therefore one would expect that neither channel will dominate the other.

Looking finally at the memory function timescales, 
the self memory function of phosphorylated Shc (ShP) has a timescale of $\timescale{\text{ShP}}=0.97$s. The ``in and out via Grb2'' and ``in and out via Grb2-SOS (GS)'' channels have timescales of 2.84s and 0.54s, respectively, and one sees that these can again be used to give an order of magnitude estimate of the full memory function timescale.

\subsection{Quantitative tests}
\label{sec:quanttests}
We conclude our discussion of the EGFR network by analysing the
quantitative accuracy of the projected equations. As before we focus
on the limit of low copy number noise ($\epsilon\to 0$) and drop the
random force terms to have a closed description of the subnetwork
dynamics. Our baseline is the solution of the full set of reaction
equations for the entire network, consisting of both subnetwork and
bulk. We compare the performance of the projected equations, including
memory terms, to two simpler approximations without memory. In the
first one we treat the subnetwork as isolated, i.e.\ all subnetwork-bulk
reactions are ignored. In the second one we assume the bulk dynamics
is fast enough for the bulk to be in steady state with respect to the
specific subnetwork concentrations at any given time \cite{Sunnaker2011}. 
In practice, this means we solve
the steady state conditions for the bulk concentrations at every time
step and substitute them into the evolution equations for the subnetwork.
All three approximation methods (projected equations without random
force, isolated subnetwork, steady state bulk) come in two versions,
one derived from the linearised dynamics and one for the full
nonlinear dynamics.

 \begin{figure}[ht!]
   \centering
\includegraphics[scale=0.8]{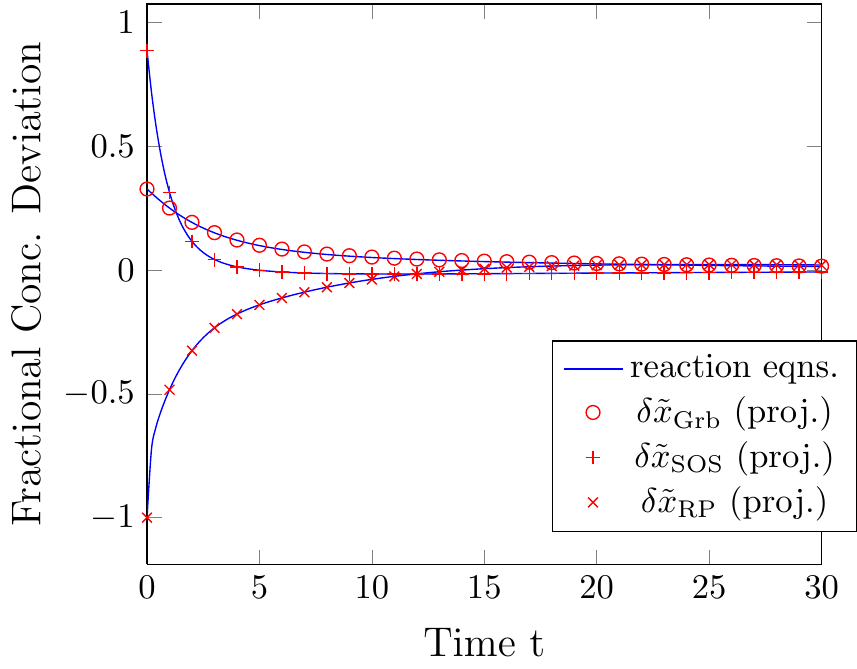}
\caption{Plots of time courses of some selected molecular species from the EGFR
  network. The fractional concentration
deviations (\ref{eq:nondimdx}) are defined so that 0 represents steady state
concentration, which is approached for long time, and $-1$ represents
zero concentration. The solutions to the nonlinear projected equations are
  visually indistinguishable from those of the full reaction
  equations. Initial conditions were chosen
  as explained in the text.}
\label{fig:egfrsoln}
 \end{figure}

Figure~\ref{fig:egfrsoln}
compares the solutions of the nonlinear projected equations to the
baseline, the nonlinear reaction equations for the entire
network. Time courses for phosphorylated EGFR (RP),
Grb2 and SOS are shown. Here and in the following, the
subnetwork initial conditions were chosen to maximise nonlinear
effects: specifically we maximised 
$\sum_s [\protnondim{s}(0)]^2$ subject to the constraint that all
conserved concentrations have the same value as at the steady state
given by $\protnondim{i}=0$, and of course that all concentrations are
non-negative ($\protnondim{s}(0)\geq -1$). The bulk was assumed to be
in steady 
state initially, i.e.\ we set $\protnondim{b}(0)$ for all bulk species.
As the figure shows, the agreement between the nonlinear
projected equations and the full dynamics is excellent, with the two
sets of time courses being visually indistinguishable. 


\begin{figure}[ht!]
  \centering
 \includegraphics[scale=0.8]{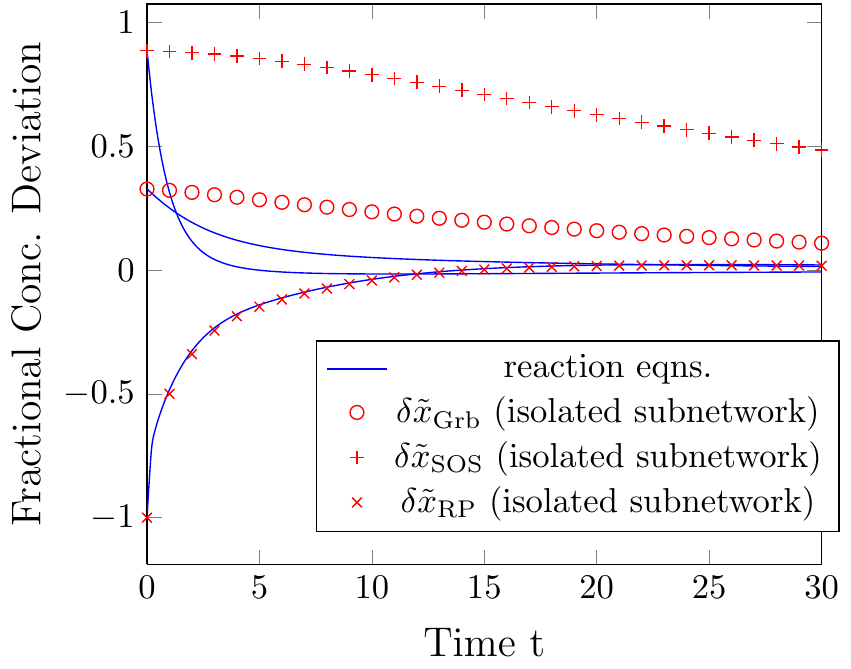}
 \caption{Comparison of time courses obtained from full reaction equations
   with time courses for isolated subnetwork. Initial conditions chosen
   as in Fig.~\ref{fig:egfrsoln}. 
 } 
  \label{fig:subsoln}
\end{figure}

To demonstrate the importance of accounting for the interactions of
the subnetwork with the bulk, we contrast in Fig.~\ref{fig:subsoln}
the solutions for the isolated subnetwork to those of the full
reaction equations: substantial differences appear, with the
relaxation to steady state predicted to occur over a much larger
timescale than in the full description.



\begin{figure}[ht!]
  \centering
 \includegraphics[scale=0.8]{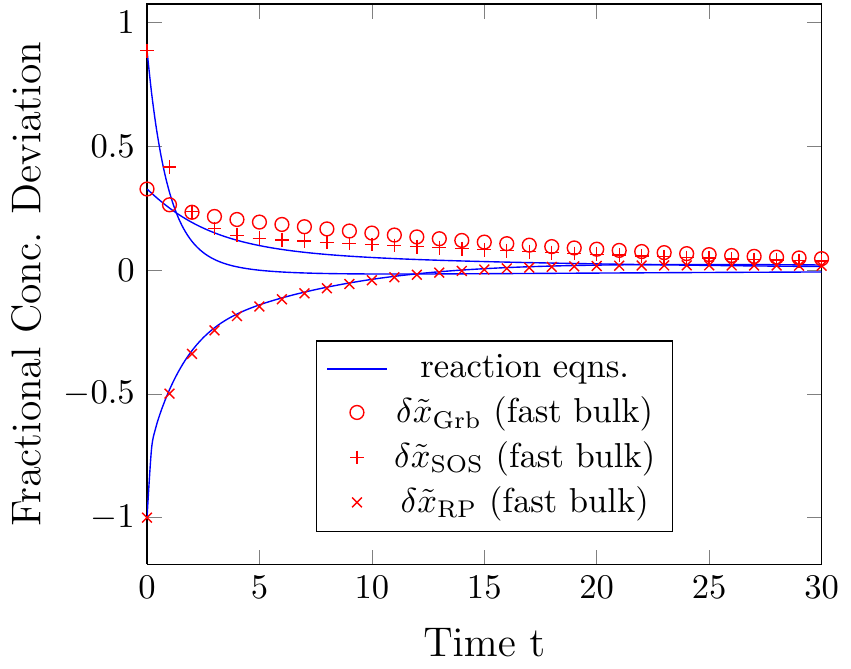}
 \caption{ Comparison of time courses obtained from full reaction
   equations with time courses found by assuming the bulk dynamics
   is fast enough for the bulk to be at steady state. Initial conditions chosen as in Fig.~\ref{fig:egfrsoln}.}
  \label{fig:bulkqss}
\end{figure}

The approximation of retaining information on the bulk network but
assuming the bulk dynamics is fast would be expected to provide a more
accurate description. This is borne out by
Fig.~\ref{fig:bulkqss}, 
though deviations from the baseline are still larger than for the
nonlinear projected equations, emphasizing the importance of keeping
track of memory effects. 

\

To develop a more quantitative picture of the performance of the
various approximations for the subnetwork dynamics, we consider the
same initial conditions as above but now scale down all the
$\protnondim{s}(0)$ by a constant factor to tune the
initial deviation of the subnetwork from steady state. The magnitude
of this deviation
will be quantified via the initial root mean squared deviation,
$\del=\{\sum_s [\protnondim{s}(0)]^2\}/\Ns$, where $\Ns$ is the number
of subnetwork species as before. The accuracy of any approximation
$\delta\hat{x}_s(t)$ for the subnetwork time courses will be measured by
\begin{equation}
  \timeerror = \frac{1}{T}\int_0^Tdt'\frac{1}{\Ns}  \sum_{s=1}^{\Ns}
  \left|\protnondim{s}(t)-\delta\hat{x}_s(t)\right|
\label{eq:time_error}
\end{equation}
This is the absolute deviation in the dimensionless
concentration of each subnetwork species, averaged over species and
also a time interval $T$ that we choose as $T=150$s 
to capture the
interesting transient regime, i.e.\ the approach to the steady state.

\begin{figure}[ht!]
  \centering
 \includegraphics[scale=0.8]{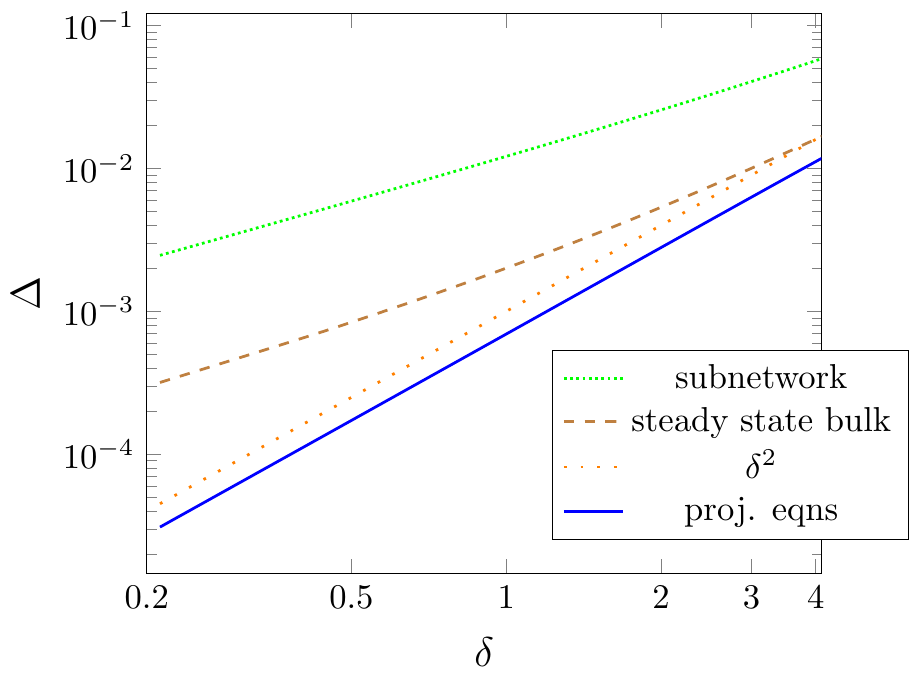}
 \caption{Plot of approximation error vs initial deviation from steady
   state in log-log representation, for three approximation methods derived
   from the linearised 
   dynamics: linear projected equations, i.e.\ incorporating memory
   terms; steady state bulk approximation, i.e.\ without memory; and
   isolated subnetwork approximation. Dotted line is proportional to
   $\del^2$ to demonstrate that the approximation error of the linear
   projected equations is only quadratic in $\del$. 
}
  \label{fig:linerror}
\end{figure}

If we now consider first the linear projected equations and compare them to
the full nonlinear reaction equations, we would expect the error to
increase quadratically with the size $\del$ of the initial deviations
from steady state, at least for small $\del$, because we are missing
the nonlinear terms but are 
correctly capturing all linear terms including the memory.
Fig.~\ref{fig:linerror} verifies this expectation, showing that the
average deviation $\timeerror$ grows only as $\del^2$.
By contrast, the simpler approximations derived from the linearised
dynamics, where we treat the subnetwork as isolated or the bulk as
fast, should show deviations from the true time courses already at
order $\del$ because they neglect linear memory
terms. Fig.~\ref{fig:linerror} is consistent with this. It
demonstrates in addition that
the errors made by the memoryless approximations are substantially larger in
absolute terms than for the 
projected equations. This demonstrates
that memory terms are essential even to describe the linearised
dynamics correctly. 

\begin{figure}[ht!]
  \centering
 \includegraphics[scale=0.8]{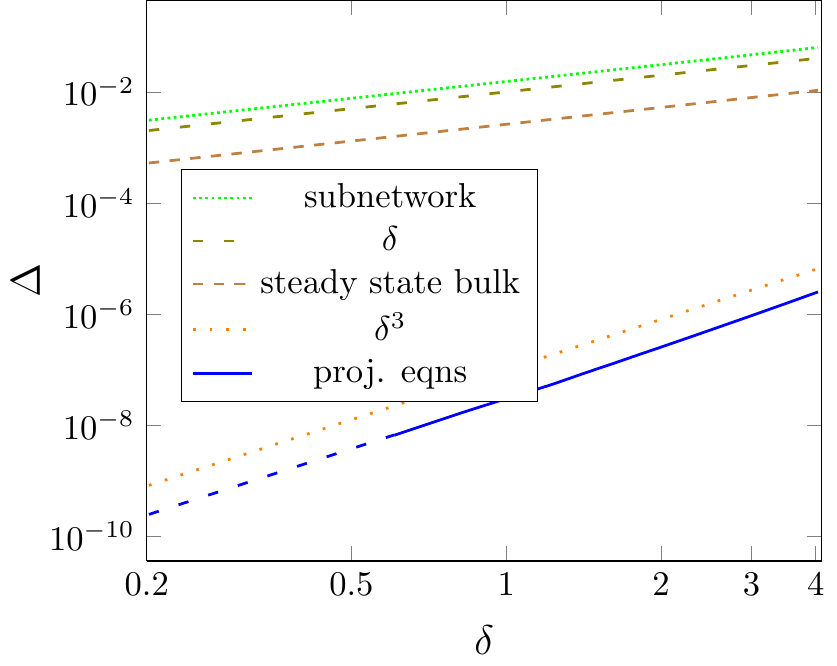}
 \caption{Plot of approximation error vs initial deviation from steady
   state in log-log representation, for three approximation methods derived
   from the nonlinear
   dynamics: nonlinear projected equations, i.e.\ incorporating memory
   terms; steady state bulk approximation, i.e.\ without memory; and
   isolated subnetwork approximation. Dotted line is proportional to
   $\del^3$ to demonstrate that the approximation error of the nonlinear
   projected equations grows only cubically in $\del$. The dashed blue line indicates the estimated size of the error in the projected equations in the region where it is too small for the given initial conditions to calculate accurately.
 }
  \label{fig:nonlinerror}
\end{figure}

In Fig.~\ref{fig:nonlinerror} we show an analogous comparison for the
nonlinear approximation methods. The memoryless approximations,
isolated subnetwork and steady state bulk, still fail to catch
memory contributions that are present already in the linearised
dynamics, and accordingly give an approximation error that grows
linearly in $\del$. 
For the nonlinear projected equations, on the other hand, the
approximation error comes only from the neglected random force terms.
As explained in Section~\ref{sec:nonlindynamics}, when the bulk is
initially in steady state then the random force will scale cubically
with the initial deviations of the subnetwork from its steady
state. One therefore expects an approximation error that is only of
order $\del^3$, and the results shown in Fig.~\ref{fig:nonlinerror}
are consistent with this. Importantly, the figure shows also that the
approximation error is very much smaller in absolute terms, by
four orders of magnitude at the largest initial deviations from steady
state and more for smaller $\del$. We regard this as conclusive
evidence that a quantitative description of subnetwork dynamics must
include memory terms. The significant reduction in error over the
linear projected equations, cf.\ Fig.~\ref{fig:linerror}, also
emphasises that for quantitative accuracy nonlinearities in the memory
have to be accounted for. By contrast, the memoryless approximations
are hardly improved by the inclusion of nonlinear terms, which tells
us that memory effects are crucial to get right first.

\section{Discussion}


\

We considered the problem of finding reduced descriptions for the
dynamics of biological subnetworks embedded in a larger bulk
environment. 
As an example we studied a subnetwork, embedded in the Shc-centred bulk, of an EGFR signalling model. Large-scale screens have identified collections of network components: interacting proteins \cite{Kerrien2012} or genes regulated in response to a perturbation or disease state \cite{Culhane2012, Sotiriou2009}, for example. The emergence of high-throughput functional imaging screens is providing further insights into potential signalling components, with direct or indirect consequences for signalling regulation and its alteration in disease states \cite{Fruhwirth2011, Carlin2011}. In the absence of mechanistic details, however, there remains the problem of understanding the dynamics of a small subnetwork of interest in the presence of a surrounding network.
While it is widely acknowledged that the presence of a
surrounding bulk systems generates {\em extrinsic noise} on the
subnetwork dynamics \cite{Swain2002, Paulsson2004}, 
we showed that in addition one
has {\em memory effects} whereby the state of the subnetwork
in the past affects its time evolution in the present. We analysed
these memory effects for a broad class of protein interaction networks
containing unary and binary reactions, but argued that they should be
much more generic.

Mathematically, our approach employed the projection method. This
allows one to obtain a set of dynamical equations for the
concentration of molecular species in a chosen subnetwork that forms
part of a larger protein interaction network. These projected
equations are closed, provided one neglects so-called random noise
terms that contain the extrinsic noise as well as contributions from
intrinsic noise. For the linearised dynamics, the projection method
gives results that are fully consistent with an explicit elimination
of the bulk variables. Non-trivially, we were able to apply the
projection method also to the full nonlinear dynamics, where in the
limit of low copy number noise we found explicit formulas for the
memory functions. These memory functions provide the weights, as a
function of the time difference, with which past subnetwork states
affect the present time evolution. We showed that they can be
calculated from appropriate matrix representations of the dynamical
(Fokker-Planck) and projection operators.

We analysed in some detail the properties of the linear and nonlinear
memory functions, including their amplitudes and timescales.
These provide insights into how the subnetwork interacts with the
bulk, with e.g.\ negative memory amplitudes requiring the existence of
ternary subnetwork-bulk-bulk or subnetwork-subnetwork-bulk complexes.

In the final results section we applied the projection method to the
EGFR network of \citet{Kholodenko99}. 
Here we illustrated how memory function amplitudes relate to
subnetwork-bulk interaction structures. To understand memory function
timescales, we used a channel decomposition. This is based on the fact
that memory is generated by the past subnetwork state first affecting
the bulk and then feeding back to the subnetwork at some later
time. We showed that accordingly each memory function can be viewed as
a sum of contributions from the different ``source'' and ``receiver''
channels in this feedback process. This allows one to identify which
channels dominate the memory effects. The interpretation of each
dominant channel can now be explored by designing experimentally
tractable studies to interrogate parts of the subnetwork.

We also gave a quantitative comparison of the accuracy of the
projected equations versus simpler memoryless approximations. This
showed that including nonlinear terms in the memoryless
approximations does little to reduce approximation error, because the
main error comes from neglecting linear memory terms. The nonlinear
projected equations, on the other hand, were significantly more
accurate than other approximations, by at least four orders of
magnitude compared to the nearest memoryless competitor.

While our analysis was focussed on memory functions, the projection
approach can also capture noise effects, via the random force
terms. In the low copy number noise limit, these terms represent
directly the extrinsic noise on the subnetwork. We plan to report
separately on the statistical properties of this extrinsic noise. The
advantage of our approach here is that we can derive the extrinsic
noise statistics, including e.g.\ temporal correlations, from the
statistics of the initial states of the bulk, rather than having to
postulate them separately. This will make it possible to test
assumptions about extrinsic noise that have been made in the extensive
literature on this subject, see e.g. \cite{Shahrezaei2008a}.


The single-peaked steady state distribution $P_{\rm ss}$ that we considered means we
do not explicitly treat systems with limit cycles or multiple states.
However, if we allow more general distributions for $P_{\rm ss}$ we would
be able to consider systems with these properties. The quantitative nature of the random force would be affected by this as ``randomness'' is always measured with respect to the assumed $P_{\rm ss}$.

Another area we are currently exploring is perturbations of the
protein interaction dynamics, e.g.\ via gene regulation. Here one can
ask, for example, what effect the perturbation of a bulk species has
on the dynamics of the subnetwork \cite{Rubin2014}. This will allow us
to study how regulation-driven changes in the concentration of bulk
species such as partial or complete knockdowns affect the subnetwork.

In future work we also hope to address the inverse problem of how
subnetwork dynamics can be used to infer information about the
structure of an unknown bulk network. We saw that memory terms appear
only in the time evolution of molecular species that lie on the
boundary of the subnetwork, in the sense that they have direct
interactions with the bulk. This can in principle be turned around
immediately, so that by measuring memory functions from dynamics
\cite{Uranagase2010} 
one could identify which molecular species
in a given subnetwork form its boundary. The next level would then
involve identifying the network structure of the reactions with the
bulk that these boundary species take part in. This would be a first step towards understanding the role of the many components which may be contained in a surrounding network, such as those identified by high-throughput functional screens \cite{Fruhwirth2011}, in coordinating a signalling response.

\


\




\section*{Acknowledgements}

We gratefully acknowledge funding from the People Programme (Marie
Curie Actions) of the European Union's Seventh Framework Programme
FP7/2007--2013/ under REA grant agreement nr.\ 290038 (PS), a BBSRC
Quota Doctoral Training Grant (KJR),
KCL-UCL Comprehensive Cancer Imaging Centre funding (CR-UK and EPSRC: C1519/ A16463 and C1519/A10331) and an FP7-HEALTH-2010 EU grant ``Imagint'' under grant agreement nr.\ 259881 (KL)
and an endowment fund from Dimbleby Cancer Care to King's College London (TN).
We would like to acknowledge
helpful discussions with Franca Fraternali and James
Monypenny.

 \appendix

\section{Protein species Table}
 \label{appendix1}

\begin{table}[!ht]
\begin{tabular}{|l|p{10cm}|}
\hline  
EGF &Epidermal Growth Factor\\
R &Extracellular domain of the monomeric EGFR\\
R$_{\text{a}}$ &EGF-EGFR complex\\
R$_{2}$ &(EGF-EGFR)/(EGF-EGFR) dimer also known as Ra:Ra dimer\\
RP&Tyrosine phosphorylated EGFR\\
PLC$\gamma$&Phospholipase C$\gamma$\\
Grb&Growth factor receptor-binding protein 2 (Grb2)\\
Shc&Src homology and collagen domain protein\\
SOS&Son of Sevenless homolog protein\\
E1&Enzyme for MM reaction between R2 and RP\\
ER&Enzyme complex for MM reaction between R2 and RP\\
E2&Enzyme for MM reaction between PLC$\gamma$ and PLC$\gamma$P\\
EP&Enzyme complex for MM reaction between PLC$\gamma$ and PLC$\gamma$P\\
E3&Enzyme for MM reaction between Shc and ShP\\
ES&Enzyme complex for MM reaction between Shc and ShP\\
RPL&RP-PLC$\gamma$\\
RPLP&RP-PLC$\gamma$P/ Phosphorylated RPL\\
RG&RP-Grb2\\
RGS&RP-Grb2-SOS\\
RSh&RP-Shc\\
RShP&RP-ShP/ Phosphorylated RSh\\
RShG&RP-Shc-Grb2\\
RShGS& RP-Shc-Grb2-SOS\\
GS& Grb2-SOS\\
ShP& Phosphorylated Shc\\
ShG&ShP-Grb2\\
ShGS&Shc-Grb2-SOS\\
PLC$\gamma$P& Phosphorylated PLC$\gamma$\\
PLC$\gamma$PI&PLC$\gamma$P translocated to membrane structures\\
\hline
\end{tabular}
\caption{Abbreviations used for EGFR network components following \citet{Kholodenko99}, including a description of each enzyme added to the system to account for the Michaelis Menten (MM) reactions. 
}
\end{table}




\bibliographystyle{model1-num-names}
\bibliography{memorypaper0514}






\end{document}